\documentclass[a4paper,10pt]{article}

\usepackage{authblk}
\usepackage{ifthen,ifpdf}

\ifpdf
  \usepackage{hyperref}
  \hypersetup{colorlinks=false,allcolors=blue}
  \usepackage{hypcap}
  \pdfpagewidth=\paperwidth
  \pdfpageheight=\paperheight
\fi

\usepackage[utf8]{inputenc}
\usepackage[T1]{fontenc}
\usepackage{amsmath, amsthm, amssymb}
\usepackage{mathtools}
\usepackage{enumerate}
\usepackage{afterpage}
\usepackage{tabularx}
\usepackage{dsfont}
\usepackage[table]{xcolor}

\usepackage{graphicx}

\usepackage{algorithm}
\usepackage{algpseudocode}
\usepackage{caption}
\DeclareCaptionLabelFormat{andtable}{#1~#2  \&  \tablename~\thetable}
\usepackage[labelformat=simple]{subcaption}

\usepackage{color}
\usepackage{units}
\usepackage{array, multirow}
\usepackage{booktabs}

\newcolumntype{M}[1]{>{\vspace{3pt}\raggedleft\arraybackslash}m{#1}}

\usepackage{siunitx}
\usepackage{tikz}
\usepackage{pgfplots}
\usepackage{pgfplotstable}
\usepgfplotslibrary{units}
\pgfplotsset{compat=1.15}
\usepackage{enumitem}
\usepackage{cite}
\usepackage{numprint}

\usepackage{anysize} 
\marginsize{2.5cm}{2.5cm}{2cm}{2cm}
\usepackage{wrapfig}
\usepackage{xspace}

\theoremstyle{plain}

\theoremstyle{remark}

\numberwithin{equation}{section}

\newcommand{\set}[1]{\left\{#1\right\}} 

\newcommand{\abq}{ABAQUS\xspace}

\newcommand{\norm}[1]{\Vert #1 \Vert}
\newcommand{\normg}[1]{\left\Vert \, #1 \, \right\Vert}

\newcommand{\tr}[1]{{\rm tr  }\left( #1 \right)}
\newcommand{\diag}[1]{{\rm diag  }\left( #1 \right)}

\renewcommand{\unit}[1]{~\rm #1}

\newcommand{\sop} {{{SO}(3)}}
\newcommand{\Sym}[1]{{\rm Sym }( #1 )}
\newcommand{\Dev}[1]{{\rm Dev }( #1 )}
\newcommand{\Sph}[1]{{\rm Sph }( #1 )}

\newcommand{\sty}[1]{\mbox{\boldmath $#1$}}
\newcommand{\styy}[1]{{\mathbb{#1}}}

\definecolorset{rgb}{KIT}{}{%
   green, 0.0, 0.58, 0.49;%
   blue, 0.0, 0.34, 0.62;%
   lead, 0.44, 0.44, 0.45;%
  darkgray, 0.53, 0.53, 0.54;%
  gray, 0.89, 0.89, 0.90;%
  lightgray, 0.93, 0.93, 0.93}
\definecolor{KITgreen}  {RGB}{  0,150,130} 
\definecolor{KITgreen70}{RGB}{ 76,181,167} 
\definecolor{KITgreen50}{RGB}{127,202,192} 
\definecolor{KITgreen30}{RGB}{178,223,217} 
\definecolor{KITgreen15}{RGB}{217,239,236} 
\definecolor{KITblue}  {RGB}{ 70,100,170} 
\definecolor{KITblue70}{RGB}{125,146,195} 
\definecolor{KITblue50}{RGB}{162,177,212} 
\definecolor{KITblue30}{RGB}{199,208,229} 
\definecolor{KITblue15}{RGB}{227,232,242} 
\definecolor{KITblack}  {RGB}{  0,  0,  0} 
\definecolor{KITblack70}{RGB}{ 77, 77, 77} 
\definecolor{KITblack50}{RGB}{128,128,128} 
\definecolor{KITblack30}{RGB}{179,179,179} 
\definecolor{KITblack15}{RGB}{217,217,217} 
\definecolor{KITpalegreen}{RGB}{130,190,60}
\colorlet{KITpalegreen70}{KITpalegreen!70}
\colorlet{KITpalegreen50}{KITpalegreen!50}
\colorlet{KITpalegreen30}{KITpalegreen!30}
\colorlet{KITpalegreen15}{KITpalegreen!15}
\definecolor{KITyellow}{RGB}{250,230,20}
\colorlet{KITyellow70}{KITyellow!70}
\colorlet{KITyellow50}{KITyellow!50}
\colorlet{KITyellow30}{KITyellow!30}
\colorlet{KITyellow15}{KITyellow!15}
\definecolor{KITorange}{RGB}{220,160,30}
\colorlet{KITorange70}{KITorange!70}
\colorlet{KITorange50}{KITorange!50}
\colorlet{KITorange30}{KITorange!30}
\colorlet{KITorange15}{KITorange!15}
\definecolor{KITbrown}{RGB}{160,130,50}
\colorlet{KITbrown70}{KITbrown!70}
\colorlet{KITbrown50}{KITbrown!50}
\colorlet{KITbrown30}{KITbrown!30}
\colorlet{KITbrown15}{KITbrown!15}
\definecolor{KITred}{RGB}{160,30,40}
\colorlet{KITred70}{KITred!70}
\colorlet{KITred50}{KITred!50}
\colorlet{KITred30}{KITred!30}
\colorlet{KITred15}{KITred!15}
\definecolor{KITlilac}{RGB}{160,0,120}
\colorlet{KITlilac70}{KITlilac!70}
\colorlet{KITlilac50}{KITlilac!50}
\colorlet{KITlilac30}{KITlilac!30}
\colorlet{KITlilac15}{KITlilac!15}
\definecolor{KITcyan}{RGB}{80,170,230}
\colorlet{KITcyan70}{KITcyan!70}
\colorlet{KITcyan50}{KITcyan!50}
\colorlet{KITcyan30}{KITcyan!30}
\colorlet{KITcyan15}{KITcyan!15}

\definecolor{boxrahmen}{gray}{0.0}
\definecolor{boxhintergrund}{gray}{0.999}
\newsavebox{\tmpbox}

\newcommand{\fa}{\sty{ a}}

\newcommand{\fe}{\sty{ e}}

\newcommand{\fn}{\sty{ n}}

\newcommand{\fp}{\sty{ p}}

\newcommand{\fx}{\sty{ x}}

\newcommand{\fz}{\sty{ z}}
\newcommand{\fA}{\sty{ A}}

\newcommand{\fE}{\sty{ E}}

\newcommand{\fN}{\sty{ N}}

\newcommand{\fQ}{\sty{ Q}}

\newcommand{\fS}{\sty{ S}}

\newcommand{\fW}{\sty{ W}}

\newcommand{\ffC}{\styy{ C}}

\newcommand{\ffI}{\styy{ I}}

\newcommand{\ffN}{\styy{ N}}

\newcommand{\ffP}{\styy{ P}}

\newcommand{\ffR}{\styy{ R}}

\newcommand{\fsigma}{\mbox{\boldmath $\sigma$}}

\newcommand{\feps}{\mbox{\boldmath $\varepsilon $}}

\newcommand{\cG}{{\cal G}}

\newcommand{\effective}[1]{\overline{#1}}
\newcommand{\fmicrostrain}{\feps}
\newcommand{\fmacrostrain}{\fE}
\newcommand{\fmicrostress}{\fsigma}
\newcommand{\fmacrostress}{\effective{\fsigma}}

\newcommand{\macrostrain}{E}
\newcommand{\microstress}{\sigma}
\newcommand{\macrostress}{\effective{\sigma}}

\newcommand*\dif{\mathop{}\!\mathrm{d}} 
\newcommand*{\eg}{e.g.,\@\xspace}
\newcommand*{\ie}{i.e.,\@\xspace}
\newcommand*{\wrt}{w.r.t.\@\xspace}
\newcommand*{\cf}{see}

\newcommand{\DMN}[1]{\mathcal{DMN}_{#1}}
\newcommand{\DMNLIN}[1]{\mathcal{DMN}^{\mathcal{L}}_{#1}}
\newcommand{\GSM}[1][]{\mathcal{GSM}^{#1}}
\newcommand{\HOM}[1]{\mathcal{M}_{#1}}

\newcommand{\BB}{\mathcal{B}}

\newcommand{\condensedpotential}{\Psi}

\newcommand{\IDsym}{\mathbb{I}_\textrm{s}}

\newcommand{\freeenergy}{\psi}
\newcommand{\dissipation}{\phi}
\newcommand{\statev}{\fz}

\newcommand{\tuckersecond}{\fA_2}

\newcommand{\discone}{{D$4$}\xspace}
\newcommand{\disctwo}{{D$10$}\xspace}
\newcommand{\discthree}{{D$31$}\xspace}

\title{An FE-DMN method for the multiscale analysis\\ of fiber reinforced plastic components}

\author[1]{Sebastian Gajek}
\author[1]{Matti Schneider}
\author[1,*]{Thomas Böhlke}

\affil[1]{Karlsruhe Institute of Technology (KIT), Institute of Engineering Mechanics}
\affil[*]{correspondence to: \texttt{thomas.boehlke@kit.edu}}

\date{\today}

\begin{document}

\maketitle 

\begin{abstract}
\noindent In this work, we propose a fully coupled multiscale strategy for components made from short fiber reinforced composites, where each Gauss point of the macroscopic finite element model is equipped with a deep material network (DMN) which covers the different fiber orientation states varying within the component. These DMNs need to be identified by linear elastic precomputations on representative volume elements, and serve as high-fidelity surrogates for full-field simulations on microstructures with inelastic constituents.\\
We discuss how to extend direct DMNs to account for varying fiber orientation, and propose a simplified sampling strategy which significantly speeds up the training process. To enable concurrent multiscale simulations, evaluating the DMNs efficiently is crucial. We discuss dedicated techniques for exploiting sparsity and high-performance linear algebra modules, and demonstrate the power of the proposed approach on an industrial-scale three-dimensional component. Indeed, the DMN is capable of accelerating two-scale simulations significantly, providing possible speed-ups of several magnitudes.\\

\end{abstract}

\section{Introduction}

\subsection{Problem setting}

\begin{figure}
	\centering
	\begin{subfigure}[t]{0.47\textwidth}
		\includegraphics[width=\textwidth]{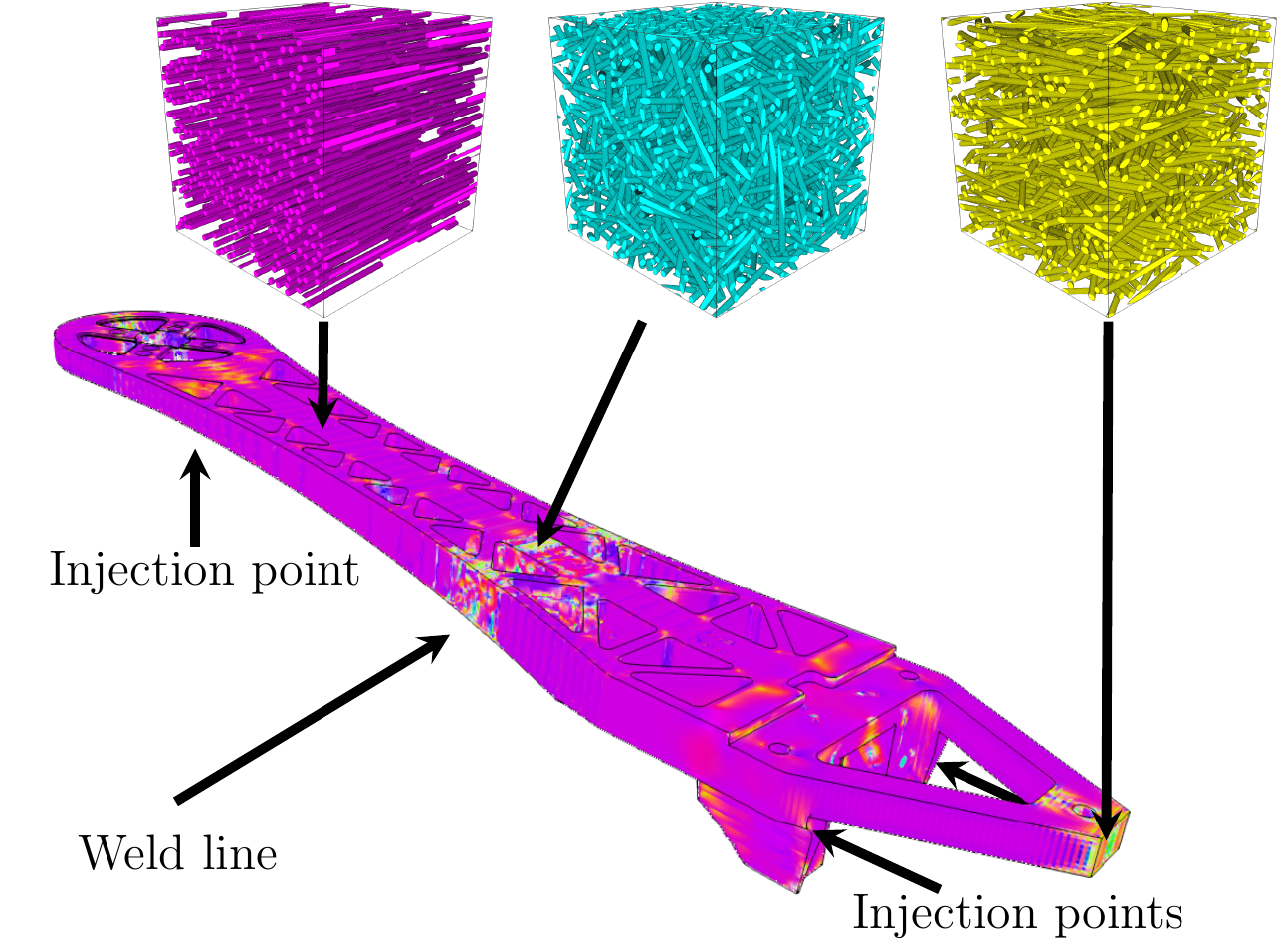}
		\caption{Injection molded quadcopter arm}
		\label{fig:motivation_fiber_orientation}
	\end{subfigure}
	\begin{subfigure}[t]{0.50\textwidth}
		\includegraphics[width=\textwidth]{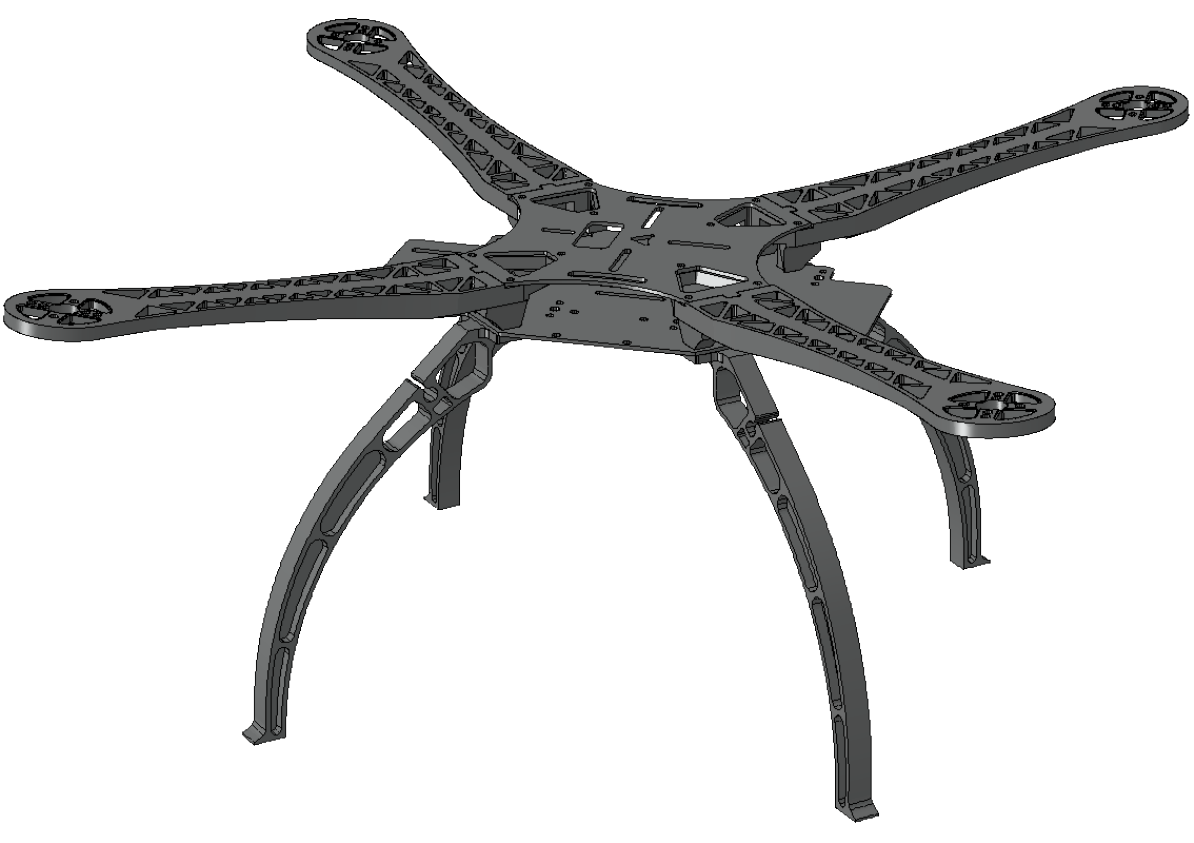}
		\caption{Quadcopter frame geometry \cite{droneFrame}}
		\label{fig:motivation_quadcopter}
	\end{subfigure}
	\caption{Benchmark component used in this work, with local fiber orientation}
\end{figure}

Injection molded short fiber reinforced components are frequently used for industrial applications as they combine favorable mechanical properties, free formability and short cycle times. As a result of the injection molding process, the fiber orientation and the fiber volume fraction may vary continuously within the component. Characterizing all possible orientation states for such materials is an arduous and expensive task, both experimentally and by simulative means.\\
To illustrate the complexity to be handled routinely, Fig.~\ref{fig:motivation_fiber_orientation} shows a quadcopter frame arm made from short fiber reinforced polyamide with a local fiber orientation determined by an injection molding simulation, \cf{} Section~\ref{sec:example} below for details. The color scale encodes the local fiber orientation tensor. Magenta corresponds to a unidirectional, cyan to an isotropic and yellow to a planar isotropic fiber orientation state, \cf{} Fig.~\ref{fig:orienation_triangle_microstructures}. We observe that, for the majority of the quadcopter arm, the fibers are almost aligned. Moreover, we encounter isotropic and planar isotropic fiber orientations in areas where weld lines have formed. As weld lines correspond to weak spots in the structure, it is critical to account for such regions accurately in mechanical simulations.

\subsection{State of the art}

Fig.~\ref{fig:motivation_quadcopter} represents an example for a component with a spatially varying complex microstructure. A monolithic finite element (FE) simulation of such a structure which \emph{resolves} the heterogeneities is not feasible with the current computational power. If the microscopic heterogeneities fluctuate on a scale that is much smaller than the size of the component, homogenization methods may be used for obtaining so-called effective material models which account for the physical mechanisms of complex and highly heterogeneous microstructures. Effective models emerge naturally by solving a partial differential equation, the cell problem, on a suitable microstructure. For linear material models, the effective properties may be precomputed and cached. In this way, it is possible to compute the mechanical response of microstructured components.\\
Treating inelastic materials is more difficult, as the internal variables live naturally on the microstructure, and cannot be "homogenized" to the macroscopic scale, in general. $\textrm{FE}^2$ methods, introduced by Renard-Marmonier~\cite{Renard1987} and subsequently refined~\cite{Smit1998,Feyel1999, Feyel2000, Feyel2003}, offer a solution by furnishing each Gauss point of the macroscopic finite element model with a microstructure on which the cell problem is solved, accounting for the evolution of the internal variables on the microscopic scale. To speed up the solution process on the microscale, FFT-based approaches~\cite{MoulinecSuquet1994,MoulinecSuquet1998} may be used, giving rise to the FE-FFT~\cite{Spahn2014,Kochmann2016, Kochmann2018} method in the concurrent multiscale context. Despite recent progress in computational efficiency and parallel computing strategies, concurrent multiscale methods with full-field models on the microscopic scale are typically too computationally demanding for industrial use.\\
To mitigate the computational burden of concurrent multiscale methods, the microscale problem is considered as a parametric partial differential equation which needs to be solved repeatedly (for slightly different input parameters), and model order reduction techniques may be utilized. Motivated by classical mean-field methods~\cite{Mori1973,Hill1965}, Dvorak and co-workers~\cite{TFADvorakBenveniste1992, TFA2, TFA3} introduced the transformation field analysis (TFA).  The TFA applies to small strain (visco-)plastic material models and assumes the inelastic strains to be piece-wise uniform on specific subdomains, accounting for the resulting elastic deformations via strain localization tensors. In this way, effective models with a finite number of internal variables arise, see also Chaboche et al.~\cite{ChabocheTFA2005}. Furthermore, Liu and co-workers \cite{SCALiu2016,SCA_CVP_2017,SCA_softening_2018} introduced the self-consistent clustering analysis (SCA), which is similar in spirit to the TFA, but exploits the Hashin-Shtrikman variational principle~\cite{Hashin1961, Hashin1962, SCAWulfinghoff2018} and is not per se limited to small strain (visco-)plasticity. Nevertheless, when considered as discretization methods, both TFA and SCA show a slow convergence rate in terms of the number of clusters~\cite{TFA4} which is rooted in the weak approximation capabilities of piecewise uniform functions~\cite{SCA_convergence}.\\
Inspired by recent variational estimates for nonlinear materials~\cite{CASTANEDA1997171}, the non-uniform transformation field analysis~\cite{NTFA1} (NTFA) relies upon non-uniform inelastic basis functions, permitting the approximation errors of the fields to be made as small as desired. However, the difficulty is transferred to prescribing suitable evolution equations~\cite{NTFA_viscoelastic1,NTFA_viscoelastic2} for the reduced inelastic strains, which should be independent of the basis, cheap to evaluate and consistent upon refinement. Possible remedies are Taylor series expansions of the force potential~\cite{NTFA_TSO1,NTFA_TSO2,NTFA_TSO3}, mixed variational principles~\cite{FritzenLeuschner,FritzenHodappLeuschner} or dedicated "reducible" models~\cite{Fatigue2020}, giving rise to the $\textrm{FE}^{2\textrm{R}}$ (R as reduced) method~\cite{FE2R}.\\
As an alternative to methods which approximate the solution of micromechanical problems on unit cells, it is possible to approximate the effective properties directly. Typically, this comes at the cost of operating on a  high-dimensional domain of interest. Data driven methods, especially artificial neural networks (ANNs), are predestined for such approximation tasks. There is a number of works considering training ANNs to approximate the effective elastic energy of a medium, \cf{} Yvonnet and coworkers~\cite{Yvonnet2009,Yvonnet2013,Le2015}. Furthermore, the regularity of the effective stress facilitates the direct approximation of the stress-strain relationship of inelastic problems, \cf{} the works of Jadid~\cite{NasserJadid1997}, Penumadu-Zhao~\cite{Penumadu1999} or Srinivasu et al.~\cite{Srinivasu2012} for different approaches. Motivated by natural language processing, recurrent neural networks (RNN) may provide a framework for incorporating history dependence into the approximation of the stress-strain relationship. For instance, Gorji and coworkers~\cite{Gorji2020} demonstrated that RNN are able to capture effects such as the Bauschinger effect, permanent softening or latent hardening in the context of elastoplasticity. Using artificial neural networks accompanied by an on-the-fly switching to a reduced order model in a two-scale simulation was investigated by Fritzen et al.~\cite{Fritzen2019}. A shortcoming of such data driven methods appears to be that their predictive quality appears low far away from the training domain, and accounting for the inherent physics, i.e., monotonicity and thermodynamic consistency, may be difficult.\\ 
Liu et al.~\cite{Liu2018, Liu2019} proposed a data driven modeling approach based on an explicit microstructure model consisting of hierarchical laminates. More precisely, for a $K$-phase microstructure, Liu et al.\ consider a $K$-ary tree structure of laminates with fixed direction of lamination and intermittent rotations. In analogy to deep artificial neural networks, they called such an identified surrogate model a deep material network (DMN). Instead of approximating the effective energy or the stress-strain relationship, DMNs are sought to approximate the effective stiffness of a fixed microstructure considered as a function of the input stiffness tensors of the constituents. For the parameter identification, they rely upon automatic differentiation and stochastic gradient descent, as typical for training artificial neural networks. After training, the hierarchical laminate can be applied to inelastic problems, even at finite strains, and the approximation accuracy is rather impressive.\\ 
Discarding the intermittent rotations, Gajek et al.~\cite{Gajek2020} introduced direct DMNs, which are based on laminates with variable direction of lamination. Direct DMNs enable a faster and more robust identification process compared to the indirect DMNs of Liu et al.~\cite{Liu2018, Liu2019}, and also compare favorably for inelastic constituents. Furthermore, Gajek et al.~\cite{Gajek2020} motivated the training on linear elastic data and generalization to the nonlinear regime by showing that, to first order in the strain rate, the effective inelastic behavior of composite materials is determined by linear elastic localization. As a byproduct, Gajek et al.~\cite{Gajek2020} established that deep material networks inherit thermodynamic consistency and stress-strain monotonicity from their phases. This property is crucial, as it ensures the effective models to inherit stabilizing numerical properties, like strong convexity, from the phases of the composite. Deep material networks were augmented by cohesive zone models by Liu~\cite{Liu2020a}.\\
Recently, Liu et al.~\cite{Liu2020b} proposed a transfer learning approach to treat fiber reinforced composites by DMNs. More precisely, they proposed to interpolate the parameters of already identified DMNs, corresponding to different fiber orientation states. In this work, we go beyond this a posteriori approach, and seek DMNs covering the entire spectrum of fiber orientations arising in such a fiber reinforced component.

\subsection{Contribution and outline}

In this article, we investigate a multiscale methodology for direct deep material networks, which covers all possible variations of the second-order fiber orientation tensors and permits concurrent multiscale simulations with DMNs at the Gauss point level. Similar to the $\textrm{FE}^2$, $\textrm{FE}$-$\textrm{FFT}$ and the $\textrm{FE}^{2\textrm{R}}$ methods, we call the ensuing method the $\textrm{FE}$-$\textrm{DMN}$ method.\\
To account for a spatially varying fiber orientation, see Section~\ref{sec:interpolated_direct_deep_material_networks}, we augment direct DMNs by the fiber orientation interpolation concept introduced by Köbler et al.~\cite{Kobler2018}. In contrast to Liu et al.~\cite{Liu2020b}, we do not consider a \emph{transfer learning} strategy, i.e., to interpolate already identified models, but propose an a priori interpolation strategy. More precisely, we investigate deep material networks which explicitly depend on the fiber orientation, and identify the optimal model parameters \emph{jointly}.\\
For this purpose, we utilize high-fidelity microstructures of fiber reinforced composites~\cite{SAM} which permit us to routinely cover the possible fiber orientations at industrial filler fraction and fiber aspect ratio. To this end, we sample the linear elastic training data from up to $31$ microstructure realizations. We also improve upon the previous sampling strategy~\cite{Liu2018, Liu2019} for the constituents' stiffness tensors used in the offline training. We show by example that it is sufficient to cover those stiffnesses which arise as possible material tangents on the microscopic scale.\\
For component scale simulations of industrial problems, it is necessary to optimize the user-defined material models based upon the identified DMNs, see Section~\ref{sec:implementation}.\\
We take special care to ensure that the interpolated DMN generalizes accurately to the inelastic regime. To this end, $78$ additional microstructure realizations were generated, exclusively for the inelastic validations. We compute the stress response of each of the $109$ generated microstructure representations by an FFT-based computational homogenization~\cite{MoulinecSuquet1994, MoulinecSuquet1998} code and compare the former to the predicted stress response of the interpolated direct DMN. The validation results show that with a maximum error of $5.5\%$, the DMN is capable of predicting the effective stress of all investigated $109$ discrete fiber orientation states sufficiently. We refer to Section~\ref{sec:identification_dmn} for details.\\
To show that our approach is applicable to state of the art engineering computations, we consider the entire process chain of a quadcopter frame, \cf{} Fig.~\ref{fig:motivation_quadcopter} and Section~\ref{sec:example}. We conduct an injection molding simulation, map the spatially varying fiber orientations upon a finite element mesh and conduct a two-scale simulation using the identified DMN surrogate model. We implement the DMN as a user-material (UMAT) subroutine in \abq only relying upon the provided software interfaces. The quadcopter consists of four arms made from injection molded short fiber reinforced polyamide, two base plates which are made from aluminum and four legs which are made from pure polyamide. The simulation model of the quadcopter consist of more than two million elements resulting in almost ten million degrees of freedom. In more than $1.9$ million elements, a deep material network is integrated implicitly at every Gauss points, accounting for the local microstructure information in the simulation.\\
Last but not least, we discuss the computational costs accompanied by our approach in Section~\ref{sec:computationl_cost}, and demonstrate that the educated guess of the potential computational power of DMNs made in the conclusion Liu et al.~\cite{Liu2020b} was too pessimistic.

\section{Direct deep material networks for variable fiber orientation}\label{sec:interpolated_direct_deep_material_networks}
\label{sec:DMN}

In this section, we extend direct deep material networks to short fiber reinforced composite microstructures parameterized by the second order fiber orientation tensor. The main technical tool is the fiber orientation interpolation technique, introduced by Köbler et al.~\cite{Kobler2018}, which we apply on the node level of the deep material network.\\ 
First, we recall the basics of direct deep material networks and the fiber orientation triangle. Subsequently, we combine both concepts.

\subsection{Direct deep material networks}\label{sec:direct_DMN}
\label{sec:DMN_dDMN}

\begin{figure}
	\centering
	\begin{subfigure}[b]{0.45\textwidth}
		\centering
		\includegraphics[width=\textwidth]{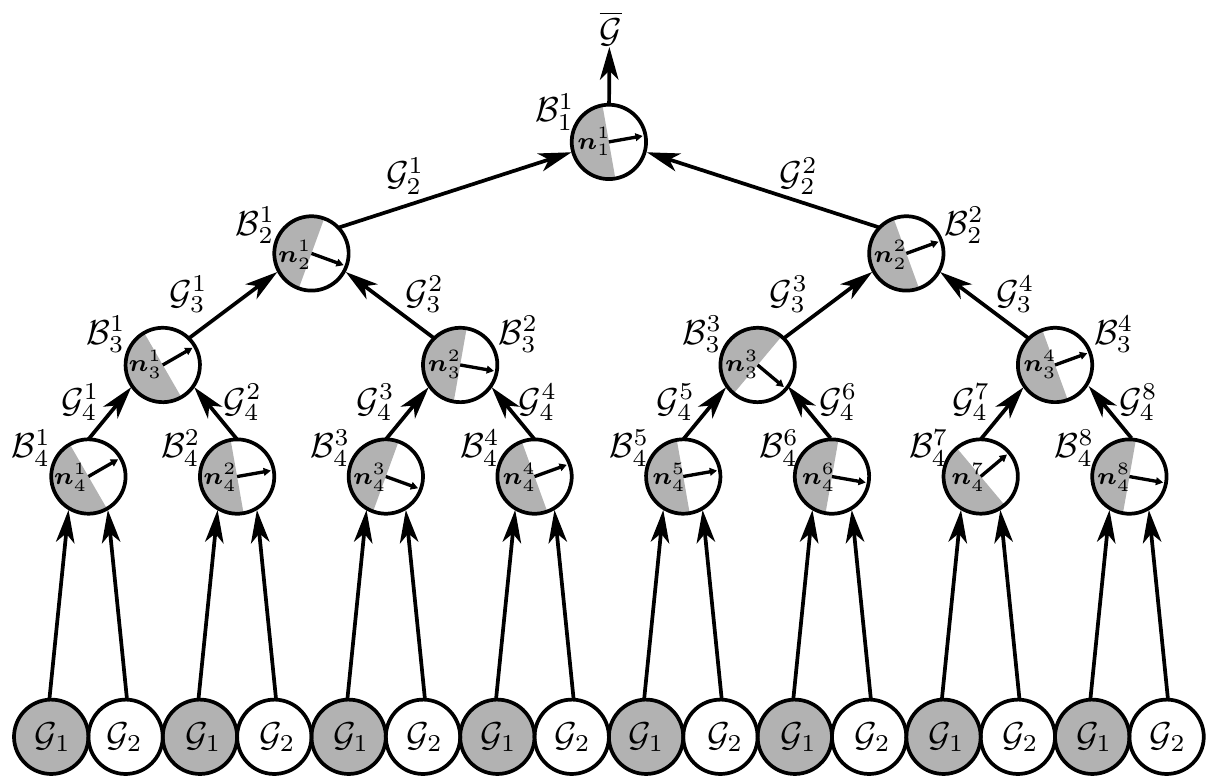}
		\caption{Material propagation}
		\label{fig:DMN_gsm}
	\end{subfigure}
	\begin{subfigure}[b]{0.45\textwidth}
		\includegraphics[width=\textwidth]{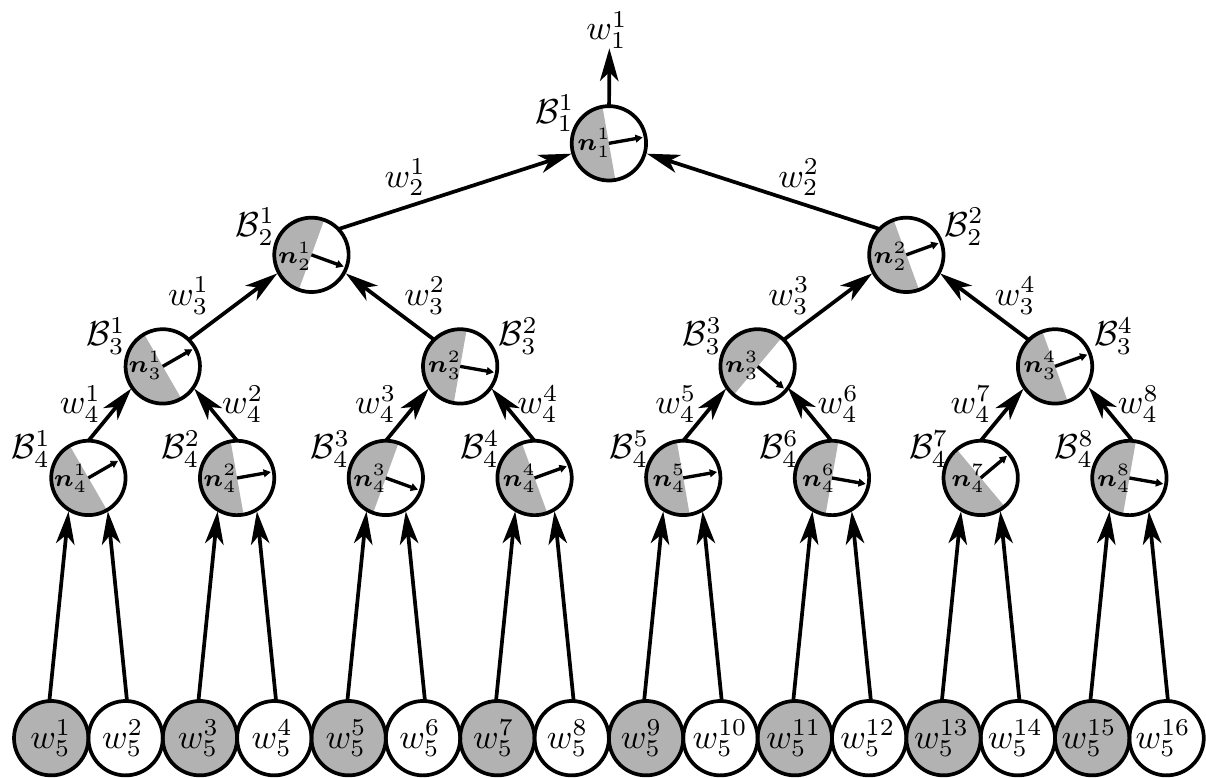}
		\caption{Weight propagation}
		\label{fig:weight_propag}
	\end{subfigure}
	\caption{A direct two-phase deep material network (DMN) of depth four (the input level is not counted)}
	\label{fig:DMN_gsm_weight_propag}
\end{figure}
Let $\GSM$ denote the set of all generalized standard materials (GSM)~\cite{HalphenNguyen}. Then, any two-phase periodic microstructure $Y \subseteq \ffR^d$ in $d$ spatial dimensions gives rise to the (nonlinear) homogenization function
\begin{equation}\label{eq:DMN_dDMN_homogenization_function}
	\HOM{Y}: \GSM \times \GSM \rightarrow \GSM, \quad (\cG_1, \cG_2) \mapsto \effective{\cG},
\end{equation}
which maps two input GSMs to the effective GSM that emerges by solving the cell problem of first order homogenization, see Gajek et al.~\cite{Gajek2020} for details. Homogenization functions may be regarded as the basic objects of studying micromechanics at small strains.\\
In general, evaluating homogenization functions requires significant computational resources. Only for special microstructures, the evaluation can be performed with minimal effort. An example for such a microstructure is given by a two-phase laminate, uniquely characterized by a direction $\fn$ of lamination
and the volume fractions $c_1$ and $c_2$ of the two phases.\\
A direct deep material network is defined as a hierarchy of such two-phase laminates, see Fig.~\ref{fig:DMN_gsm_weight_propag}. The concept was originally introduced by Liu and co-workers~\cite{Liu2018, Liu2019} for laminates with \emph{fixed} direction of lamination, but an additional rotation layer, and simplified by Gajek et al.~\cite{Gajek2020}. By combining laminates in a hierarchical manner, the resulting homogenization function
\begin{equation}
	\DMN{Y}: \GSM \times \GSM \rightarrow \GSM
\end{equation}
may be rather complex and, by a judicious choice of the involved laminates, may be used as an approximation of the homogenization function \eqref{eq:DMN_dDMN_homogenization_function} corresponding to the original microstructure $Y$, which is significantly less demanding to evaluate.\\
On a more formal level, a direct DMN is a perfect, ordered, rooted binary tree of depth $K$, where a two-phase laminate $\BB^i_k$ is assigned to each node of the tree. We reserve the letter $k$ for labeling the depth of a node, whereas the horizontal index is consistently indexed by the letter $i$. Our layer count only comprises the laminate layers, and the input is counted separately. Thus, the DMN comprises $2^K-1$ laminate nodes. For a two-phase DMN of depth $K$, the homogenization function $\DMN{Y}$
\begin{equation}
	\effective{\cG} = \DMN{Y}(\cG_1, \cG_2)
\end{equation}
is defined recursively by traversing the binary tree from the leaves, at level $K$, to the root
\begin{equation}
	\effective{\cG} = \cG^1_1 \quad \textrm{with} \quad \cG^i_k = \BB^i_k(\cG^{2i-1}_{k+1}, \cG^{2i}_{k+1}), \quad k  = 1 \dots K, \; i = 1 \dots 2^{k-1}.
\end{equation}
Input materials are assigned in an alternating fashion, \ie
\begin{equation}
	\cG^i_{K+1} = \left\{
	\begin{array}{l r}
		\cG_1, \quad i \textrm{ odd},\\
		\cG_2, \quad i \textrm{ even},
	\end{array}
	\right.
\end{equation}
holds. We refer to Fig.~\ref{fig:DMN_gsm} for a schematic.\\
Liu et al.~\cite{Liu2018, Liu2019} noticed that parameterizing the involved laminates by the volume fractions $c_1$ and $c_2$ is not optimal. Indeed, if one of the volume fractions is zero, the entire corresponding sub-tree will have no further influence. It is more convenient to parameterize the laminates' volume fractions by assigning pairs of weights $w^{2i-1}_{K+1}$ and $w^{2i}_{K+1}$ to each laminate on the input level $K$. These weights should be non-negative and sum to unity. Then, by traversing the binary tree from the leaves to the root, the weights on the $k$-th level are computed by the sum of weights of the respective laminates on the previous level, \ie
\begin{equation}
	w^i_k = w^{2i-1}_{k+1} + w^{2i}_{k+1}
\end{equation}
holds, \cf{} Fig.~\ref{fig:weight_propag}. The volume fractions $c_1$ and $c_2$ of each laminate $\BB^i_k$ are then computed by normalization
\begin{equation}
	c_1 = \frac{w^{2i-1}_{k+1}}{w^{2i-1}_{k+1} + w^{2i}_{k+1}} \quad \textrm{and} \quad c_2 = 1 - c_1.
\end{equation}
For fixed tree topology, a direct DMN is uniquely determined by the directions of lamination, one for each laminate, and the weights of the input layer. These free parameters are identified based on linear elastic precomputations, the so-called \textbf{offline training}, see Section \ref{sec:offline_training}. Once the free parameters are identified, the DMN can be applied to nonlinear and inelastic materials. This \textbf{online evaluation} is described in Section \ref{sec:online_evaluation}.

\subsection{The fiber orientation triangle} 
\label{sec:fiber_orientation_tensor}

\begin{figure}[H]
	\begin{subfigure}{0.29\textwidth}
		\includegraphics[width=\textwidth]{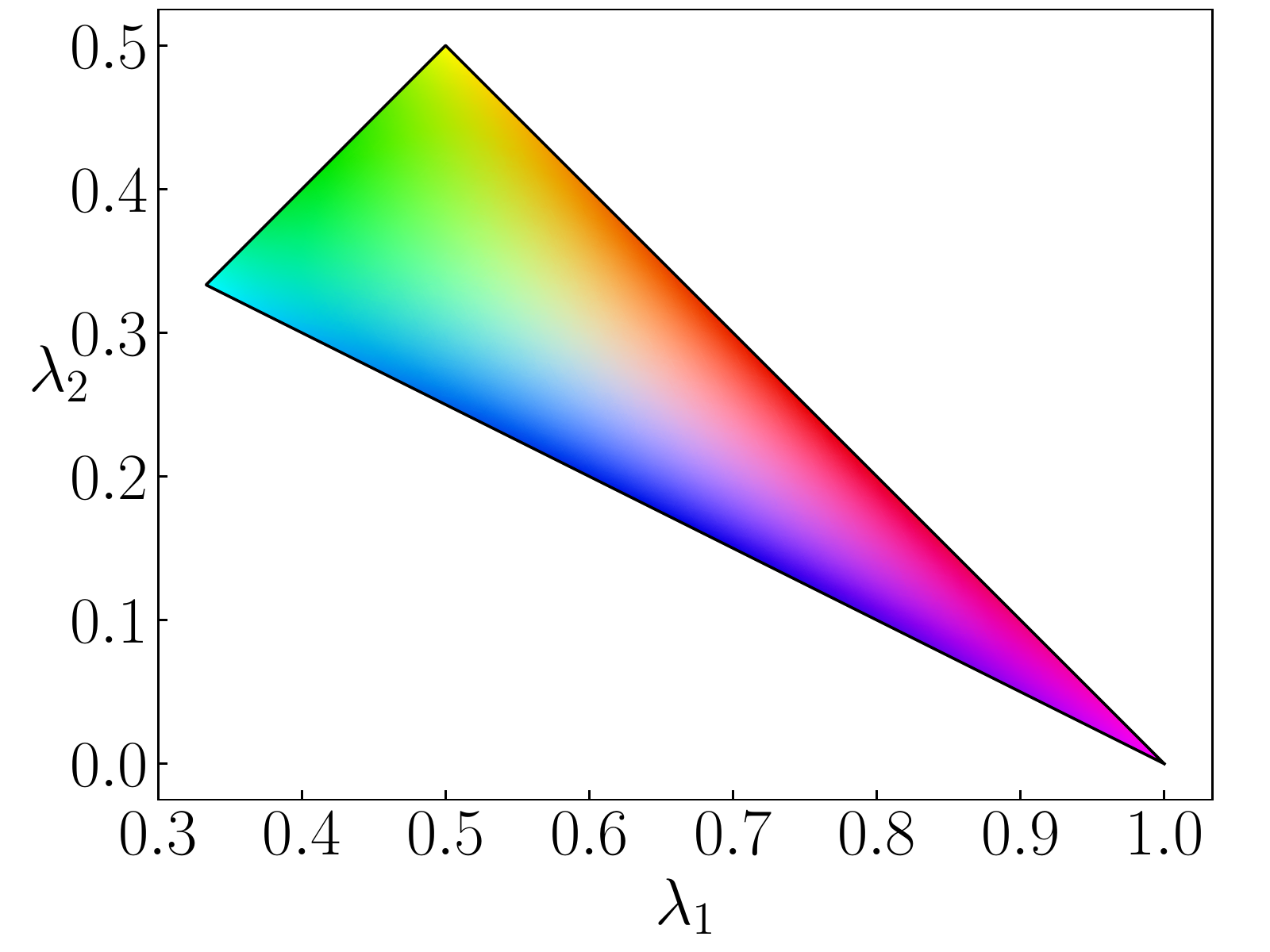}
		\label{fig:orienation_triangle}
	\end{subfigure}
	\begin{subfigure}{0.23\textwidth}
		\centering
		\includegraphics[width=\textwidth]{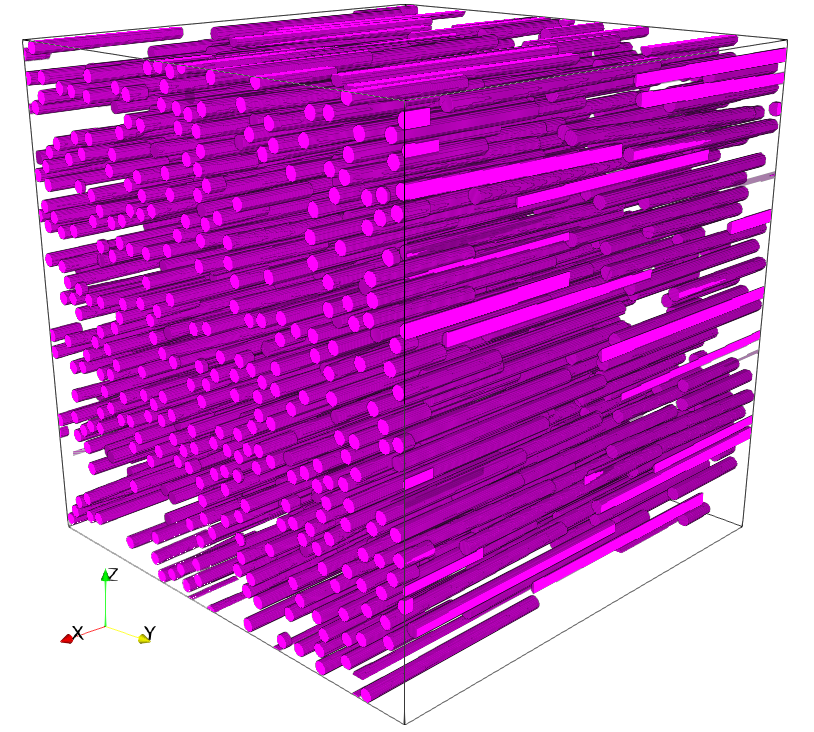}
		\caption{$\lambda_1=1.0$, $\lambda_2=0.0$}
		\label{fig:ms_000}
	\end{subfigure}
	\begin{subfigure}{0.23\textwidth}
		\centering
		\includegraphics[width=\textwidth]{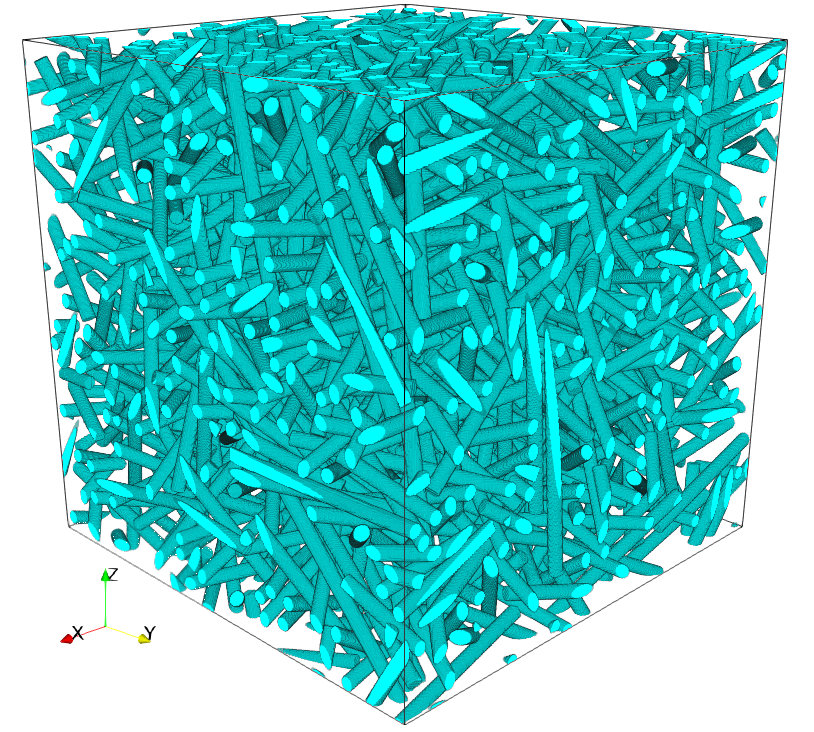}
		\caption{$\lambda_1=0.\overline{33}$, $\lambda_2=0.\overline{33}$}
		\label{fig:ms_001}
	\end{subfigure}
	\begin{subfigure}{0.23\textwidth}
		\centering
		\includegraphics[width=\textwidth]{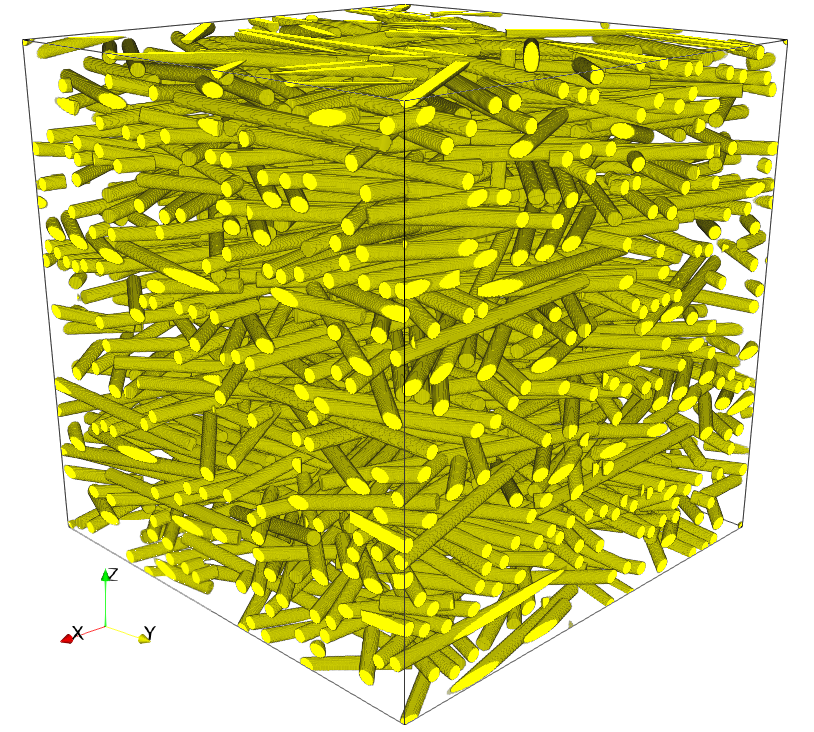}
		\caption{$\lambda_1=0.495$, $\lambda_2=0.495$}
		\label{fig:ms_002}	
	\end{subfigure}
	\caption{Fiber orientation reference triangle showing the two largest eigenvalues of the fiber orientation tensor. The three extreme cases, \ie (a) unidirectional, (b) isotropic and (c) planar isotropic fiber orientation are shown.}
	\label{fig:orienation_triangle_microstructures}
\end{figure}

Suppose a fiber orientation state is given in terms of a fiber orientation distribution (FOD) function $\rho$, which specifies the probability to find fibers in direction $\fp$. Advani-Tucker~\cite{AdvaniTucker1987} introduced the second order fiber orientation tensor~\cite{AdvaniTucker1987}
\begin{equation}
	\tuckersecond = \int \fp \otimes \fp\, \rho(\fp)\, \dif A(\fp)
\end{equation}
as a compact measure for the current fiber orientation state. Despite its limited information content, its compact form makes it the typical quantity of interest for commercial injection molding simulations~\cite{Kennedy2013}. Higher-order moments of the FOD are then estimated by closure approximations, \cf{} Montgomery-Smith et al.~\cite{MontgomerySmith2011}.\\
The tensor $\tuckersecond$ is symmetric and positive definite with unit trace. Consequently, only five independent parameters are involved. In terms of an eigenvalue decomposition
\begin{equation}\label{eq:eigen_decomposition_A}
	\tuckersecond = \fQ \, \diag{\lambda_1, \lambda_2, \lambda_3} \fQ^T,
\end{equation}
where the matrix $\fQ \in \sop$ is orthogonal and the eigenvalues $\lambda_1 \geq \lambda_2 \geq \lambda_3$ are sorted in a descending order, the fiber orientation tensor $\tuckersecond$ may be described by two parameters $\lambda_1$ and $\lambda_2$ which satisfy the inequalities
\begin{equation}\label{eq:triangle_inequal}
	\frac{1}{3} \leq \lambda_1 \leq 1 \quad \textrm{and} \quad 1-2\lambda_{1} \leq \lambda_2 \leq \lambda_{1}.
\end{equation}
Thus, up to an orthogonal transformation, every tensor $\tuckersecond$ corresponds to a point in the triangle described by the inequalities \eqref{eq:triangle_inequal}, see Fig.~\ref{fig:orienation_triangle_microstructures}. In this article, we follow K{\"o}bler et al.~\cite{Kobler2018} and use the CMYK coloring scheme for encoding different fiber orientations as shown in Fig.~\ref{fig:orienation_triangle_microstructures}.
The manufacturing process induces local variations of the fiber orientation of short fiber reinforced thermoplastic components. Thus, for component-scale simulations, these variations need to be accounted for by the material models. If the fiber orientation state is described in terms of the second-order fiber orientation tensor $\tuckersecond$, a family of effective material models, one for each such tensor $\tuckersecond$, needs to be supplemented.\\
By general covariance considerations, two fiber orientation states which differ only by an orthogonal transformation should give rise to effective material responses which differ only by this orthogonal transformation. Consequently, if the considered fiber orientation states are parameterized by the second order fiber orientation tensor, the essentially different fiber orientation states will be parameterized by the points inside the fiber orientation triangle \eqref{eq:triangle_inequal}.\\
Thus, the material models to be identified are parameterized by a two-dimensional continuum. Furthermore, a certain continuity of the effective response of the material model, considered as a function of the fiber orientation tensor, is expected. Indeed, changing the fiber orientation only slightly is expected to change the effective mechanical response only slightly, as well, at least for non-critical loading. Unfortunately, the typical multiscale approach based on representative volume elements is unable to leverage this continuity. Indeed, although the effective material response depends continuously on the fiber orientation tensor, the representative volume element does not. Indeed, due to the stochastic nature of such fiber-filled volume elements, infinitely many \emph{different} representative volume elements may be used to give rise to the same effective response.\\
In particular, this reasoning has the following implication. Suppose that we furnish each point $(\lambda_1,\lambda_2)$ in the fiber orientation triangle \eqref{eq:triangle_inequal} with a corresponding representative volume element $Y_{\lambda_1\lambda_2}$. Even if all these elements have the same size, the function $(\lambda_1,\lambda_2)\mapsto Y_{\lambda_1\lambda_2}$ will not be continuous in any useful way.\\
As an alternative, K{\"o}bler et al.~\cite{Kobler2018} proposed a fiber orientation interpolation procedure on the level of effective stresses. This idea avoids the difficulty of interpolating internal variables which live in different locations for different microstructures. However, this approach comes at a price: the number of stress evaluations is tripled by this approach. Indeed, for any fiber orientation state, the stress response associated to the three corners of the interpolating triangle needs to be evaluated.

\subsection{Fiber orientation interpolation of deep material networks}
\label{sec:fiber_orientation_interpolation}

\begin{table}[h!]
	\begin{center}
		\begin{tabular}{l l l l l}
			\hline
			Linear & $M=3$ & $\phi_1 = \varphi_1$ & $\phi_2 = \varphi_2$ & $\phi_3 = \varphi_3$\\
			\hline
			\multirow{2}{*}{Tri-linear} & \multirow{2}{*}{$M=4$}  & $\phi_1 = \varphi_1 - 9 \varphi_1 \varphi_2 \varphi_3$ & $\phi_2 = \varphi_2 - 9 \varphi_1 \varphi_2 \varphi_3$ & $\phi_3 = \varphi_3 - 9 \varphi_1 \varphi_2 \varphi_3$ \\
			& & $\phi_4 = 27 \varphi_1 \varphi_2 \varphi_3$ &  & \\
			\hline
			\multirow{2}{*}{Quadratic} & \multirow{2}{*}{$M=6$}  & $\phi_1 = \varphi_1 \left(2 \varphi_1 - 1\right)$ & $\phi_2 = \varphi_2 \left(2 \varphi_2 - 1\right)$ & $\phi_3 = \varphi_3 \left(2 \varphi_3 - 1\right)$ \\
			& & $\phi_{4} = 4 \varphi_1 \varphi_2$ & $\phi_{5} = 4 \varphi_1 \varphi_3$ & $\phi_{6} = 4 \varphi_2 \varphi_3$\\
			\hline
		\end{tabular}
	\end{center}
	\caption{Shape functions used for orientation interpolation}
	\label{tab:shape_functions}
\end{table}
Due to their specific structure, deep material networks may overcome the difficulties mentioned at the end of the previous section. Indeed, for fixed tree topology, the internal variables of the individual phases live on identical locations, also for different DMNs. Indeed, in any case, the internal variables are tied to the materials on the lowest level of the tree, \cf{} Fig.~\ref{fig:DMN_gsm_weight_propag}.\\
Of course, if DMNs are identified \emph{independently} for each point in the fiber orientation triangle \eqref{eq:triangle_inequal}, the parameters of the DMN need not depend continuously on the fiber orientation tensor $\tuckersecond$. Still, it appears reasonable to identify the DMN's parameters \emph{jointly} over all fiber orientations in the fiber orientation triangle.\\
More precisely, we consider DMNs which are parameterized by points $(\lambda_1,\lambda_2)$ inside the fiber orientation triangle. As the DMN's weights are directly linked to the constituent volume fractions of the underlying microstructure~\cite{Liu2018, Liu2019,Gajek2020}, we will seek weights which are even \emph{independent} of the fiber orientation. This appears reasonable, and we refer to Milton~\cite{Milton1986} and Torquato~\cite[Sec. 20.2.2]{torquato2005random} for background material.\\
In order to interpolate the lamination directions $\fn^i_k$ on the fiber orientation triangle, we parameterize each normal $\fn^i_k \in \fS^2$ by spherical coordinates
\begin{equation}
	\fn^i_k = \left[%
	\begin{array}{c}
		\sin\left(\alpha^i_k\right) \cos\left(\beta^i_k\right)\\
		\sin\left(\alpha^i_k\right) \sin\left(\beta^i_k\right)\\
		\cos\left(\alpha^i_k\right)\\
	\end{array}%
	\right]
\end{equation}
with angles $\alpha^i_k \in [0, \pi]$ and $\beta^i_k \in [0, 2\pi]$. Then, we interpolate the angles $\alpha^i_k$ and $\beta^i_k$ on the fiber orientation triangle \eqref{eq:triangle_inequal} by a global (finite element) shape function. Please note the difference to Köbler et al.~\cite{Kobler2018}, who rely upon linear interpolation on a subtriangulation of the fiber orientation triangle.\\
Particularly compact expressions for the finite element shape functions are obtained by transforming the parameters $\lambda_1$ and $\lambda_2$ to barycentric coordinates $\varphi_1$, $\varphi_2$ and $\varphi_3$, \ie via solving the linear system
\begin{equation}
	\left[%
	\begin{array}{c c c}
		1 & 1/3 & 1/2\\
		0 & 1/3 & 1/2\\
		1 &   1 &  1 \\
	\end{array}%
	\right]
	\left[%
	\begin{array}{c}
		\varphi_{1}\\
		\varphi_{2}\\
		\varphi_{3}\\
	\end{array}%
	\right]=
	\left[%
	\begin{array}{c}
		\lambda_{1}\\
		\lambda_{2}\\
		1          \\
	\end{array}%
	\right],
\end{equation}
\cf{} \eg Vince~\cite{Vince2017}. We collect the parameters of the polynomial shape functions in a vector $\vec{\phi} = \left[\phi_1, \dots, \phi_M\right]$, where $M$ denotes the number of shape functions. Then, the interpolated angles may be expressed as
\begin{equation}\label{eq:interpolating_the_angles}
	\alpha^i_k(\lambda_1, \lambda_2) = \vec{p}^{\; i \; T}_k \vec{\phi}(\lambda_1, \lambda_2) \quad \textrm{and} \quad \beta^i_k(\lambda_1, \lambda_2) = \vec{q}^{\; i \; T}_k \vec{\phi}(\lambda_1, \lambda_2)
\end{equation}
in terms of the parameter vectors $\vec{p}=\left[p_1, \dots, p_M \right] \in \ffR^M$ and $\vec{q}=\left[q_1, \dots, q_M\right]  \in \ffR^M$. In this article, we investigate linear, tri-linear and quadratic shape functions, see Tab.~\ref{tab:shape_functions}.\\
To sum up, extending DMNs to account for varying fiber orientation reduces to increasing the number of unknown parameters. Indeed, instead of identifying the angles $\alpha^i_k$ and $\beta^i_k$, the parameter vectors $\vec{p}^{\; i}_k$ and $\vec{q}^{\; i}_k$ are sought, in addition to the unknown weights $w^i_{K+1}$. 

\section{Implementation}
\label{sec:implementation}

In this section, we explain how to evaluate linear and nonlinear homogenization functions of an interpolated direct DMN efficiently.

\subsection{Offline training} 
\label{sec:offline_training}

\begin{figure}[H]
	\centering
	\includegraphics[width=0.48\textwidth]{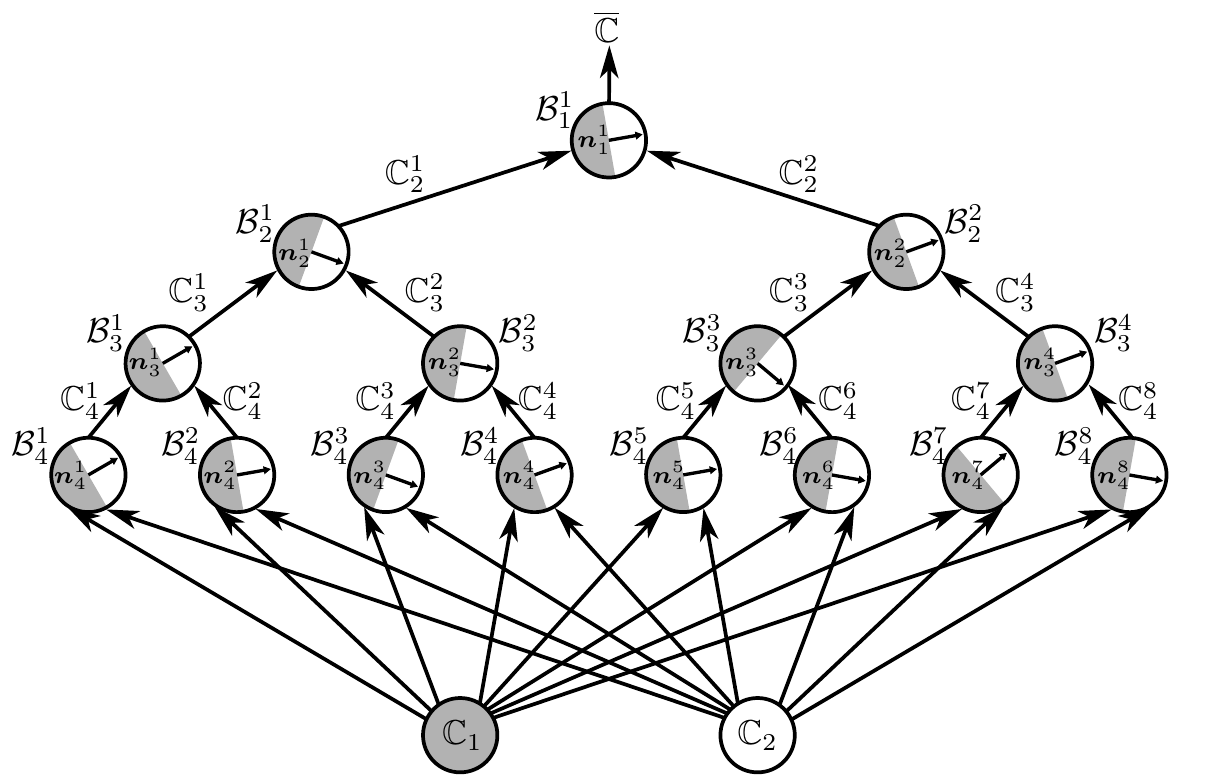}
	\caption{Schematic illustration of the stiffness propagation in a two-phase DMN of depth four}
	\label{fig:stiffness_propag}
\end{figure}

\begin{figure}[H]
	\centering
	\begin{subfigure}{0.48\textwidth}
		\includegraphics[width=\textwidth]{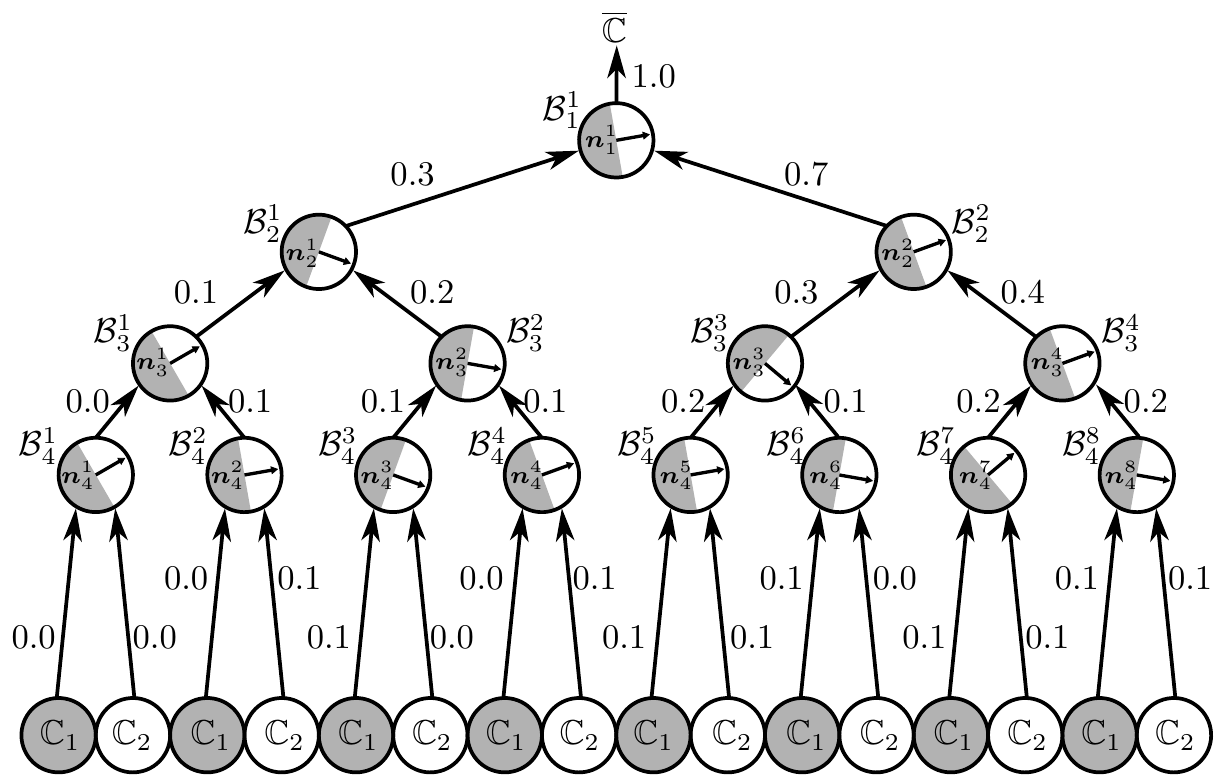}
		\caption{Perfect binary tree}
	\end{subfigure}
	\hfill
	\begin{subfigure}{0.48\textwidth}
		\includegraphics[width=\textwidth]{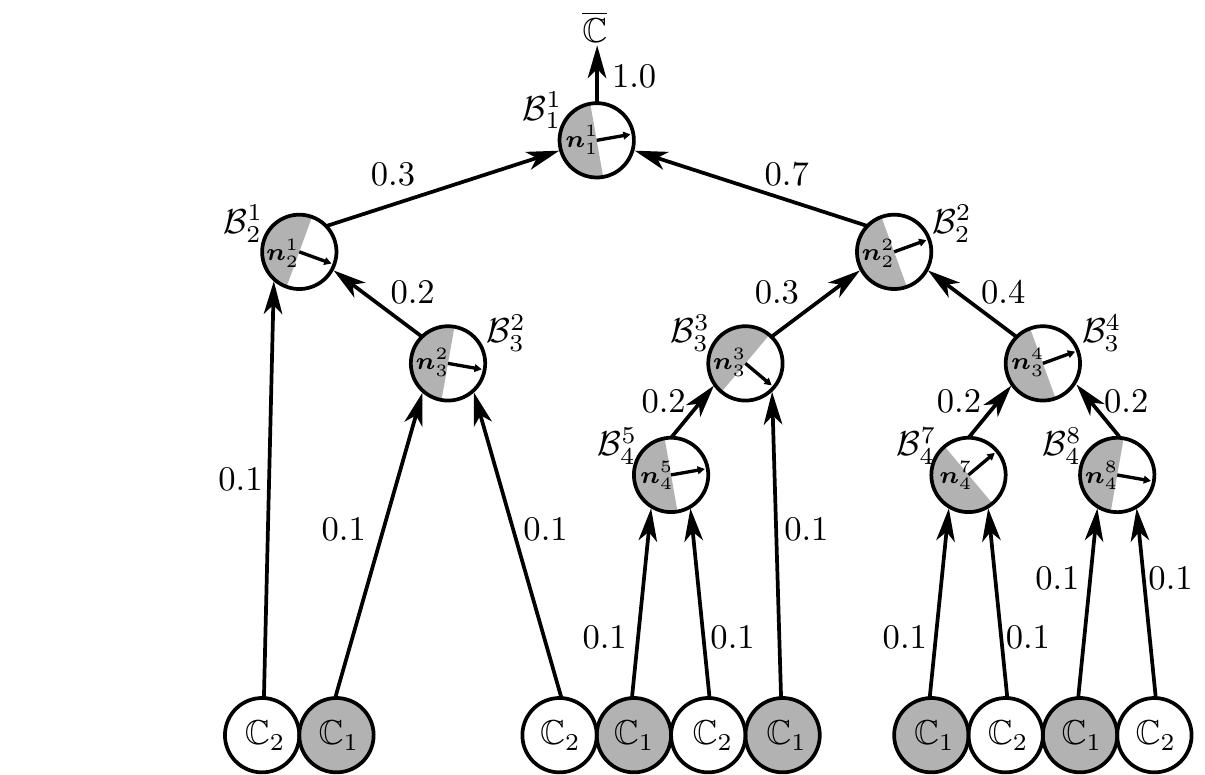}
		\caption{Binary tree with removed laminates}
	\end{subfigure}
	\caption{Binary tree compression to speed up offline training and online evaluation}
	\label{fig:tree_compression}
\end{figure}

The goal of the offline training is to identify the free parameters of the DMN, namely the weights and the vectors $\vec{p}^{\; i}_k$ and $\vec{q}^{\; i}_k$ used for interpolating the angles \eqref{eq:interpolating_the_angles}, which we collect in "long" vectors 
\begin{equation}
	\vec{p} = \left[\vec{p}^{\; 1}_K, \vec{p}^{\; 2}_K,\ldots \vec{p}^{\; {2^{K-1}}}_K, \vec{p}^{\; 1}_{K-1}, \vec{p}^{\; 2}_{K-1}, \ldots, \vec{p}^{\; 2^{K-2}}_{K-1}, \ldots, \vec{p}^{\; 1}_2, \vec{p}^{\; 2}_2, \vec{p}^{\; 1}_1\right] \in \left(\ffR^M\right)^{2^K-1}
\end{equation}
and
\begin{equation}
	\vec{q} = \left[\vec{q}^{\; 1}_K, \vec{q}^{\; 2}_K,\ldots \vec{q}^{\; {2^{K-1}}}_K, \vec{q}^{\; 1}_{K-1}, \vec{q}^{\; 2}_{K-1}, \ldots, \vec{q}^{\; 2^{K-2}}_{K-1}, \ldots, \vec{q}^{\; 1}_2, \vec{q}^{\; 2}_2, \vec{q}^{\; 1}_1\right] \in \left(\ffR^M\right)^{2^K-1}.
\end{equation}
We insert the parameters of laminates on level $K$ first and add the parameters of laminates for decreasing level index in their corresponding order. We enforce the non-negativity constraint on the weights
\begin{equation}
	w^i_{K+1} \geq 0
\end{equation}
by defining $w^i_{K+1} = \langle v_i \rangle_{+}$ in terms of unconstrained weights $v_i$, $i=1,\dots,2^K$. Here, $\langle \cdot \rangle: \ffR \rightarrow \ffR_{\geq0}$, $x \mapsto \max(0, x)$ denotes the Macauley bracket. By collecting $v_i$ in a vector $\vec{v} = \left[v_1, \dots, v_{2^K}\right] \in \ffR^{2^K}$, we represent the DMN's linear elastic homogenization function in the form
\begin{equation} \label{eq:DMN_effective_stiffness_parameter}
	\effective{\ffC} = \DMNLIN{\Lambda}\left(\ffC_1, \ffC_2,  \lambda_1, \lambda_2, \vec{p}, \vec{q}, \vec{v}\right). 
\end{equation}
$\DMNLIN{\Lambda}$ maps the input stiffnesses $\ffC_1$ and $\ffC_2$, the fiber orientation parameters $\lambda_1$ and $\lambda_2$ and the unknown fitting parameters $\vec{p}$, $\vec{q}$ and $\vec{v}$ to the DMN's effective stiffness. The specific binary tree structures of the DMN can be exploited to evaluate $\DMNLIN{\Lambda}$ efficiently. To this end, we assign the input stiffnesses $\ffC_1$ and $\ffC_2$ to the laminates of level $K$ in an alternating fashion. The directions of lamination are interpolated on the fiber orientation triangle based on the given parameters $\lambda_1$, $\lambda_2$, $\vec{p}$ and $\vec{q}$, \cf{} Section~\ref{sec:fiber_orientation_interpolation}. The input stiffnesses are homogenized at level $K$ for each laminate independently. In the next step, the homogenized stiffnesses serve as the input for level $K-1$ and so forth, until the root is reached, giving rise to the DMN's effective stiffness $\effective{\ffC}$. The process of propagating stiffnesses from the $K$-th level to the root is illustrated in Fig.~\ref{fig:stiffness_propag}. 
Thus, computing the effective stiffness of a DMN reduces to computing a sequence of effective stiffnesses of two-phase laminates. According to Section 9.5 in Milton's book \cite{Milton2002}, the linear elastic homogenization function of a laminate
\begin{equation}\label{eq:laminate_homogenization_linear}
	\ffC^i_k = \BB^i_k(\ffC^{2i-1}_{k+1}, \ffC^{2i}_{k+1})
\end{equation}
may be determined by solving the equation
\begin{equation}
	\left(\ffP(\fn^i_k) + \alpha \left[ \ffC^i_k - \alpha \IDsym \right]^{-1}\right)^{-1}  = c_1 \left(\ffP(\fn^i_k) + \alpha \left[ \ffC^{2i-1}_{k+1} - \alpha \IDsym \right]^{-1}\right)^{-1} + c_2 \left(\ffP(\fn^i_k) + \alpha \left[ \ffC^{2i}_{k+1} - \alpha \IDsym \right]^{-1}\right)^{-1}
\end{equation}
for the effective stiffness $\ffC^i_k$. Here, $\IDsym: \Sym{d} \rightarrow \Sym{d}$ denotes the identity on $\Sym{d}$, the set of symmetric $d \times d$ matrices, and  $\ffP: \Sym{d} \rightarrow \Sym{d}$ stands for a projection operator, which depends on the direction of lamination $\fn^i_k$, and reads
\begin{equation}
	\left(\ffP(\fn)\right)_{mnop} =  \frac{1}{2}(n_m \delta_{no} n_p + n_n \delta_{mo} n_p + n_m \delta_{np} n_o + n_n \delta_{mp} n_o) - n_m n_n n_o n_p
\end{equation}
in Cartesian coordinates. Here, $\delta$ denotes the Kronecker symbol and $\alpha$ is a parameter which needs to be chosen either sufficiently large or suitably small, \cf{} Kabel et al.~\cite{Kabel2015}.\\
Keeping the former in mind, we turn our attention to the offline training. We sample $N_s$ quadruples of input stiffnesses and fiber orientations $(\ffC^s_1, \ffC^s_2, \lambda^s_1, \lambda^s_2)$, generate the respective microstructures and compute the effective stiffnesses $\effective{\ffC}^s$.  We denote the generated training data by as sequence of quintuples $\set{\left( \effective{\ffC}^s, \ffC^s_{1}, \ffC^s_{2}, \lambda^s_1, \lambda^s_2 \right)}_{s=1}^{N_s}$ where $s$ enumerates the sample index. The actual sampling process will be discussed in Section~\ref{sec:material_sampling}. For the moment, we assume the training data to be given and fixed.\\
We express the offline training as an optimization problem
\begin{equation} \label{eq:optim_problem}
	J\left(\vec{p}, \vec{q}, \vec{v}\right) \longrightarrow \min_{\vec{p}, \vec{q}, \vec{v}}
\end{equation}
involving the objective function
\begin{equation} \label{eq:loss_function}
	J\left(\vec{p}, \vec{q}, \vec{v}\right) = \frac{1}{N_b} \sqrt[q]{\sum_{i=1}^{N_b} \left(\frac{\normg{\effective{\ffC}^s - \DMNLIN{\Lambda}\left(\ffC^s_{1}, \ffC^s_{2},  \lambda^s_{1}, \lambda^s_{2}, \vec{p}, \vec{q}, \vec{v},\right)}_p}{\normg{ \effective{\ffC}^s }_p}\right)^q} + \lambda \left( \sum_{i=1}^{2^K} \langle v_i \rangle_{+}  - 1\right)^2.
\end{equation}
The quadratic penalty term encodes the mixing constraint
\begin{equation}
	\sum_{i=1}^{2^K} w^i_{K+1} = 1.
\end{equation}
We implemented the offline training in PyTorch~\cite{paszke2017automatic}, see Gajek et al.~\cite{Gajek2020}, making use of the framework's automatic differentiation capabilities to solve the regression problem \eqref{eq:optim_problem} by means of accelerated stochastic gradient descent methods using mini batches of size $N_b$. An epoch $j$ consists of evaluating \ref{eq:DMN_effective_stiffness_parameter} for all samples of the respective mini batch, evaluating the loss function~\eqref{eq:loss_function}, computing the gradients ${\partial J}/{\partial \vec{p}}\left(\vec{p}_{j}, \vec{q}_{j}, \vec{v}_{j}\right)$, ${\partial J}/{\partial \vec{q}}\left(\vec{p}_{j}, \vec{q}_{j}, \vec{v}_{j}\right)$ and ${\partial J}/{\partial \vec{v}}\left(\vec{p}_{j}, \vec{q}_{j}, \vec{v}_{j}\right)$ by means of automatic differentiation and updating the fitting parameters 
\begin{equation}
	\vec{p}_{j+1} = \vec{p}_{j} - \alpha_{p} \frac{\partial J}{\partial \vec{p}}\left(\vec{p}_{j}, \vec{q}_{j}, \vec{v}_{j}\right), \quad \vec{q}_{j+1} = \vec{q}_{j} - \alpha_{q} \frac{\partial J}{\partial \vec{q}}\left(\vec{p}_{j}, \vec{q}_{j}, \vec{v}_{j}\right) \quad \textrm{and} \quad \vec{v}_{j+1} = \vec{v}_{j} - \alpha_{v} \frac{\partial J}{\partial \vec{v}}\left(\vec{p}_{j}, \vec{q}_{j}, \vec{v}_{j}\right).
\end{equation}
During offline training, it may happen that a portion of weights $w^i_{K+1}$ becomes equal to zero, and remains zero due to the vanishing gradient. Liu-Wu~\cite{Liu2019} removed such sub-trees from the binary tree by deleting nodes and merging the respective subtrees. In this work, we follow Liu-Wu and compress the binary tree to speed-up the training, and eventually, the online evaluation. Fig.~\ref{fig:tree_compression} shows a schematic of how to remove laminates from the binary tree. The former illustrates a weighted tree with edge weights corresponding to the propagated weights $w^i_{K+1}$. Remember that the volume fractions of the individual laminates are computed from these weights by normalization.
During training, superfluous laminates are removed from the binary tree to minimize the number of computed laminate homogenizations~\ref{eq:laminate_homogenization_linear}. The former happens dynamically during every forward pass. For the example in Fig.~\ref{fig:tree_compression}, only nine out of $15$ laminate homogenization functions are computed in a forward pass, resulting in a speed-up of about $40\%$ compared to a perfect binary tree. 

\subsection{Online evaluation}  
\label{sec:online_evaluation}

Eventually, we seek to employ DMNs to speed up a two-scale simulation, \ie for every Newton iteration and at every Gauss point of a finite element model, a deep material network needs to be integrated implicitly. At the same time, we want to account for arbitrary eigenvalues $\lambda_1$ and $\lambda_2$ of the fiber orientation tensor at every Gauss point of the macro simulation. In this work, we extend the fast and flexible solution technique, introduced in Gajek et al.~\cite{Gajek2020}, to compute the effective stress of a two-phase DMN.\\
We restrict to the two-potential framework of small-strain isothermal generalized standard materials (GSM) \cite{HalphenNguyen}. In $d$ spatial dimensions, a GSM is a quadruple ($Z$, $\freeenergy$, $\dissipation$, $\statev_0$) $\in \GSM$ comprising a (Banach) vector space $Z$ of internal variables, a free energy density $\freeenergy: \Sym{d} \times Z \rightarrow \ffR$, a dissipation potential $\dissipation: Z \rightarrow \ffR \cup \set{+\infty}$ and $\statev_0 \in Z$ serves as the initial condition. We assume that the dissipation potential $\dissipation$ is proper, convex, lower semi-continuous and satisfies $\dissipation(0)=0$ as well as $0 \in \partial \dissipation(0)$, where $\partial \dissipation$ denotes the sub-differential of the convex function $\dissipation$. For the complete documentation of the GSM model structure in a continuous setting, see, \eg Gajek et al.~\cite{Gajek2020}.\\
Suppose that the two GSMs $\mathcal{G}_1 = (Z_1, \freeenergy_1, \dissipation_1, \statev_{0,1})$ and $\mathcal{G}_2 = (Z_2, \freeenergy_2, \dissipation_2, \statev_{0,2})$ are given. A time discretization of both phases $i\in\{1,2\}$ by the implicit Euler method gives rise to the formulae for discretized stress and Biot's equation, respectively, 
\begin{equation}
	\fmicrostress^{n+1}_i = \frac{\partial \freeenergy_i}{\partial \fmicrostrain} \left(\fmicrostrain^{n+1}_i, \statev^{n+1}_i\right) \quad \textrm{and} \quad  \frac{\partial \freeenergy_i}{\partial \statev} \left(\fmicrostrain^{n+1}_i, \statev^{n+1}_i\right) + \partial \dissipation_i\left(\frac{\statev^{n+1}_i - \statev^n_i}{\triangle t}\right) \ni 0.
\end{equation}
Here, $\triangle t = t^{n+1} - t^{n}$ denotes the time increment and the superscript $n$ refers to the $n$-th time step at time $t_n$. Due to the time discretization and freezing of the internal variables $\statev^n_i$ at time $t_n$, each GSM reduces to a nonlinear elastic material, \cf{} Lahellec-Suquet \cite{Lahellec2007_2}. The condensed free energy ${\condensedpotential_i: \Sym{d} \times Z_i \rightarrow \ffR}$,
\begin{equation}
	\condensedpotential_i\left(\fmicrostrain^{n+1}_i, \statev^n_i\right) = \inf_{\statev^{n+1}_i \in Z_i}\left(\freeenergy_i\left(\fmicrostrain^{n+1}_i, \statev^{n+1}_i\right) + \triangle t \, \dissipation_i\left(\frac{\statev^{n+1}_i - \statev^n_i}{\triangle t}\right) \right)
\end{equation}
is solely dependent on the input strain $\fmicrostrain^{n+1}_i$ and the internal variables $\statev^{n}_i$ of the last (converged) time step. Then, the stress response reads
\begin{equation}
	\fmicrostress^{n+1}_i = \frac{\partial \condensedpotential_i}{\partial \fmicrostrain} \left(\fmicrostrain^{n+1}_i, \statev^{n}_i\right).
\end{equation}
For the sake of readability, we omit explicit reference to time step $n+1$. First, let us collect the lamination directions of all laminates in a single vector $\vec{\fn} \in (\ffR^d)^{2^K-1}$ with the same ordering that we used for the vectors $\vec{p}$ and $\vec{q}$, \ie
\begin{equation}\label{eq:normal_ordering}
	\vec{\fn} = \left[n_K^1, n_K^2,\ldots n_K^{2^{K-1}}, n_{K-1}^1, n_{K-1}^2, \ldots, n_{K-1}^{2^{K-2}}, \ldots, n_2^1, n_2^2, n_1^1\right].
\end{equation}
We consider the displacement jump vector $\vec{\fa}\in (\ffR^d)^{2^K-1}$, which inherits its ordering from $\vec{\fn}$ and the vector of strains $\vec{\fmicrostrain} = \left[\fmicrostrain_1,\fmicrostrain_2,\ldots,\fmicrostrain_{2^K}\right] \in (\Sym{d})^{2^K}$. By introducing the gradient operator $\fA_{\lambda_1\lambda_2}:(\ffR^d)^{2^K-1} \rightarrow (\Sym{d})^{2^K}$, we express the phase strains 
\begin{equation}
	\vec{\fmicrostrain} = \vec{\fmacrostrain} + \fA_{\lambda_1\lambda_2} \vec{\fa}
\end{equation}
\wrt the macro strain $\fmacrostrain$ and the unknown displacement jumps $\vec{\fa}$. Here, the shorthand notation $\vec{\fmacrostrain} = \left[\fmacrostrain,\fmacrostrain,\ldots,\fmacrostrain\right] \in (\Sym{d})^{2^K}$ is used. Indeed, for the work at hand, the gradient operator, which encodes the DMN's topology and lamination directions, depends on the fiber orientation parameters $\lambda_1$ and $\lambda_2$. To illustrate this concept, consider the following example. For a two-phase DMN of depth three, $\fA_{\lambda_1\lambda_2}$ takes the following form 
\begin{equation}\label{eq:gradient_operator_DMN}
	\fA_{\lambda_1\lambda_2} = \begin{bsmallmatrix}
		- c^2_3 N^1_{3, \lambda_1\lambda_2} & 0 & 0 & 0 & -c^2_2 N^1_{2, \lambda_1\lambda_2} & 0 & -c^2_1 N^1_{1, \lambda_1\lambda_2}\\
		\phantom{-}c^1_3 N^1_{3, \lambda_1\lambda_2} & 0 & 0 & 0 & -c^2_2 N^1_{2, \lambda_1\lambda_2} & 0 & -c^2_1 N^1_{1, \lambda_1\lambda_2}\\
		0 & -c^4_3 N^2_{3, \lambda_1\lambda_2} & 0 & 0 & \phantom{-}c^1_2 N^1_{2, \lambda_1\lambda_2} & 0 & -c^2_1 N^1_{1, \lambda_1\lambda_2}\\
		0 & \phantom{-}c^3_3 N^2_{3, \lambda_1\lambda_2} & 0 & 0 & \phantom{-}c^1_2 N^1_{2, \lambda_1\lambda_2} & 0 & -c^2_1 N^1_{1, \lambda_1\lambda_2}\\
		0 & 0 & -c^6_3 N^3_{3, \lambda_1\lambda_2} & 0 & 0 & -c^4_2 N^2_{2, \lambda_1\lambda_2} & \phantom{-}c^1_1 N^1_{1, \lambda_1\lambda_2}\\
		0 & 0 & \phantom{-}c^5_3 N^3_{3, \lambda_1\lambda_2} & 0 & 0 & -c^4_2 N^2_{2, \lambda_1\lambda_2} & \phantom{-}c^1_1 N^1_{1, \lambda_1\lambda_2}\\
		0 & 0 & 0 & -c^8_3 N^4_{3, \lambda_1\lambda_2} & 0 & \phantom{-}c^3_2 N^2_{2, \lambda_1\lambda_2} & \phantom{-}c^1_1 N^1_{1, \lambda_1\lambda_2}\\
		0 & 0 & 0 & \phantom{-}c^7_3 N^4_{3, \lambda_1\lambda_2} & 0 & \phantom{-}c^3_2 N^2_{2, \lambda_1\lambda_2} & \phantom{-}c^1_1 N^1_{1, \lambda_1\lambda_2}
	\end{bsmallmatrix}
\end{equation}
with the symmetrization operators \wrt the lamination direction $\fn^i_{k, \lambda_1\lambda_2} = \fn^i_k(\lambda_1, \lambda_2)$
\begin{equation}
	\fN^i_{k, \lambda_1\lambda_2}\fa = \frac{1}{2}\left( \fa \otimes \fn^i_{k, \lambda_1\lambda_2} +  \fn^i_{k, \lambda_1\lambda_2} \otimes \fa \right)
\end{equation}
as building blocks. Since we defined $\fn^i_k(\lambda_1\lambda_2)$ to depend on the parameters $\lambda_1$ and $\lambda_2$ explicitly, \cf{} Section~\ref{sec:fiber_orientation_interpolation}, the gradient operator depends on the fiber orientation, as well. We account for this situation in our notation, \ie we write $\fA_{\lambda_1\lambda_2}$ and $\fN^i_{k, \lambda_1\lambda_2}$. For the application at hand, \ie integrating a DMN at every Gauss point during a two-scale simulation, the fiber orientation parameters $\lambda_1$ and $\lambda_2$ are held fixed. Indeed, we assume that the microstructure does not evolve under the applied load.\\
Let us define the vector of internal variables of the last converged time step $\vec{\statev}^{\, n} = \left[\statev_1^n, \statev_2^n, \statev_3^n, \dots, \statev_{2^K}^n\right] \in \mathcal{Z} := Z_1 \oplus Z_2 \oplus \cdots \oplus Z_1 \oplus Z_2$ and let $\overline{\condensedpotential}:(\Sym{d})^{2^K}\!\!\!\!\times\mathcal{Z} \rightarrow \ffR$ be the averaged condensed free energy of the flattened laminate 
\begin{equation}
	\overline{\condensedpotential}(\vec{\fmicrostrain}, \vec{\statev}^{\, n}) = \sum_{i=1}^{2^K} w^i_{K+1} \condensedpotential_{i}(\fmicrostrain_i, \statev_i^n) \quad \textrm{where} \quad \condensedpotential_i= \left\{ \begin{array}{rl}
		\condensedpotential_1, & i\textrm{ odd,}\\
		\condensedpotential_2, & i\textrm{ even,}
	\end{array}
	\right.
\end{equation}
alternating between the two given condensed free energies $\condensedpotential_1$ and $\condensedpotential_2$. Then, we wish to solve the Euler-Lagrange equation of the DMN
\begin{equation}\label{eq:euler_lagrange}
	\fA^T_{\lambda_1\lambda_2} \fW\vec{\fmicrostress}(\vec{\fmacrostrain} + \fA_{\lambda_1\lambda_2} \vec{\fa}, \vec{\statev}^{\, n}) = 0
\end{equation}
for the unknown displacement jumps $\vec{\fa}$, where
\begin{equation}
	\vec{\fmicrostress} = \left[\fmicrostress_1, \dots, \fmicrostress_{2^K} \right] \in (\Sym{d})^{2^K} \quad \textrm{with} \quad \fmicrostress_i = \frac{\partial \condensedpotential_i}{\partial \fmicrostrain}(\fmicrostrain_i, \statev_i^n),
\end{equation}
is the vector of phase stresses. The strain-wise ``mass'' matrix $\fW:\Sym{d}^{2^K} \rightarrow \Sym{d}^{2^K}$ 
\begin{equation}\label{eq:mass_matrix}
	\fW(\vec{\fmicrostrain}) = \left(w^1_{K+1}\fmicrostrain_1,w^2_{K+1}\fmicrostrain_2,\ldots,w^{2^K}_{K+1}\fmicrostrain_{2^K}\right),
\end{equation}
associates the weight $w^i_{K+1}$, $i=1\dots2^K$, to the corresponding phase strain $\fmicrostrain_i$ . We solve the Euler-Lagrange equation \eqref{eq:euler_lagrange} by Newton's method. For an initial guess ${\vec{\fa}_0 \in (\ffR^d)^{N-1}}$, the displacement jump vector $\vec{\fa}$ is iteratively updated $\vec{\fa}_{j+1} = \vec{\fa}_{j} + s_j\,\triangle \vec{\fa}_j$, where the increment $\triangle\vec{\fa}_j \in \left(\ffR^d\right)^{2^K-1}$ solves the linear system
\begin{equation} \label{eq:newton_complex}
	\left[\fA^T_{\lambda_1\lambda_2} \fW \frac{\partial \vec{\fmicrostress}}{\partial \vec{\fmicrostrain}}(\vec{\fmacrostrain} + \fA_{\lambda_1\lambda_2}\vec{\fa_j}, \vec{\statev}^{\, n}) \fA_{\lambda_1\lambda_2}\right] \triangle \vec{\fa}_j = - \fA^T_{\lambda_1\lambda_2} \fW \vec{\fmicrostress}(\vec{\fmacrostrain} + \fA_{\lambda_1\lambda_2}\vec{\fa_j}, \vec{\statev}^{\, n}).
\end{equation}
A step size $s_j\in(0,1]$ strictly less than unity may arise by backtracking. The Jacobian $\partial \vec{\fmicrostress} / \partial \vec{\fmicrostrain}(\vec{\fmacrostrain} + \fA_{\lambda_1\lambda_2}\vec \fa_j, \vec{\statev}^{\, n})$ is a block-diagonal matrix containing the algorithmic tangents of the DMN's input materials, \ie
\begin{equation}
	\frac{\partial \vec{\fmicrostress}}{\partial \vec{\fmicrostrain}}(\vec{\fmicrostrain}, \vec{\statev}^{\, n}) = \textrm{block-diag}\left(\frac{\partial^2 \condensedpotential_1}{\partial \fmicrostrain \partial \fmicrostrain}(\fmicrostrain_1, \statev_1^n), \dots, \frac{\partial^2 \condensedpotential_{2^K}}{\partial \fmicrostrain \partial \fmicrostrain}(\fmicrostrain_{2^K}, \statev_{2^K}^n) \right).
\end{equation} 
Upon convergence, the phase strains $\vec{\fmicrostrain} = \vec{\fmacrostrain} + \fA_{\lambda_1\lambda_2} \vec{\fa}$ and, subsequently, the effective stress
\begin{equation}
\fmacrostress = \sum_{i=1}^{2^K} w^i_{K+1} \fmicrostress_i(\fmicrostrain_i, \statev_i^n)
\end{equation}
are computed by averaging. To determine the algorithmic tangent of the deep material network, for a start, the linear system
\begin{equation}\label{eq:lin_system_da_dE}
	\left[\fA^T_{\lambda_1\lambda_2} \fW \frac{\partial \vec{\fmicrostress}}{\partial \vec{\fmicrostrain}}(\vec{\fmacrostrain} + \fA_{\lambda_1\lambda_2}\vec{\fa_j}, \vec{\statev}^{\, n}) \fA_{\lambda_1\lambda_2}\right]\frac{\partial \vec{\fa}}{\partial \fmacrostrain} = -\fA^T_{\lambda_1\lambda_2} \fW \frac{\partial \vec{\fmicrostress}}{\partial \fmacrostrain}(\vec{\fmacrostrain} + \fA_{\lambda_1\lambda_2} \vec{\fa}, \vec{\statev}^{\, n}) 
\end{equation}
is solved for $\partial \vec{\fa}/\partial \fmacrostrain$. Then, the algorithmic tangent may be represented in the form 
\begin{equation}
	\ffC^\textrm{algo} \equiv \frac{\partial \fmacrostress}{\partial \fmacrostrain} = \left[\ffI_\textrm{s}, \ffI_\textrm{s}, \dots, \ffI_\textrm{s}\right]^T \fW \left[ \frac{\partial \vec{\fmicrostress}}{\partial \fmacrostrain}(\vec{\fmacrostrain} + \fA_{\lambda_1\lambda_2} \vec{\fa}, \vec{\statev}^{\, n}) + \frac{\partial \vec{\fmicrostress}}{\partial \vec{\fmicrostrain}}(\vec{\fmacrostrain} + \fA_{\lambda_1\lambda_2} \vec{\fa}, \vec{\statev}^{\, n}) \fA_{\lambda_1\lambda_2} \frac{\partial \vec{\fa}}{\partial \fmacrostrain} \right],
\end{equation}
where $\partial \vec{\fmicrostress} / \partial {\fmacrostrain}$ denotes the vector of algorithmic tangents
\begin{equation}
	\frac{\partial \vec{\fmicrostress}}{\partial \fmacrostrain}(\vec{\fmicrostrain}, \vec{\statev}^{\, n}) = \left[\frac{\partial^2 \condensedpotential_1}{\partial \fmicrostrain \partial \fmicrostrain}(\fmicrostrain_1, \statev_1^n), \dots, \frac{\partial^2 \condensedpotential_{2^K}}{\partial \fmicrostrain \partial \fmicrostrain}(\fmicrostrain_{2^K}, \statev_{2^K}^n) \right]
\end{equation}
and $\left[\ffI_\textrm{s}, \ffI_\textrm{s}, \dots, \ffI_\textrm{s}\right] \in \Sym{d}^{2^K}$, $\IDsym: \Sym{d} \rightarrow \Sym{d}$,  is a vector of the identity operators on $\Sym{d}$. By comparing equation \eqref{eq:lin_system_da_dE} to equation \eqref{eq:newton_complex}, we observe that both problems share the same linear operator, but with different right hand sides. When using a direct solver, \eg a Cholesky decomposition, it is recommended to reuse the matrix decomposition for reasons of efficiency.\\
To reduce the number of degrees of freedom and, thus, to speed up the solution process, we exploit that some weights become zero during training as explained in the previous section. We learned that in the offline training, we can dynamically build a binary tree with simplified topology but identical effective behavior. This is also true for the offline evaluation of the DMN. The DMN's topology is encoded by the gradient operator $\fA_{\lambda_1\lambda_2}$. Deleting laminate blocks from the binary tree is equivalent to deleting the associated rows and columns of $\fA_{\lambda_1\lambda_2}$. For the example shown in Fig.~\ref{fig:tree_compression}, we obtain a (reduced) gradient operator of the form
\begin{equation}\label{eq:gradient_operator_DMN_compressed}
	\fA_{\lambda_1\lambda_2} = \begin{bsmallmatrix}
		0 & 0 & 0 & 0 & 0 & 0 & -c^2_2 N^1_2 & 0 & -c^2_1 N^1_1\\
		0 & 0 & 0 & -c^4_3 N^2_3 & 0 & 0 & \phantom{-}c^1_2 N^1_2 & 0 & -c^2_1 N^1_1\\
		0 & 0 & 0 & \phantom{-}c^3_3 N^2_3 & 0 & 0 & \phantom{-}c^1_2 N^1_2 & 0 & -c^2_1 N^1_1\\
		-c^{10}_4 N^5_4 & 0 & 0 & 0 & -c^6_3 N^3_3 & 0 & 0 & -c^4_2 N^2_2 & \phantom{-}c^1_1 N^1_1\\
		\phantom{-}c^{9}_4 N^5_4 & 0 & 0 & 0 & -c^6_3 N^3_3 & 0 & 0 & -c^4_2 N^2_2 & \phantom{-}c^1_1 N^1_1\\
		0 & 0 & 0 & 0 & \phantom{-}c^5_3 N^3_3 & 0 & 0 & -c^4_2 N^2_2 & \phantom{-}c^1_1 N^1_1\\
		0 & -c^{14}_4 N^7_4 & 0 & 0 & 0 & -c^8_3 N^4_3 & 0 & \phantom{-}c^3_2 N^2_2 & \phantom{-}c^1_1 N^1_1\\
		0 & \phantom{-}c^{13}_4 N^7_4 & 0 & 0 & 0 & -c^8_3 N^4_3 & 0 & \phantom{-}c^3_2 N^2_2 & \phantom{-}c^1_1 N^1_1\\
		0 & 0 & -c^{16}_4 N^8_4 & 0 & 0 & \phantom{-}c^7_3 N^4_3 & 0 & \phantom{-}c^3_2 N^2_2 & \phantom{-}c^1_1 N^1_1\\
		0 & 0 & \phantom{-}c^{15}_4 N^8_4 & 0 & 0 & \phantom{-}c^7_3 N^4_3 & 0 & \phantom{-}c^3_2 N^2_2 & \phantom{-}c^1_1 N^1_1
	\end{bsmallmatrix}
\end{equation}
with a reduced size, where we dropped the subscript $\left(\cdot\right)_{\lambda_1\lambda_2}$ of $\fN^i_{k, \lambda_1\lambda_2}$ for readability.

\section{Identifying a DMN surrogate model} 
\label{sec:identification_dmn}

This section is dedicated to the identification of the DMN surrogate model. We discuss the pre-processing steps, \ie finding the necessary resolution and the appropriate size of the volume elements, and investigate the discretization of the fiber orientation triangle.  Subsequently, we explain the sampling of the training data, the offline training and the validation of the DMN surrogate model on the fiber orientation triangle. All computations were performed on a workstation equipped with two AMD EPYC $7642$ with $48$ physical cores each, enabled SMT and $1024 \unit{GB}$ of DRAM.

\subsection{Short glass fiber reinforced polyamide} 
\label{sec:mat_param}

For the work at hand, we focus on a short glass fiber reinforced polyamide. We consider E-glass fibers with a length of $L_\textrm{f} = 200 \unit{\mu m}$ and a diameter of $D_\textrm{f} = 10 \unit{\mu m}$. The glass fibers are assumed to be isotropic, linear elastic. The fiber volume fraction is set to $c_\textrm{f} = 16 \%$ corresponding to a fiber mass fraction of approx.\ $30 \%$. The matrix is assumed to be governed by $J_2$-elastoplasticity, \cf{} Chapter 3 in Simo-Hughes \cite{SimoHughes1998} with an exponential-linear hardening
\begin{equation}
	\microstress_\textrm{Y} = \microstress_0 + k_{\infty} \varepsilon_\textrm{p} + (\microstress_{\infty} - \microstress_0) \left(1 - \exp\left(-\frac{k_0 - k_{\infty}}{\microstress_{\infty} - \microstress_0}  \, \varepsilon_\textrm{p}\right)\right).
\end{equation}
The mechanical properties used in the simulation, taken from Doghri et al.\ \cite{Doghri2011}, are summarized in Tab.~\ref{tab:materialParameters_FRP}.
\begin{table}[h!]
	\begin{center}
		\begin{tabular}{l l l l l l l}
			\hline
			Matrix & $E=2.1 \unit{GPa}$ & $\nu=0.3$ & $\microstress_0=29 \unit{MPa}$ & $\microstress_{\infty}= 61.7 \unit{MPa}$ & $k_0=10.6 \unit{GPa}$ & $k_{\infty}=139 \unit{MPa}$\\
			Fibers & $E=72 \unit{GPa}$  & $\nu=0.22$\\
			\hline
		\end{tabular}
	\end{center}
	\caption{Material parameters for the short glass fiber reinforced polyamide \cite{Doghri2011}}
	\label{tab:materialParameters_FRP}
\end{table}
We rely upon the Sequential Addition and Migration (SAM) method~\cite{SAM} for generating periodic volume elements with prescribed volume fraction and second order fiber orientation tensor. The fiber length $L_\textrm{f}$, fiber diameter $D_\textrm{f}$, fiber volume fraction $c_\textrm{f}$ and the axis aligned fiber orientation tensor $\tuckersecond$, \ie $\lambda_1$ and $\lambda_2$, serve as input parameters for the SAM method. Please note that we only consider fiber orientation states with $\lambda_3 \ge 0.01$, as purely planar fiber orientation states cannot be generated at high filler content essentially for geometric reasons, see Schneider~\cite{SAM} for a discussion.

\subsection{On the necessary resolution and the size of the RVE}
\label{sec:res_RVE_study}

For a start, we study the resolution necessary to obtain accurate effective properties in the purely elastic case. For this purpose, we consider cubic microstructures with an edge length of $L=384 \unit{\mu m}$, \ie roughly twice the fiber length of $L_\textrm{f} = 200 \unit{\mu m}$. We compute the effective stiffness with the help of an FFT-based computational micromechanics code~\cite{MoulinecSuquet1994,MoulinecSuquet1998} as described in Schneider~\cite{BB2018}, using the staggered grid discretization \cite{willot2014fourier, willot2015fourier} and the conjugate gradient solver \cite{Zeman2010,BrisardDormieux2010}.\\
We consider the extreme orientations individually, \ie unidirectional, isotropic and planar isotropic fiber orientation as shown in Fig.~\ref{fig:orienation_triangle_microstructures}, and vary the resolution from $1.7$ to $13.3$ voxels per fiber diameter in equidistant steps. This corresponds to volume element discretizations with $64^3$ to $512^3$ voxels. We measure the error relative to the effective stiffness $\effective{\ffC}$ and choose a resolution of $20$ voxels per fiber diameter, \ie a discretized by $768^3$ voxels, as the reference.
\begin{figure}[H]
	\centering
	\begin{subfigure}{\textwidth}
		\centering
		\includegraphics[height=5.0mm]{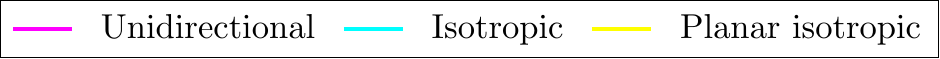}
	\end{subfigure}
	\begin{subfigure}{0.45\textwidth}
		\centering
		\includegraphics[width=0.98\textwidth]{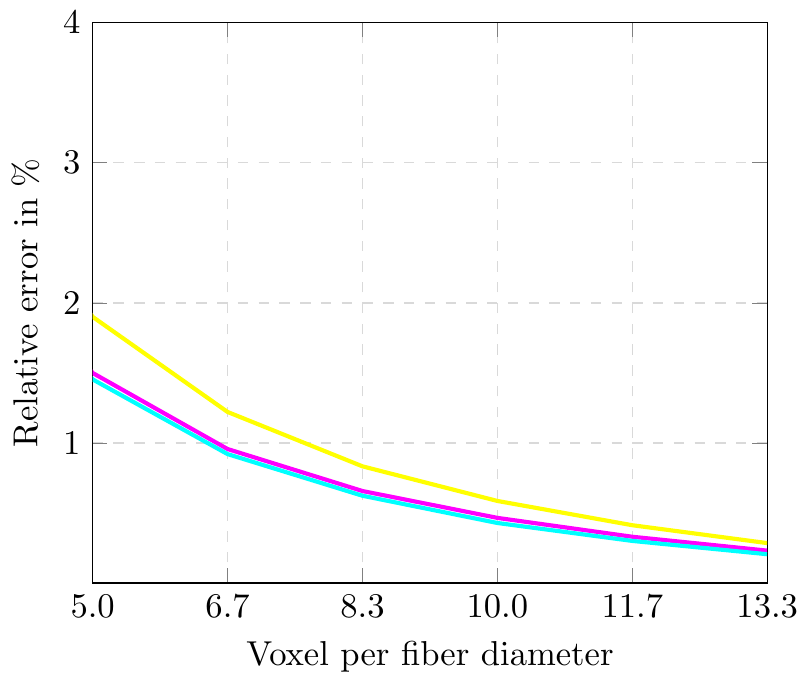}
		\caption{Relative error vs. resolution}
		\label{fig:res_study}
	\end{subfigure}
	\begin{subfigure}{0.45\textwidth}
		\centering
		\includegraphics[width=\textwidth]{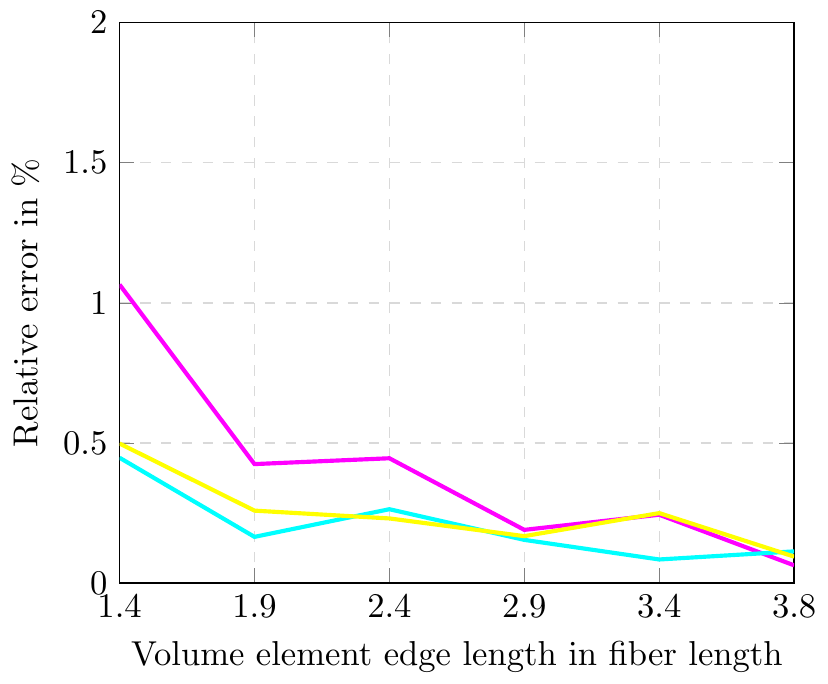}
		\caption{Relative error vs. volume element size}
		\label{fig:RVE_study}
	\end{subfigure}
	\caption{Relative error of the effective stiffness}
	\label{fig:res_RVE_study}
\end{figure}
Fig.~\ref{fig:res_study} shows the relative error of the effective stiffness computed by the Frobenius norm of the corresponding Voigt matrices. For the crudest resolution, \ie five voxels per fiber diameter, the relative error is well below $2\%$ for all three considered fiber orientations. Notice that the relative error for the volume element with planar isotropic fiber orientations is consistently larger than the error for the unidirectional and isotropic orientations. As expected, the relative error decreases with increasing resolution. At a resolution of $6.7$ voxels per fiber diameter, the relative errors of the unidirectional and isotropic fiber orientation fall below $1\%$. For $8.3$ voxels per fiber diameter, the relative error of the planar isotropic fiber orientation is below $1\%$, as well. For this article, we consider a resolution of $6.7$ voxels per fiber diameter as sufficient, \ie relative errors below $1\%$ for isotropic and unidirectional fiber orientation and an error slightly above $1\%$ for the planar isotropic case. We fix this resolution and focus on finding the size of a representative volume element.
For a resolution of $6.7$ voxels per fiber diameter, we investigate volume elements with edge lengths $L$ ranging from $1.44$ up to $3.84$ fiber lengths corresponding to volume element discretizations with $192^3$ up to $512^3$ voxels.\\ 
To obtain our reference, we generated a volume element with edge lengths of $7.68$ fiber lengths discretized with $1024^3$ voxels. As for studying the necessary resolution, we again consider the relative error in the effective stiffness as our error measure. For the volume elements with edge lengths of $1.9$ fiber length and above, the relative error is well below $0.5\%$ and does not further decrease significantly for increasing volume element size. For this reason, we consider volume elements with an edge length of $L=384\unit{\mu m}$ as sufficient. To sum up, we finally choose a resolution of $6.7$ voxel per fiber length, \ie a voxel size of $1.5\unit{\mu m}$ and a discretization with $256^3$ voxels for the article at hand.

\subsection{Discretization of the fiber orientation triangle}
\label{sec:orientation_discretization}

\begin{figure}[H]
	\centering
	\begin{subfigure}{0.32\textwidth}
		\centering
		\includegraphics[width=\textwidth]{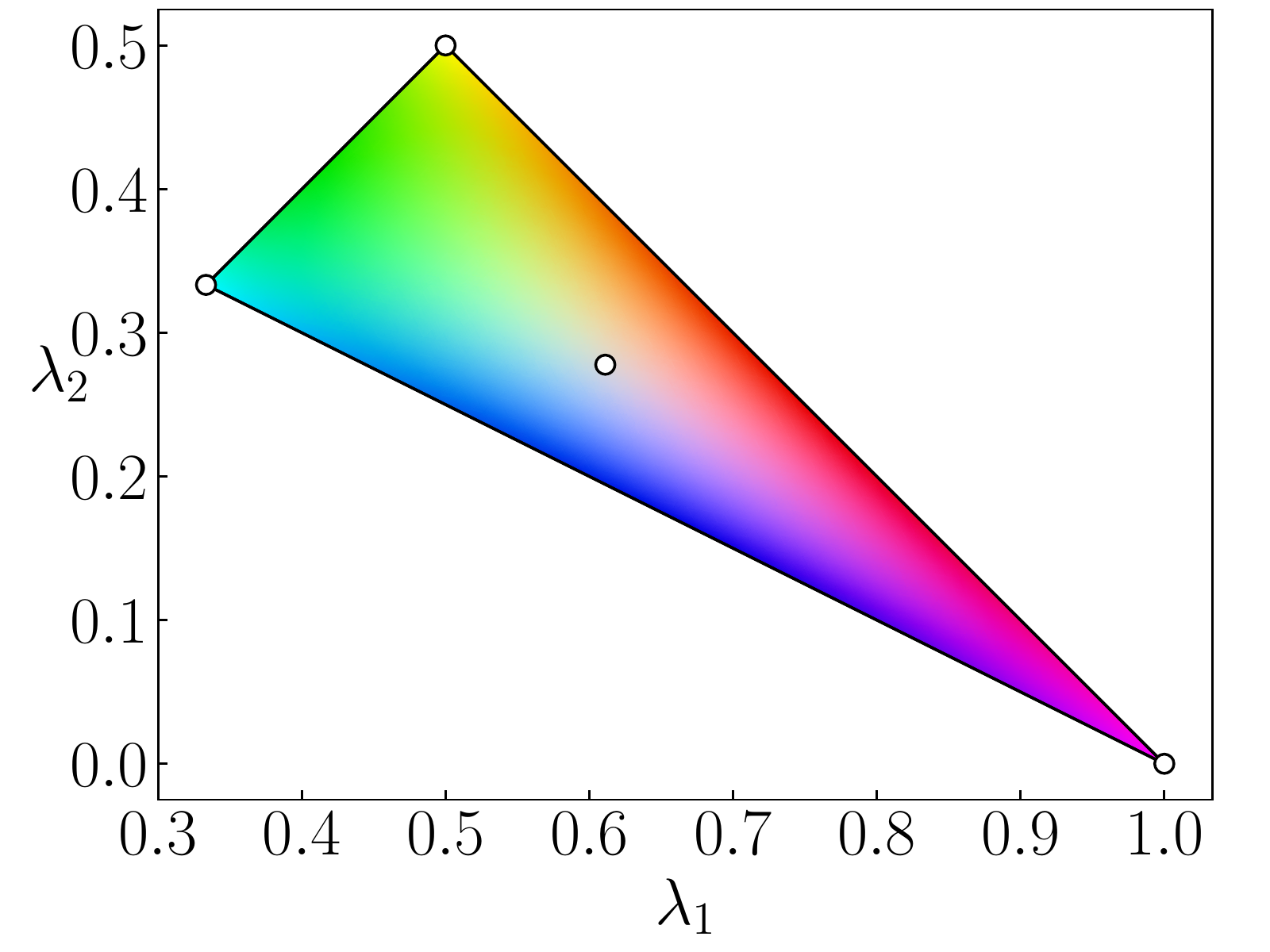}
		\caption{\discone}
		\label{fig:ori_triangle_0}
	\end{subfigure}
	\begin{subfigure}{0.32\textwidth}
		\centering
		\includegraphics[width=\textwidth]{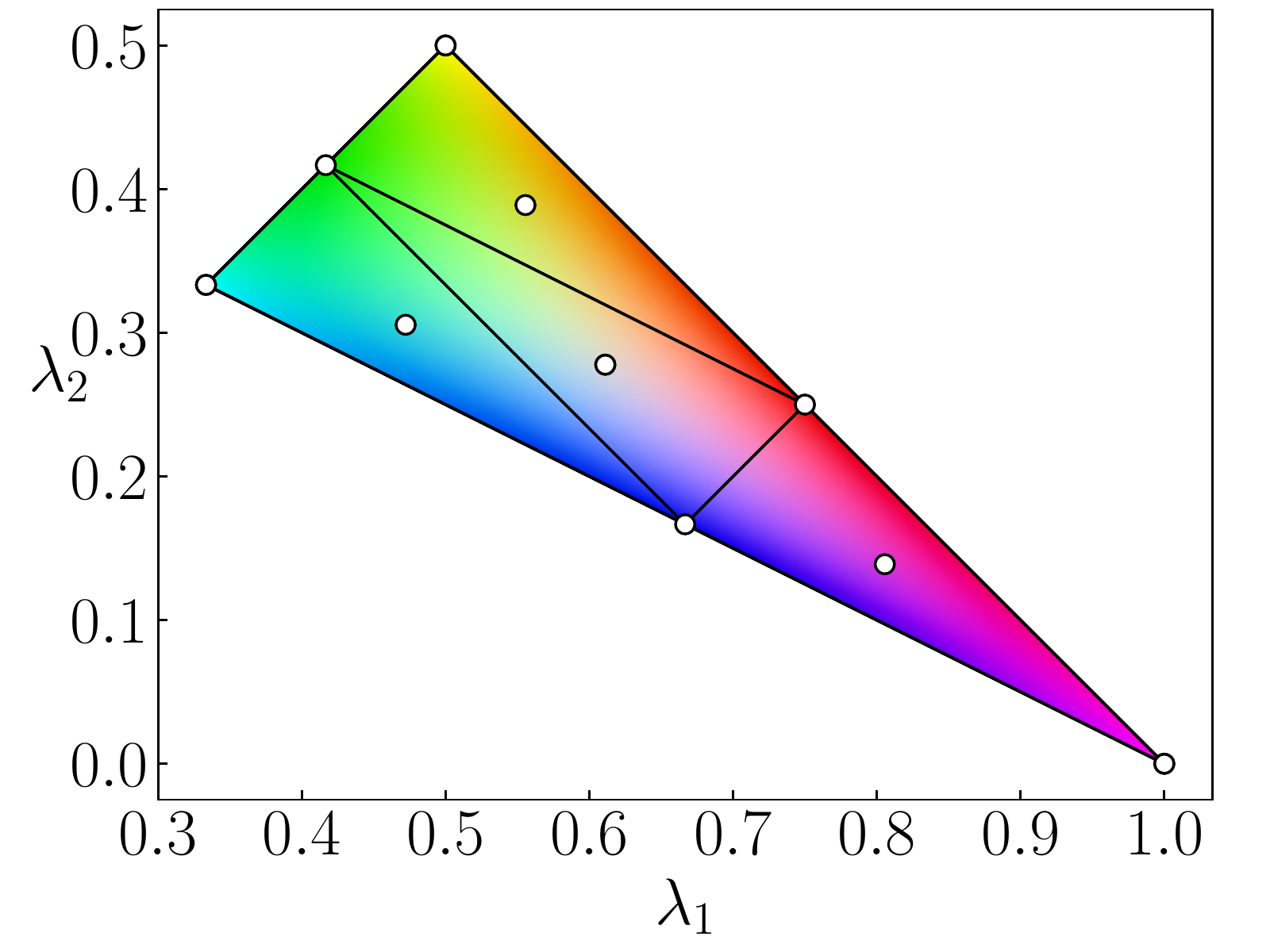}
		\caption{\disctwo}
		\label{fig:ori_triangle_1}
	\end{subfigure}
	\begin{subfigure}{0.32\textwidth}
		\centering
		\includegraphics[width=\textwidth]{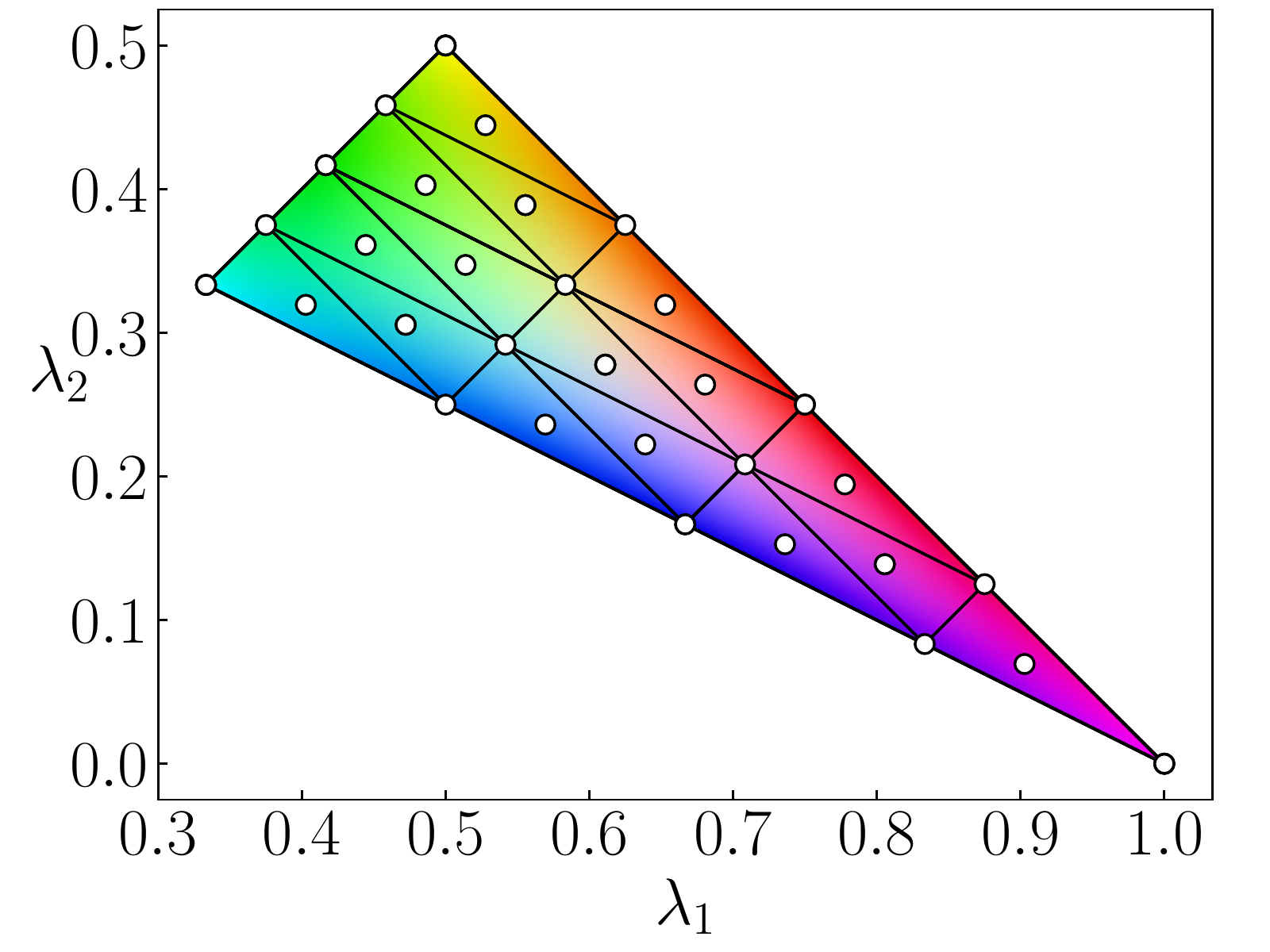}
		\caption{\discthree}
		\label{fig:ori_triangle_2}
	\end{subfigure}
		\caption{The investigated fiber orientation discretizations: (a) four, (b) ten and (c) 31 sampling points}
		\label{fig:ori_disc}
\end{figure}

To generate the linear elastic training data, we seek to sample the space of input stiffnesses and fiber orientations uniformly. Apart from sampling the input stiffnesses, it is possible to sample $\lambda_1$ and $\lambda_2$ as well, \eg via a low-discrepancy sequence such as the Sobol sequence~\cite{Sobol} or via Latin hypercube sampling~\cite{LatinHypercube}. Due to the high dimension of the input space, we follow the former approach for generating tuples of input stiffnesses. The considered fiber orientations are parameterized by a two-dimensional space, and traditional methods are more efficient.\\
We discretize the fiber orientation triangle by partitioning it into four self-similar triangles, which may be subsequently partitioned, as well. We select the three points on the vertices of each triangle plus the centers of the triangles as sampling points for the parameters $\lambda_1$ and $\lambda_2$. We start with the full orientation triangle, \cf{} Fig.~\ref{fig:ori_triangle_0}.  The four sampling points, illustrated by four hollow circles, are the three corners and the center of the orientation triangle. After the first splitting, the orientation triangle comprises four triangles and ten points, \cf{} Fig.~\ref{fig:ori_triangle_1}.  After dividing the triangles one more time, we arrive at $31$ sampling points. For each of these points in fiber orientation space, we generate a single volume element using the SAM method~\cite{SAM} with edge lengths $L=384 \unit{\mu m}$ and a discretization with $256^3$ voxels, \cf{} Section~\ref{sec:res_RVE_study}. In this work, we consider the discretizations shown in Fig.~\ref{fig:ori_disc}, \ie we discretize the orientation triangle with four, ten and $31$ sampling points which we call \discone, \disctwo and \discthree, respectively.
Choosing the sampling points in a hierarchical manner permits us to re-use already generated volume elements for the next finer discretization. For instance, going from \disctwo to \discthree only requires generating $21$ new volume elements. This restriction is not imposed by the SAM~\cite{SAM} algorithm since generating volume elements with the given resolution and edge length is just a matter of milliseconds to seconds. Being able to reuse results from coarser discretizations comes in handy for the inelastic validations we perform in Section~\ref{sec:res_online_evaluation}. In Section~\ref{sec:res_online_evaluation}, we generate $78$ additional volume elements to validate the DMN outside of its training regime and to check if the DMN generalizes sufficiently on the orientation space. For this purpose, we test several different load paths for each volume element, so being able to reuse already computed full field solutions comes in handy to keep the validation effort manageable.

\subsection{Material sampling}
\label{sec:material_sampling}

We turn our attention to the sampling of the input stiffnesses. For the sampling, we take into account that the glass fibers are linear elastic and the polyamide matrix is governed by $J_2$-elastoplasticity. For that reason, we assume that the samples $\ffC_1$, corresponding to the glass fibers, are isotropic, \ie the equation
\begin{equation}
	\ffC_1 = 3 K_1 \, \ffP_1 + 2 G_1 \, \ffP_2
\end{equation}
holds, where $\ffP_1: \Sym{d} \rightarrow \Sph{d}$ and $\ffP_2: \Sym{d} \rightarrow \Dev{d}$ project onto the spherical and deviatoric subspaces of $\Sym{d}$, respectively. Secondly, we assume that the samples corresponding to the polyamide matrix are isotropic minus a rank-one perturbation, \ie
\begin{equation}
	\ffC_2 = 3 K_2 \, \ffP_1 + 2 G_2 \left(\ffP_2 - a \, \fN' \otimes \fN'\right). 
\end{equation}
Here, the tensor $\fN' \in \mathcal{N}$ is normalized and deviatoric, \ie
\begin{equation}
	\mathcal{N} = \set{\fN \in \Sym{d} \ | \ \tr{\fN} = 0, \ \normg{\fN}_\textrm{F}=1}.
\end{equation}
The structure of the second stiffness $\ffC_2$ encompasses the possible algorithmic tangents of $J_2$-elastoplasticity, \cf{} Chapter 3 in Simo-Hughes~\cite{SimoHughes1998}. Sampling $\ffC_1$ and $\ffC_2$ in this way contrasts to Liu et al.~\cite{Liu2018}, Liu and Wu \cite{Liu2019} and Gajek et al.~\cite{Gajek2020}, who sampled (axis aligned) orthotropic stiffnesses.\\
The set of all considered positive definite stiffness tuples $\left(\ffC_1, \ffC_2\right)$ may be parameterized via
\begin{equation}\label{eq:sample_set}
	\left(K_1, G_1, K_2, G_2, a, \fN' \right) \in \ffR^4_{>0} \times \left[0, 1\right) \times \mathcal{N},
\end{equation}
where $K_i$ and $G_i$ have the dimensions of a Young's modulus and $a$ and $\fN'$ are dimensionless. Since the latter set is unbounded, we restrict to the subset of elements $\left(K_1, G_1, K_2, G_2, a, \fN' \right)$ with
\begin{equation}
	K_1 = 1 \unit{GPa}, \quad G_1 = 10^{e_1} \unit{GPa}, \quad K_2 = 10^{e_2} \unit{GPa}, \quad G_2 = 10^{e_3} \unit{GPa}
\end{equation}
and exponents $e_1, e_2, e_3 \in \left[-3, 3\right]$. By fixing the compression modulus $K_1$, we removed the redundancy due to homothetic rescaling via $\left(\ffC_1, \ffC_2\right) \mapsto \left(\lambda \ffC_1, \lambda \ffC_2\right)$ for $\lambda > 0$. For parameterizing the tensor $\fN'$, we make use of an eigenvalue decomposition $\fN' = \fQ \fN \fQ^T$ with an orthogonal $\fQ \in \sop$ and a diagonal $\fN \in \mathcal{N}$ matrix. We parameterize the tensor $\fN$ by spherical coordinates
\begin{equation}\label{eq:N_spherical}
	\fN = \textrm{diag}\left(\sin\left(\alpha\right) \cos\left(\beta\right), \sin\left(\alpha\right) \sin\left(\beta\right), \cos\left(\alpha\right)\right)
\end{equation}
ensuring the condition $\normg{\fN}_\mathrm{F} = 1$ to hold. To account for the vanishing trace, $\tr{\fN} = 0$, we eliminate the angle $\alpha$ in equation \eqref{eq:N_spherical} and arrive at the parameterization
\begin{equation}
	\fN = \frac {1}{\sqrt {{\frac {\cos \left( \beta \right) \sin \left( \beta \right) +1}{2\,\cos \left( \beta \right) \sin \left( \beta \right) +1}}}} \mathrm{diag} \left(-\frac{\sqrt{2} \cos{\left(\beta\right)}}{2 \cos{\left(\beta\right)} + 2 \sin{\left(\beta\right)}}, -\frac{\sqrt{2} \sin{\left(\beta\right)}}{2 \cos{\left(\beta\right)} + 2 \sin{\left(\beta\right)}}, \frac{1}{\sqrt{2}}\right)
\end{equation}
in terms of an single remaining angle $\beta\in\left[0,2\pi\right]$. As in Gajek et al.~\cite{Gajek2020}, the special orthogonal group is parameterized via an axis-angle representation
\begin{equation}
	\fQ: \ffR^3 \rightarrow \ffR^3, \ \fx \mapsto \cos\left(\theta\right)\fx + \sin\left(\theta\right) \fn\times \fx + (1-\cos\left(\theta\right))(\fn\cdot \fx) \fn,
\end{equation}
for the axis $\fn = \left( \sin\left(\psi\right) \cos\left(\varphi\right), \sin\left(\psi\right) \sin\left(\varphi\right),\cos\left(\psi\right) \right)$, and where the conditions $\theta - \sin\left(\theta\right) \in \left[0, \pi\right]$, $\psi \in \left[0, \pi\right]$ and $\varphi \in \left[0, 2\pi\right]$ hold, see Miles~\cite{Miles1965}.\\
To sum up, we consider the following eight degrees of freedom
\begin{equation} \label{eq:param_sampling}
	\left( a, e_1, e_2, e_3, \beta, \theta, \psi, \phi \right)
\end{equation}
with their respective domains specified above. To sample the input space evenly, we sample the parameters \eqref{eq:param_sampling} by the Sobol sequence and, subsequently, construct the stiffness tensors $(\ffC_1, \ffC_2)$. 

\subsection{Offline training}
\label{sec:offline_training_example}

\begin{figure}[H]
	\centering
	\begin{subfigure}{\textwidth}
		\centering
		\includegraphics[height=5.0mm]{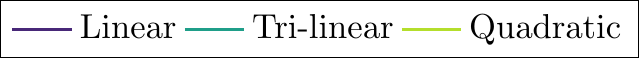}
	\end{subfigure}
	\begin{subfigure}[t]{.49\textwidth}	
		\includegraphics[width=\textwidth]{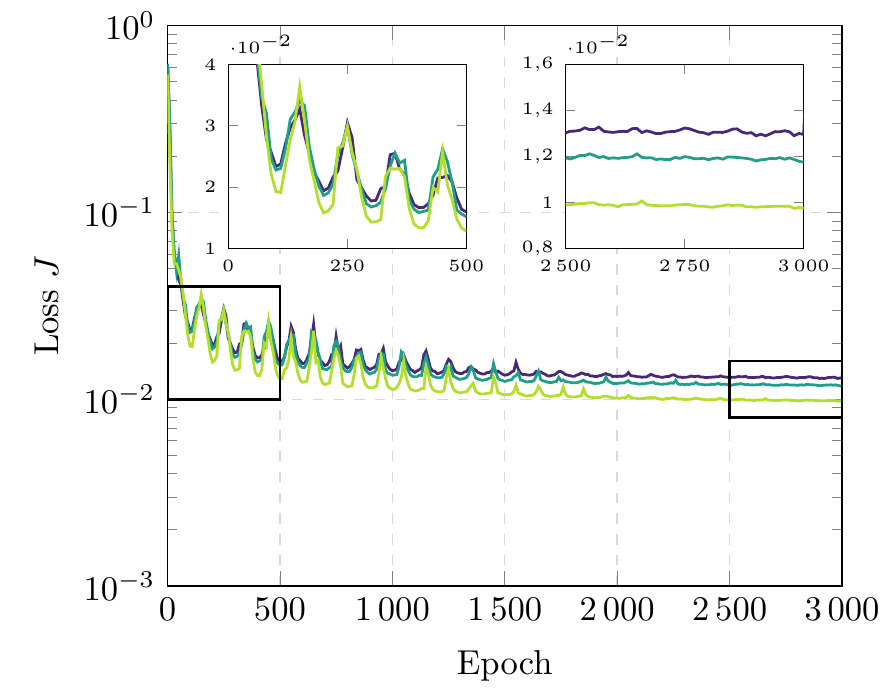}
		\caption{Loss function}
		\label{fig:convergence_offline_FRP2}
	\end{subfigure}
	\begin{subfigure}[t]{.49\textwidth}
		\includegraphics[width=0.98\textwidth]{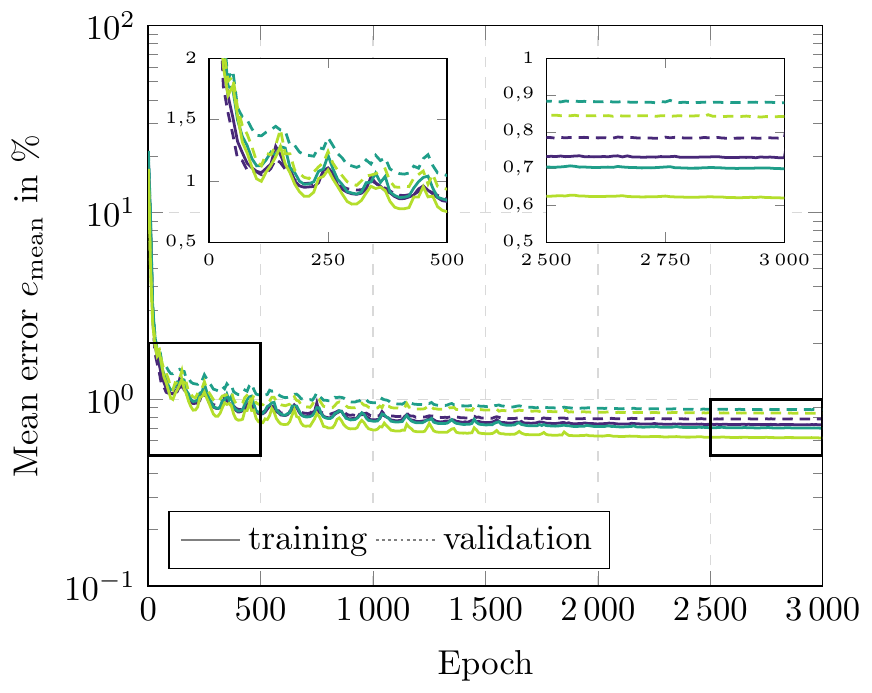}
		\caption{Mean error}
		\label{fig:convergence_offline_FRP3}	
	\end{subfigure}
	\caption{Learning rate, loss function, mean training and validation errors during training for \discthree}
	\label{fig:convergence_offline_FRP}	
\end{figure}

For the offline training, we generate $N_s$ pairs of stiffnesses $\left(\ffC^s_{1}, \ffC^s_{2} \right)$ by the protocol described in Section \ref{sec:material_sampling}. Then, we assign each stiffness tuple to one of the previously generated volume elements in a cyclic fashion. For instance, for the orientation discretization \discone, we assign $\left(\ffC^1_{1}, \ffC^1_{2} \right)$ to the volume element with fiber orientation $\left( \lambda^1_1, \lambda^1_2 \right)$, $\left(\ffC^2_{1}, \ffC^2_{2} \right)$ to the volume element with $\left( \lambda^2_1, \lambda^2_2 \right)$ and $\left(\ffC^5_{1}, \ffC^5_{2} \right)$ to the volume element with $\left( \lambda^1_1, \lambda^1_2 \right)$ and so forth. For every quadruple $\left( \ffC^s_{1}, \ffC^s_{2}, \lambda^s_1, \lambda^s_2 \right)$, we compute the associated effective stiffness $\effective{\ffC}^s$ with the help of an FFT-based computational micromechanics code\cite{MoulinecSuquet1994,MoulinecSuquet1998}. The generated data $\set{\left( \effective{\ffC}^s, \ffC^s_{1}, \ffC^s_{2}, \lambda^s_1, \lambda^s_2 \right)}_{i=1}^{N_s}$ serves as training data for identifying the DMN. The number of samples depends on the discretization of the orientation triangle and is summarized in Table~\ref{tab:samples}. For \discone, we generate $800$ samples in total which corresponds to $200$ samples per volume element. To keep the sampling and training effort manageable, we reduce the number of generated samples to $100$ and $50$ per microstructure when increasing the number of discrete orientations to ten and $31$, respectively. For \discone, \disctwo and \discthree, we randomly split the pre-computed samples into a training and validation set, comprising $90\%$ and $10\%$ of the samples. We train the deep material network on mini-batches with a batch size of $N_b=32$ samples. More precisely, we draw the batches randomly from the training set and drop the last batch, should the remaining batch size be smaller than $32$.
\begin{table}[h!]
	\begin{center}
		\begin{tabular}{l r r r r}
			\hline
					   &   \discone & \disctwo & \discthree\\
			\hline
			Total                  &   $800$            &   $\numprint{1000}$           &   $\numprint{1550}$\\
			Per microstructure     &   $200$            &   $100$            &   $50$\\
			\hline
			Training set           &   $720$            &   $900$            &   $\numprint{1395}$\\ 
			Validation set           &   $80$            &   $100$            &   $155$\\ 
			\hline
		\end{tabular}
	\end{center}
	\caption{Number of generated samples and training and validation set sizes}
	\label{tab:samples}
\end{table}\\
We consider deep material networks with eight layers. In general, eight layers are necessary to achieve a sufficient approximation quality, in particular for inelastic computations~\cite{Liu2018, Liu2019, Gajek2020}. We train for $\numprint{3000}$ epochs and rely upon the AMSGrad method~\cite{Adam, AMSGrad} combined with the warm restart technique suggested by Loshchilov-Hutter~\cite{Loshchilov2016}. Furthermore, we make use of a modulation of the learning rate between a minimum learning rate $\alpha_\textrm{min}$ and a maximum learning rate $\alpha_\textrm{max}$, \ie
\begin{equation}
	\alpha: \ffN \rightarrow \ffR, \quad m \mapsto \gamma^m\left(\alpha_\textrm{min} + \frac{1}{2}\left( \alpha_\textrm{max} - \alpha_\textrm{min} \right) \left( 1 + \cos\left(\pi \frac{m}{M}\right) \right)\right),
\end{equation}
where $2M$ corresponds to the period and $M=50$ is chosen. Additionally, we decay the learning rate at a geometric rate with $\gamma=0.999$.\\
Since gradient descent is sensitive \wrt the proper choice of the step size, we determine the learning rates $\alpha_{p}$, $\alpha_{q}$ and $\alpha_{v}$ by a learning rate sweep as introduced by Smith-Topin~\cite{Smith2019}. The resulting learning rates are almost identical for all three parameter groups, and we set $\alpha_\textrm{max} = 1.5 \cdot 10^{-2}$. The minimum learning rate is chosen to be an order of magnitude smaller than the maximum learning rate, \ie $\alpha_\textrm{min} = 1.5 \cdot 10^{-3}$. We sample the initial weights $\vec{v}$ from a uniform distribution on $[0, 1]$ and rescale the weights to sum to unity. The entries of the parameter vectors $\vec{p}$ and $\vec{q}$ are sampled from a uniform distribution on $[0, 2\pi]$.\\
The penalty parameter of the objective function~\eqref{eq:loss_function} is set to $\lambda=10^3$. Additionally, we set the exponents to $p=1$ and $q=10$, \cf{} Gajek et al.~\cite{Gajek2020}, \ie we enforce the maximum of the component-wise mean error to be minimized. To assess the accuracy of the fit, we define the sample-wise error
\begin{equation}
	e_s = \frac{\normg{\DMNLIN{\Lambda}\left(\ffC_1, \ffC_2,  \lambda_1, \lambda_2\right) - \effective{\ffC}^s}_1}{\normg{\effective{\ffC}^s}_1},
\end{equation}
where $\norm{\cdot}_1$ refers to the Frobenius-1 norm defined by the $\ell^1$-norm of the stiffness components in Voigt notation. Additionally, we define the maximum and mean errors of all samples
\begin{equation}
	e_\text{max} = \max_s\left(e_s\right) \quad \textrm{and} \quad e_\textrm{mean} = \frac{1}{N_s} \sum_{s=1}^{N_s} e_s,
\end{equation}
where $N_s$ denotes the number of elements in the training or validation set, depending on the considered scenario.\\
In Fig.~\ref{fig:convergence_offline_FRP}, the training progress for the \discthree orientation discretization and the investigated linear, tri-linear and quadratic orientation interpolations is shown. 
In the first $500$ epochs, the effect of the learning rate modulation becomes apparent. The loss function and mean error fluctuate noticeably. In the last $500$ epochs, the decay of the learning rate ensures convergence of the trained parameters. Conforming to intuition, increasing the degrees of freedom, \ie choosing a tri-linear or quadratic orientation interpolation over a linear orientation interpolation, decreases the loss function at convergence.  This trend carries over to the mean training errors, as well. For the mean validation error, however, the linear orientation interpolation provides the best mean validation error. Such overfitting phenomena are not uncommon for training deep neural networks, where increasing the degrees of freedom not necessarily yields better generalization and validation results. As training progresses, no increasing validation errors can be observed for linear, tri-linear and quadratic orientation interpolation. Thus, no significant model over-fitting is observed during training.\\
In Table~\ref{tab:training_results}, we summarized the training results for the investigated orientation discretization and interpolation. Additionally, we listed the number of non-zero weights $w^i_{K+1}$ at the end of training. In general, we observe the following trends: Mean and maximum training errors will decrease if a tri-linear or quadratic orientation interpolation is chosen instead of a linear interpolation. The same observation holds for the mean validation error. The maximum validation error, on the other hand, does not necessarily decrease by introducing additional fitting parameters. Furthermore, the maximum validation error shows significant fluctuations. Indeed, for \discone and the linear interpolation, it reaches almost 12\%. Since the maximum validation error is dominated by a single sample, \cf{} Liu et al.~\cite{Liu2018, Liu2019} or Gajek et al.~\cite{Gajek2020}, we consider the mean validation error to be a more appropriate indicator of the quality of the training results.\\
If we go from \discone to \disctwo and \discthree, the loss function as well as the training and validation errors will increase, in general. Indeed, the DMN has to predict the effective behavior of significantly more volume elements with different fiber orientations and this result does not come unexpected.
\begin{table}
  \centering
	\begin{tabular}{l l c r r r r c}
				\hline
				 & & \multicolumn{1}{c}{$J$}	 & \multicolumn{1}{c}{$e^\textrm{tr}_\textrm{mean}$} 	& \multicolumn{1}{c}{$e^\textrm{tr}_\textrm{max}$} 	& \multicolumn{1}{c}{$e^\textrm{val}_\textrm{mean}$}  & \multicolumn{1}{c}{$e^\textrm{val}_\textrm{max}$} & \multicolumn{1}{c}{active weights}\\
				\hline
				\multirow{3}{*}{\discone} & Linear	    & 	 $7.378 \cdot 10^{-3}$   & $0.462 \%$	& $1.236 \%$ & $0.669 \%$ & $7.088 \%$ & $71\%$ \\
				                                     & Tri-linear	& 	 $6.456 \cdot 10^{-3}$   & $0.413 \%$	& $1.016 \%$ & $0.572 \%$ & $3.960 \%$ & $73\%$ \\
				                                     & Quadratic	& 	 $6.736 \cdot 10^{-3}$   & $0.431 \%$	& $1.054 \%$ & $0.563 \%$ & $2.948 \%$ & $71\%$ \\
                \hline
				\multirow{3}{*}{\disctwo} & Linear	    & 	 $1.026 \cdot 10^{-2}$   & $0.638 \%$	& $1.610 \%$ & $0.982 \%$ & $11.982 \%$ & $70\%$ \\
				                                     & Tri-linear	& 	 $9.109 \cdot 10^{-3}$   & $0.587 \%$	& $1.266 \%$ & $0.698 \%$ & $4.933 \%$ & $70\%$ \\
				                                     & Quadratic	& 	 $8.156 \cdot 10^{-3}$   & $0.541 \%$	& $1.127 \%$ & $0.665 \%$ & $2.969 \%$ & $68\%$ \\
	            \hline
	            \multirow{3}{*}{\discthree} & Linear	    & 	 $1.296 \cdot 10^{-2}$   & $0.730 \%$	& $2.173 \%$ & $0.783 \%$ & $3.935 \%$ & $67\%$ \\
	                                                 & Tri-linear	& 	 $1.175 \cdot 10^{-2}$   & $0.700 \%$	& $1.928 \%$ & $0.879 \%$ & $6.018 \%$ & $65\%$ \\
	                                                 & Quadratic	& 	 $9.757 \cdot 10^{-3}$   & $0.620 \%$	& $1.361 \%$ & $0.841 \%$ & $7.051 \%$ & $68\%$ \\			                                     
				\hline
	\end{tabular}
	\caption{Training results of the short glass fiber reinforced polyamide}
	\label{tab:training_results}
\end{table}

\subsection{Online evaluation} 
\label{sec:res_online_evaluation}

\begin{figure}[H]
	\centering
	\begin{subfigure}{\textwidth}
		\centering
		\includegraphics[height=5.0mm]{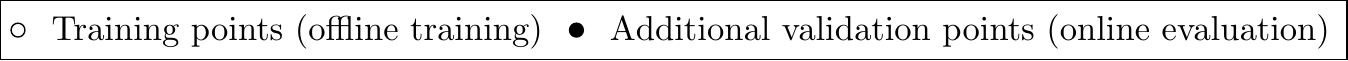}
	\end{subfigure}
	\begin{subfigure}{0.32\textwidth}
		\centering
		\includegraphics[width=\textwidth]{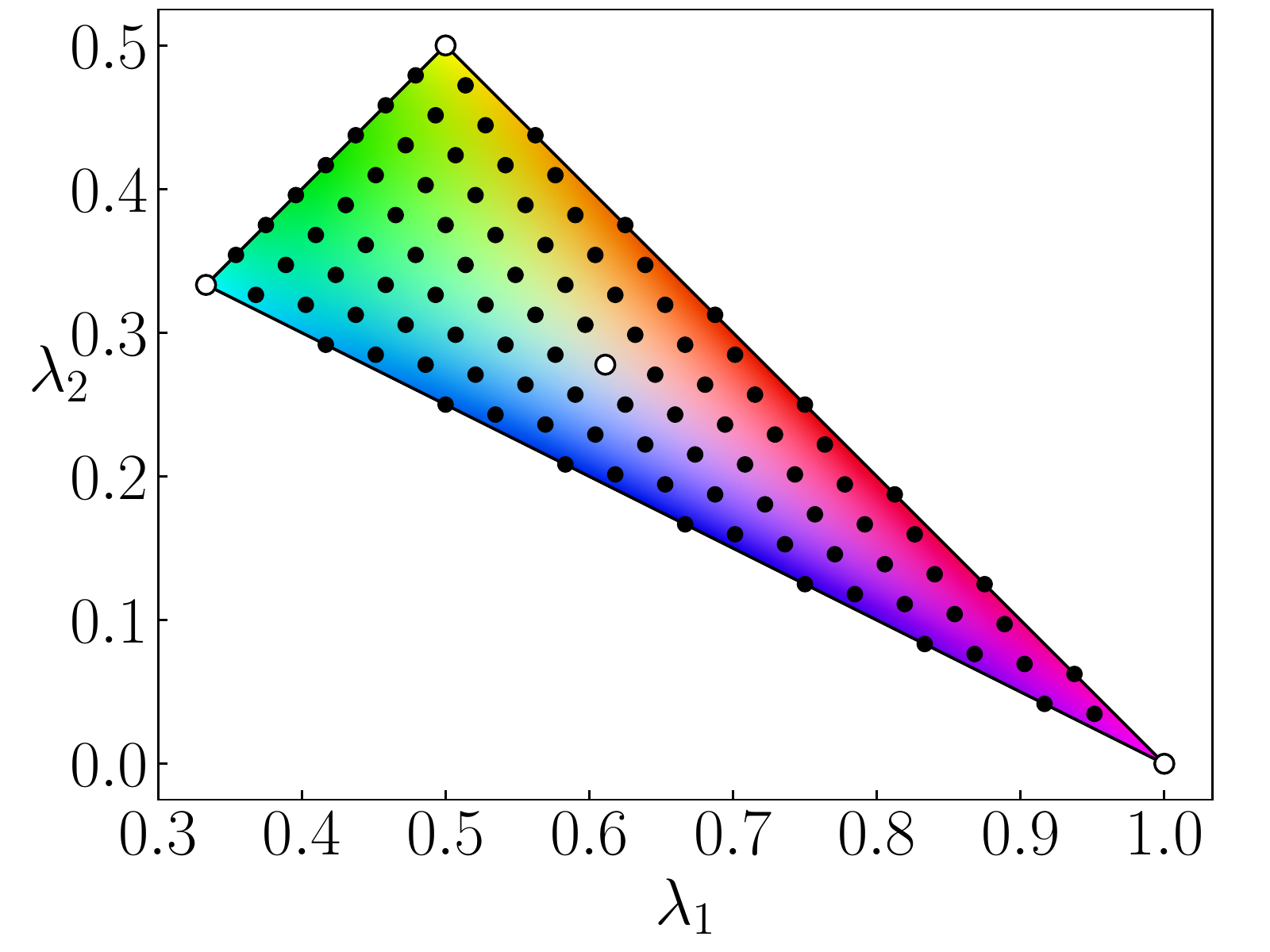}
		\caption{\discone}
		\label{fig:ori_triangle_0_validation}
	\end{subfigure}
	\begin{subfigure}{0.32\textwidth}
		\centering
		\includegraphics[width=\textwidth]{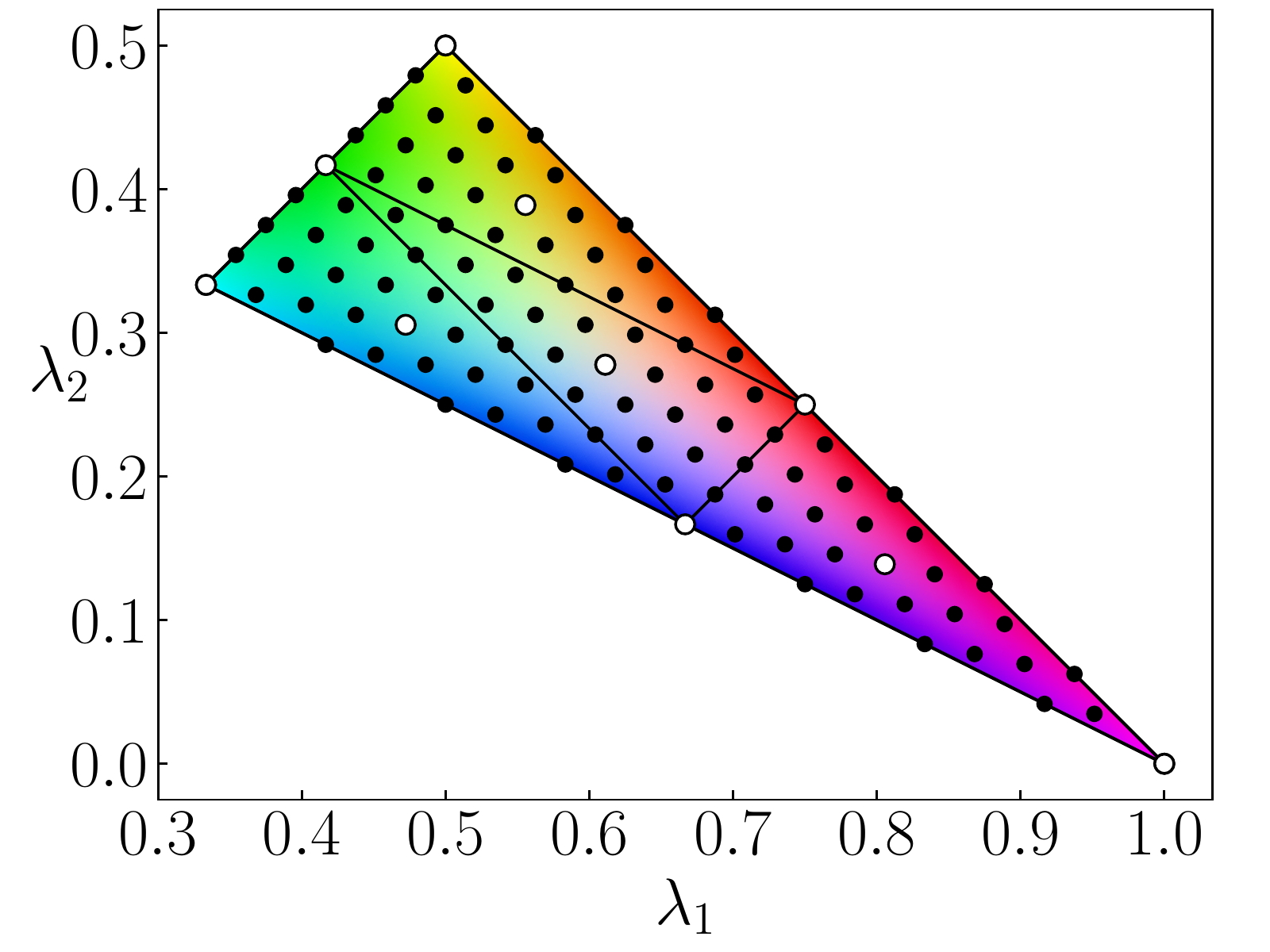}
		\caption{\disctwo}
		\label{fig:ori_triangle_1_validation}
	\end{subfigure}
	\begin{subfigure}{0.32\textwidth}
		\centering
		\includegraphics[width=\textwidth]{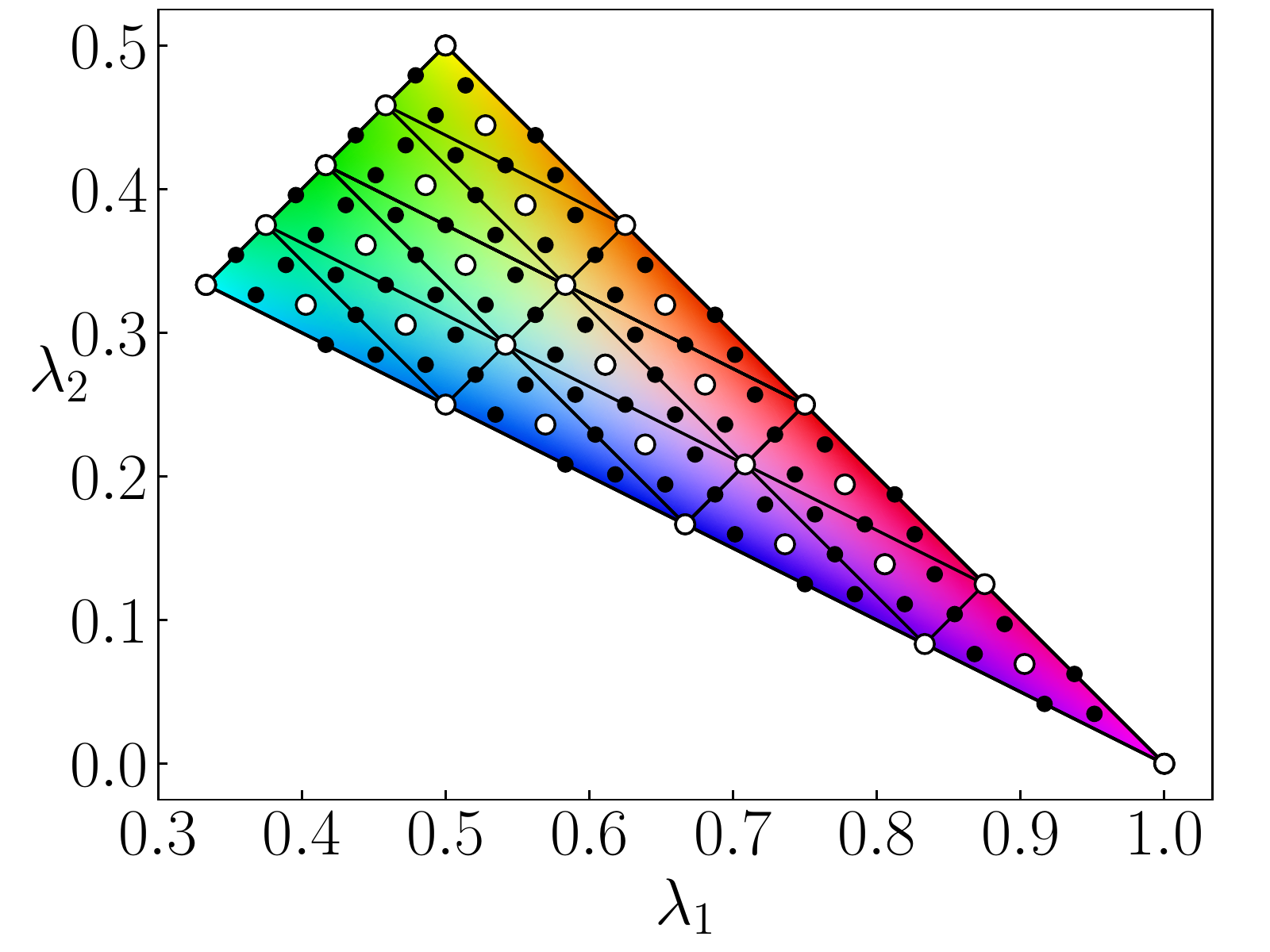}
		\caption{\discthree}
		\label{fig:ori_triangle_2_validation}
	\end{subfigure}
	\caption{Investigated fiber orientation discretizations comprising volume elements used in offline training and additionally generated volume elements for the  online validation: (a) four training and $105$ validation, (b) ten training and $99$ validation and (c) $31$ training and $78$ validation points}
	\label{fig:ori_disc_validation}
\end{figure}

\begin{figure}[H]
	\begin{subfigure}{\textwidth}
		\centering
		\includegraphics[height=5.0mm]{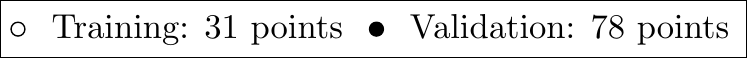}
	\end{subfigure}
	\centering
	\begin{subfigure}{0.48\textwidth}
		\centering
		\includegraphics[width=\textwidth]{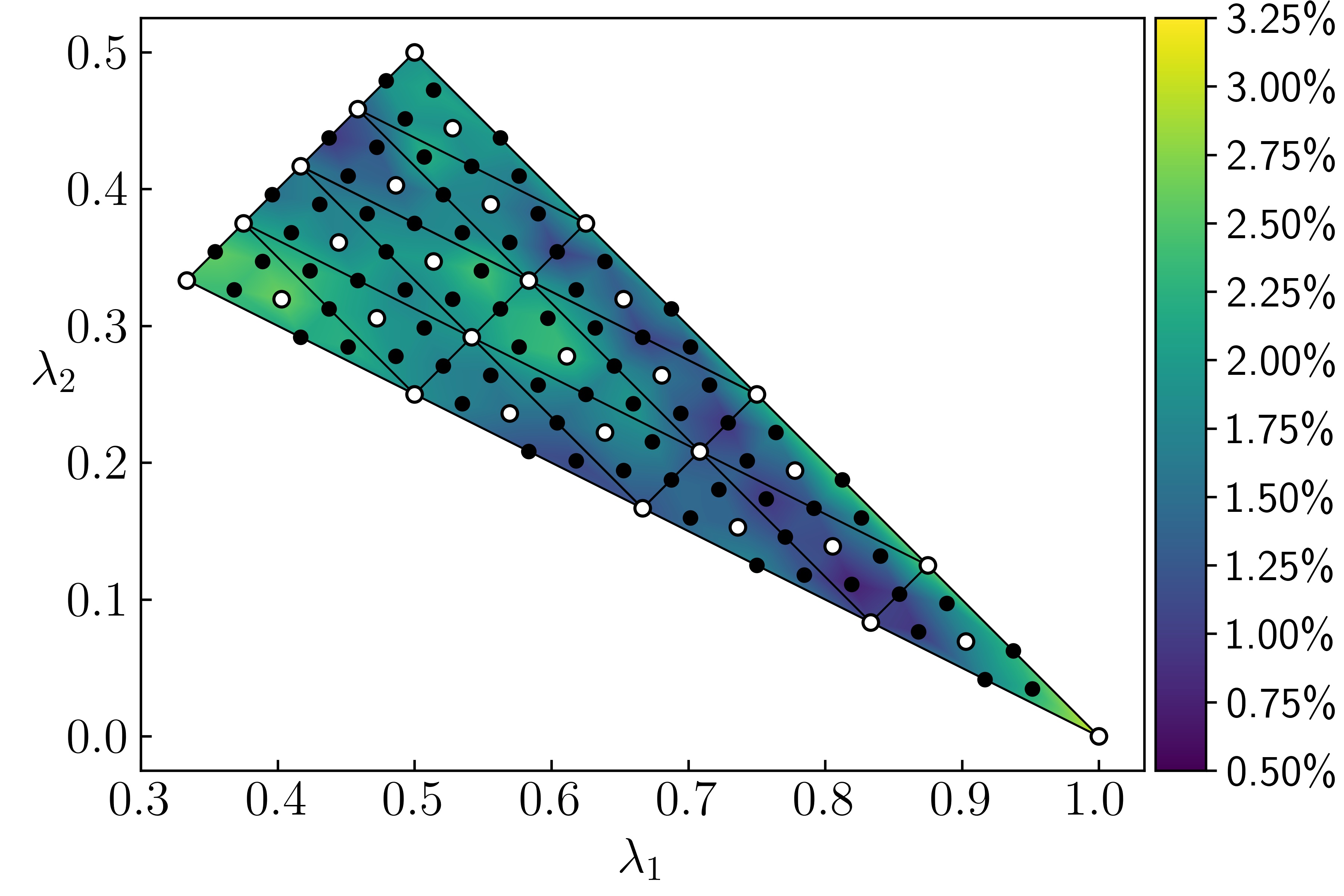}
		\caption{$\eta^\textrm{mean}_{\lambda_1\lambda_2}$}
		\label{fig:error_inelastic_contour_1}
	\end{subfigure}
	\begin{subfigure}{0.48\textwidth}
		\centering
		\includegraphics[width=\textwidth]{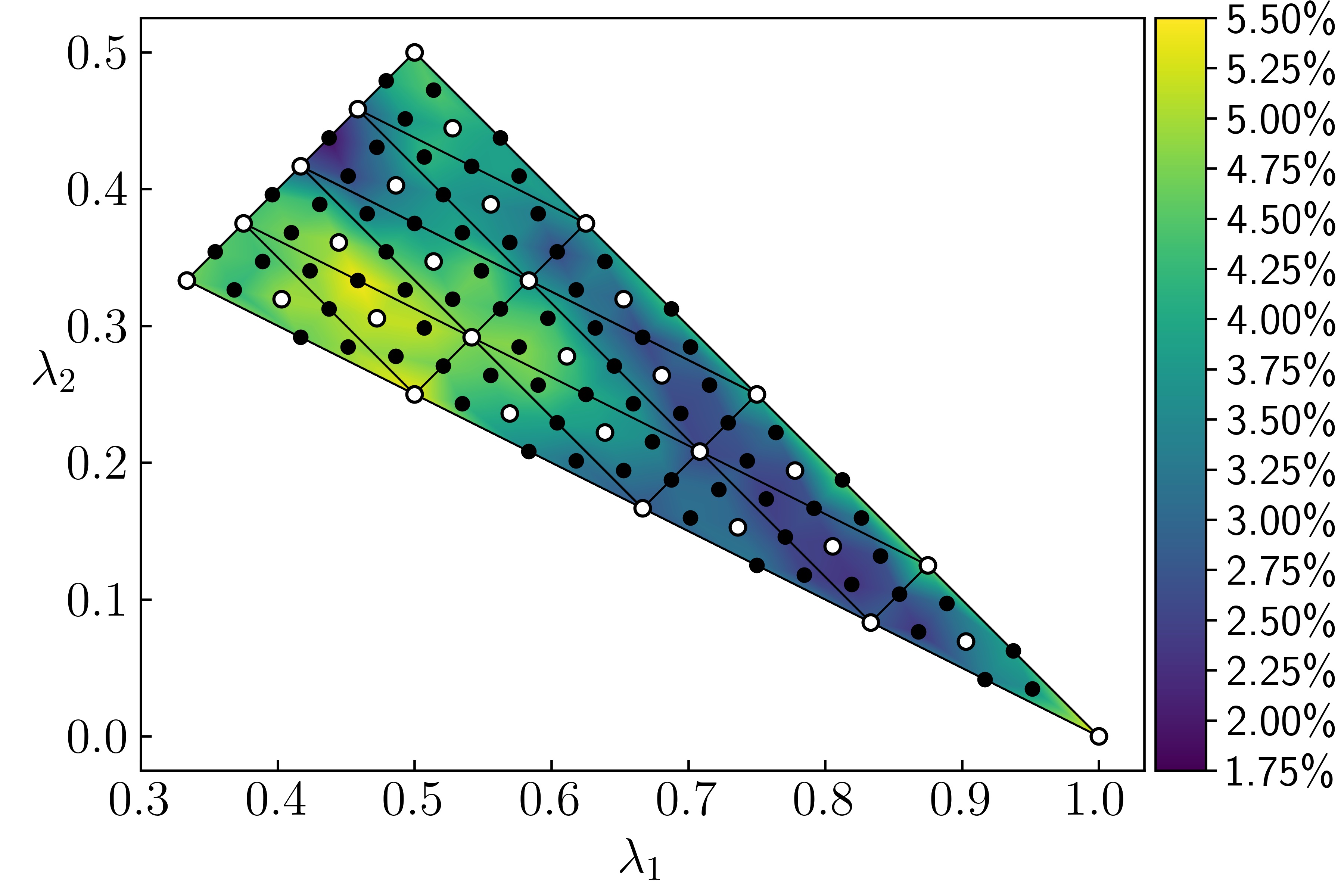}
		\caption{$\eta^\textrm{max}_{\lambda_1\lambda_2}$}
		\label{fig:error_inelastic_contour_2}
	\end{subfigure}
	\caption{Mean and maximum error for \discthree and a linear orientation interpolation}
	\label{fig:error_inelastic_contour}
\end{figure}

\begin{figure}[H]
	\begin{subfigure}{\textwidth}
		\centering
		\includegraphics[height=5.0mm]{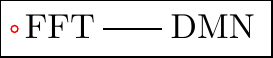}
	\end{subfigure}\\
	\begin{subfigure}[t]{0.32\textwidth}
		\includegraphics[width=\textwidth]{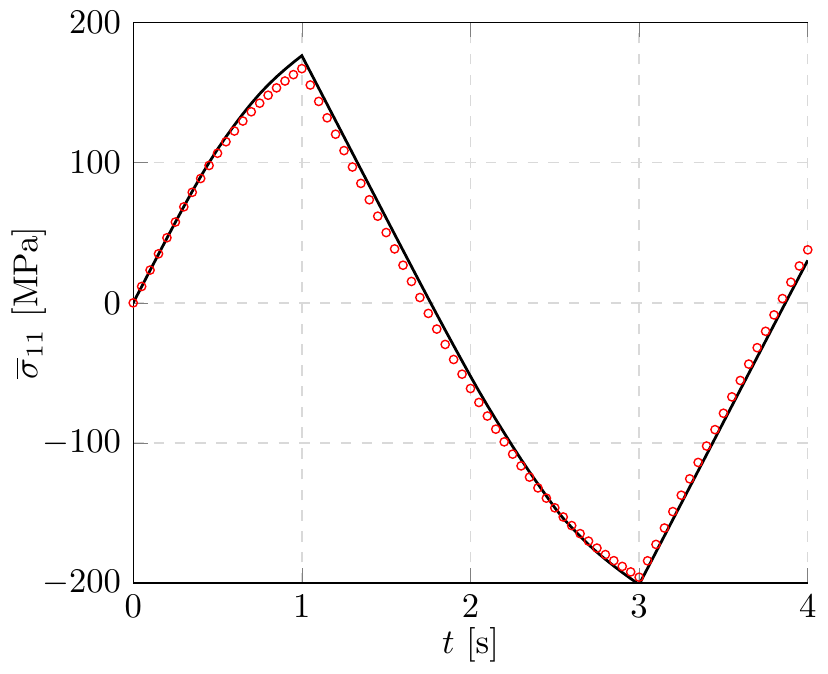}
		\caption{Unidirectional, \cf{} Fig.~\ref{fig:ms_000}}
		\label{fig:effective_stress_comparison_000}
	\end{subfigure}
	\begin{subfigure}[t]{0.32\textwidth}
		\includegraphics[width=\textwidth]{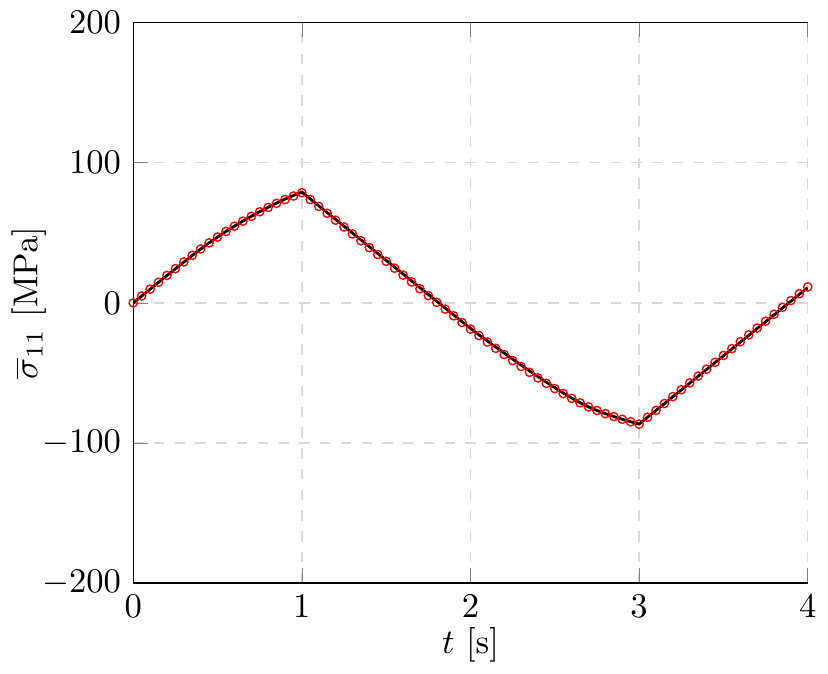}
		\caption{Isotropic, \cf{} Fig.~\ref{fig:ms_001}}
		\label{fig:effective_stress_comparison_001}
	\end{subfigure}
	\begin{subfigure}[t]{0.32\textwidth}
		\includegraphics[width=\textwidth]{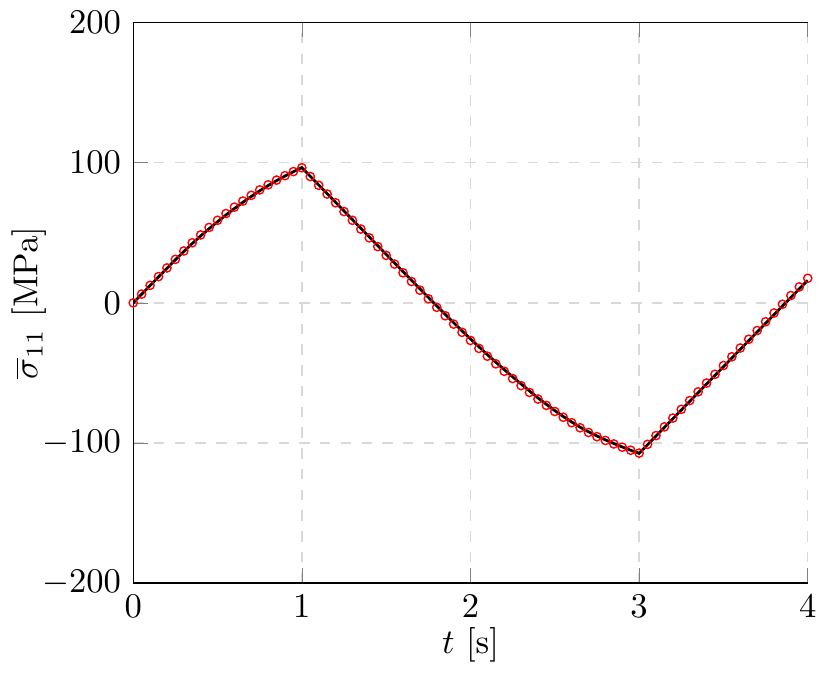}
		\caption{Planar isotropic, \cf{} Fig.~\ref{fig:ms_002}}
		\label{fig:effective_stress_comparison_002}
	\end{subfigure}
	\caption{Comparing the effective stresses under uniaxial extensions in $11$-direction for different fiber orientations}
	\label{fig:effective_stress_comparison}
\end{figure}

\begin{figure}[H]
	\centering
	\begin{subfigure}{\textwidth}
		\centering
		\includegraphics[height=5.0mm]{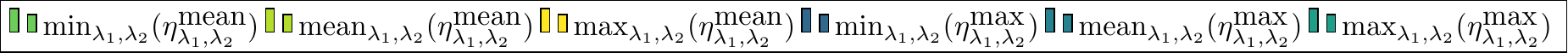}
	\end{subfigure}
	\begin{subfigure}[b]{.32\textwidth}	
		\includegraphics[height=\textwidth]{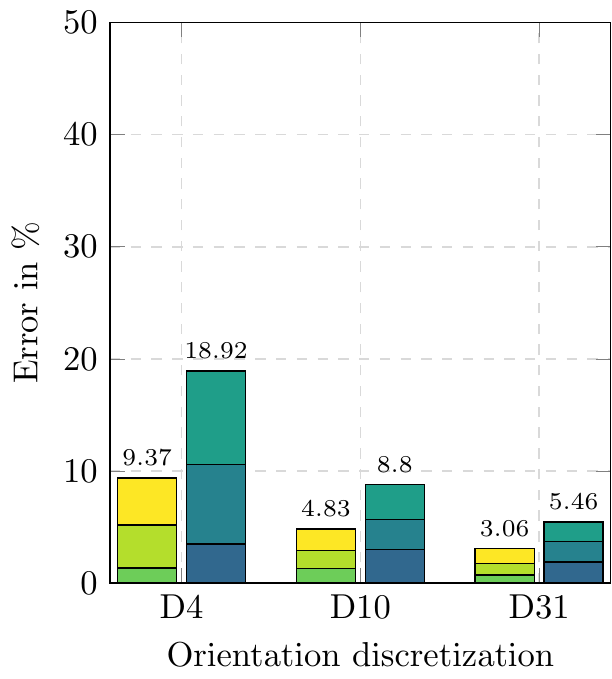}
		\caption{Linear interpolation}
	\end{subfigure}
	\begin{subfigure}[b]{.32\textwidth}	
		\includegraphics[height=\textwidth]{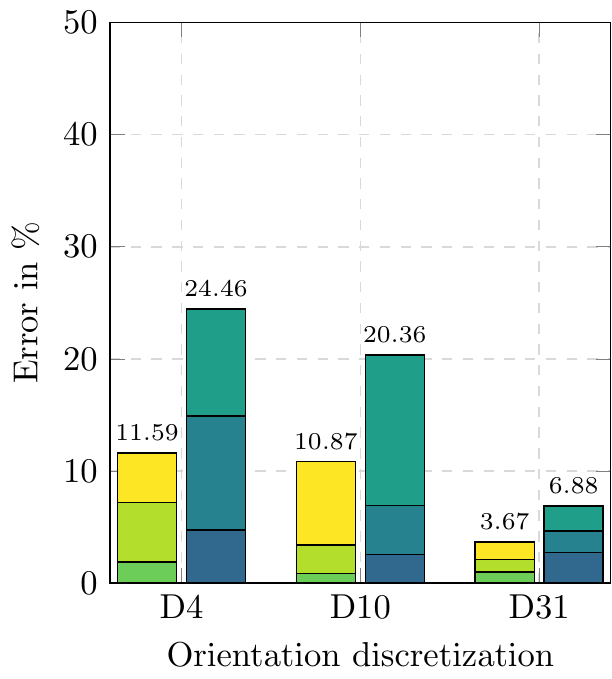}
		\caption{Tri-linear interpolation}
	\end{subfigure}
	\begin{subfigure}[b]{.32\textwidth}	
		\includegraphics[height=\textwidth]{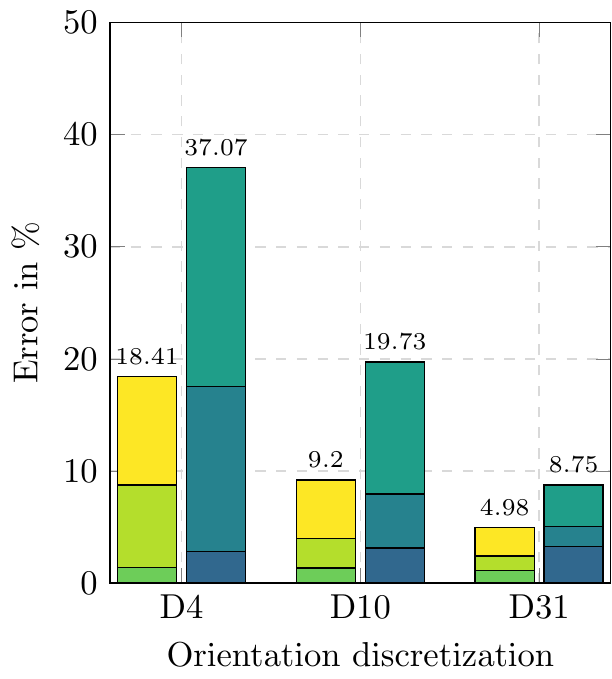}
		\caption{Quadratic interpolation}
	\end{subfigure}
	\caption{Minimum, mean and maximum errors \wrt mean $\eta^\textrm{mean}_{\lambda_1\lambda_2}$ and maximum $\eta^\textrm{max}_{\lambda_1\lambda_2}$ error.}
	\label{fig:error_offline}
\end{figure}

We implemented Newton's method, as described in Section~\ref{sec:online_evaluation}, as a user-material subroutine in \abq. The pseudo-code for the implementation can be found in Gajek et al.~\cite{Gajek2020}. In terms of implementation, the major difference compared to Gajek et al.~\cite{Gajek2020} is that, for the work at hand, the gradient operator depends on the fiber orientation parameters $\lambda_1$ and $\lambda_2$, \cf{} Section~\ref{sec:online_evaluation}. This does not infer any additional challenges, since the microstructure characteristics do not change during computation. Both parameters $\lambda_1$ and $\lambda_2$ are fixed during the online evaluation, and after assembling  $\fA_{\lambda_1\lambda_2}$, the effective stress and algorithmic tangent are computed by Newton's method in the proposed manner.\\
Our goal in Section~\ref{sec:example} is to employ deep material networks for a two-scale simulation using \abq. For this purpose, we seek to speed up the inelastic computations as much as possible. As a first step, the elimination procedure described in Section~\ref{sec:implementation} proves effective. The deleting is performed in an upstream pre-processing step after the offline training to avoid unnecessary computational overhead. Secondly, we exploit the sparsity pattern of all involved linear operators, both, the sparsity pattern of the gradient operator and the Jacobian $\partial \vec{\fmicrostress} / \partial \vec{\fmicrostrain}$, containing the algorithmic tangents of the phases. To this end, we rely upon the library Eigen3~\cite{eigen3} for all linear algebra operations and use sparse matrices whenever possible. For Newton's method, we use the following convergence criterion
\begin{equation}
	\frac{\normg{\fA^T_{\lambda_1\lambda_2} \fW\vec{\fmicrostress}(\vec{\fmacrostrain}^{\, n+1} + \fA_{\lambda_1\lambda_2} \vec{\fa}^{\, n+1}, \vec{\statev}^{\, n})}_\textrm{F}}{(2^K - 1) \normg{\fmacrostress^{\, n+1}}_\textrm{F}} \le \textrm{tol},
\end{equation}
where we set the tolerance $\textrm{tol}$ to $10^{-12}$ and $\norm{\cdot}_\textrm{F}$ refers to the Frobenius norm defined by the $\ell^2$-norm of the involved matrices in Voigt notation. We solve the linear system by means of a sparse Cholesky decomposition.\\
Before employing the DMN in a two-scale simulation, we turn our attention to validating the predicted effective stress of the deep material network. To evaluate the approximation error in a quantitative way, we introduce the following error measures. For fixed orientation parameters $\lambda_1$ and $\lambda_2$, we define the relative error in the stress component $(i,j)$ as
\begin{equation}\label{eq:rel_stress_errors}
	\eta_{ij, \lambda_1\lambda_2}(t) = \frac{\left|\macrostress^{\textrm{DMN}}_{ij, \lambda_1\lambda_2}(t) - \macrostress^{\textrm{FFT}}_{ij, \lambda_1\lambda_2}(t)\right|}{\underset{t \in \mathcal{T}}{\max}\left|\macrostress^{\textrm{FFT}}_{ij, \lambda_1\lambda_2}(t)\right|},
\end{equation}
where $\mathcal{T}=[0,T]$ denotes the considered time interval. Furthermore, the mean and the maximum error are defined by
\begin{equation}
	\eta^\textrm{mean}_{\lambda_1\lambda_2} = \underset{i,j \in \{1,2,3\}}{\max} \frac{1}{T} \int_{0}^{T} \eta_{ij, \lambda_1\lambda_2}(t) \dif t
\end{equation}
and
\begin{equation}
	\eta^\textrm{max}_{\lambda_1\lambda_2} = \underset{i,j \in \{1,2,3\}}{\max} \underset{t \in \mathcal{T}}{\max} \,\, \eta_{ij, \lambda_1\lambda_2}(t),
\end{equation}
respectively. To investigate whether deep material networks are capable of accurately interpolating the effective stress outside of the training regime, additional volume elements were generated. We use \discone as our point of departure and subdivide the orientation triangle three more times yielding $105$ additional sampling points on the orientation triangle. For \disctwo and \discthree, we subdivide the orientation triangle two and one more times, giving rise to $99$ and $78$ additional sampling points, respectively. In Fig.~\ref{fig:ori_disc_validation}, for all three discretizations, the sampling points used for obtaining the training data (hollow circles) and the sampling points exclusively used for the inelastic validations (black filled circles) are shown. For every additional sampling point, we generated a volume element using SAM~\cite{SAM}. We keep using the volume elements already generated for offline training such that we have $109$ generated volume elements in total for every \discone, \disctwo and \discthree. 
Using the material parameters summarized in Section~\ref{sec:mat_param}, we investigate six independent uniaxial strain loadings $\fmacrostrain = \macrostrain \, \fe_i \otimes \fe_j$. In the respective strain direction, a full hysteresis with a strain amplitude of $\macrostrain = 2.5\%$ is computed in $80$ equidistant load steps. As reference, we compute the volume elements' effective stress  $\fmacrostress^{\textrm{FFT}}$ by means of an FFT-based computational micromechanics code and use an Eyre-Milton solver~\cite{EyreMilton1999,EyreMilton2019}.\\ 
For each of those $109$ fiber orientation states, we compute six independent load paths and evaluate the previously defined error measures. The resulting mean and maximum errors for all orientations are shown in Fig.~\ref{fig:error_inelastic_contour}, for \discthree and linear orientation interpolation. Comparing Fig.s~\ref{fig:error_inelastic_contour_1} and \ref{fig:error_inelastic_contour_2}, we observe that the mean error fluctuates less on the orientation triangle than the maximum error, in particular in the vicinity of the isotropic fiber orientation. The maximum error attains its maximum value of around $5.5\%$ relative error close to the isotropic orientation. The mean error, on the other hand, fluctuates less and attains its maximum for the unidirectional case. Still, with a maximum error of $5.5\%$, the DMN is capable of predicting the effective stress of all investigated $109$ discrete fiber orientation states with sufficient accuracy.
Fig.~\ref{fig:effective_stress_comparison} gives an impression of how the mean and maximum errors shown in Fig.~\ref{fig:error_inelastic_contour} translate into actual stress-time curves. For the three extreme cases and an uniaxial extension in the $11$-direction, the $11$-component of the effective stress predicted by the DMN and computed by an FFT-solver is shown. For isotropic and planar isotropic fiber orientations, the DMN's effective stress $\macrostress^{\textrm{DMN}}_{11}$ and the full field solution $\macrostress^{\textrm{FFT}}_{11}$ are almost indistinguishable. Even for the unidirectional fiber orientation, which exhibits a maximum error of about $5\%$, $\macrostress^{\textrm{DMN}}_{11}$ and $\macrostress^{\textrm{FFT}}_{11}$ show a good agreement.
Fig.~\ref{fig:error_offline} summarizes the minimum, mean and maximum of the individual error measures with respect to the orientation parameters $\lambda_1$ and $\lambda_2$ for \discone, \disctwo, \discthree and linear, tri-linear and quadratic orientation interpolation.
Fig.~\ref{fig:error_offline}, which compares $\numprint{5886}$ computed DMN load paths with $654$ full field simulations, permits us to draw the following conclusions. For fixed type of orientation interpolation, a finer orientation discretization reduces the mean and maximum errors. The dependence of the mean and maximum error on the polynomial degree is more complicated. Compared to the linear orientation interpolation, both, the mean and the maximum error increase significantly for tri-linear and quadratic interpolation. The former holds true for \discone, \disctwo and \discthree. Similar to the offline training, the loss and training errors decrease for higher order interpolations. Only the validation errors shows a slight increase. This suggests an overfitting of the online phase, so that the linear approaches to the orientation interpolation is recommended, in the end.

\section{A computational example} 
\label{sec:example}

\begin{figure}[H]
	\centering
	\begin{subfigure}{0.49\textwidth}
		\centering
		\includegraphics[width=\textwidth]{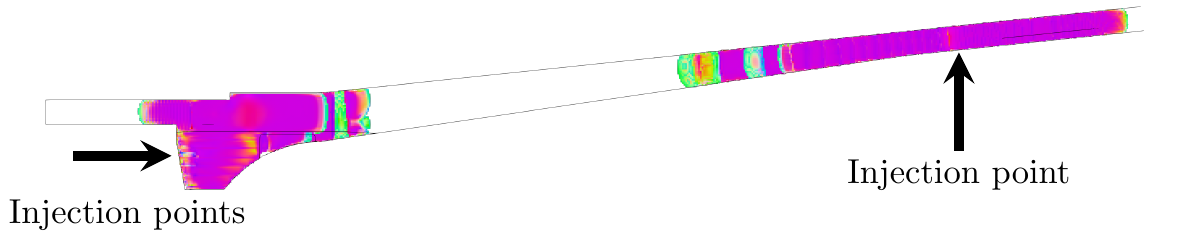}
	\end{subfigure}
	\begin{subfigure}{0.49\textwidth}
		\centering
		\includegraphics[width=\textwidth]{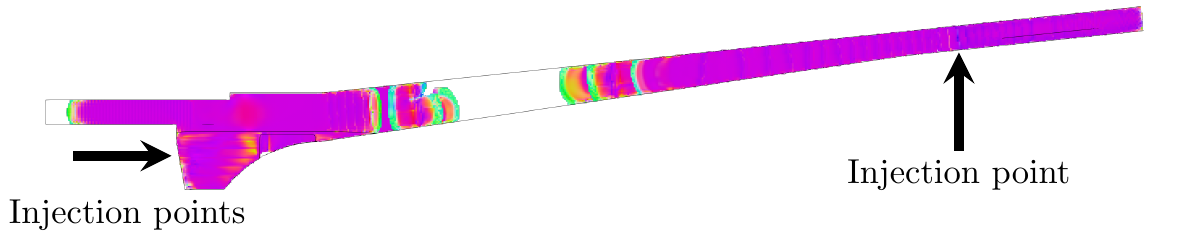}
	\end{subfigure}
	\begin{subfigure}{0.49\textwidth}
		\centering
		\includegraphics[width=\textwidth]{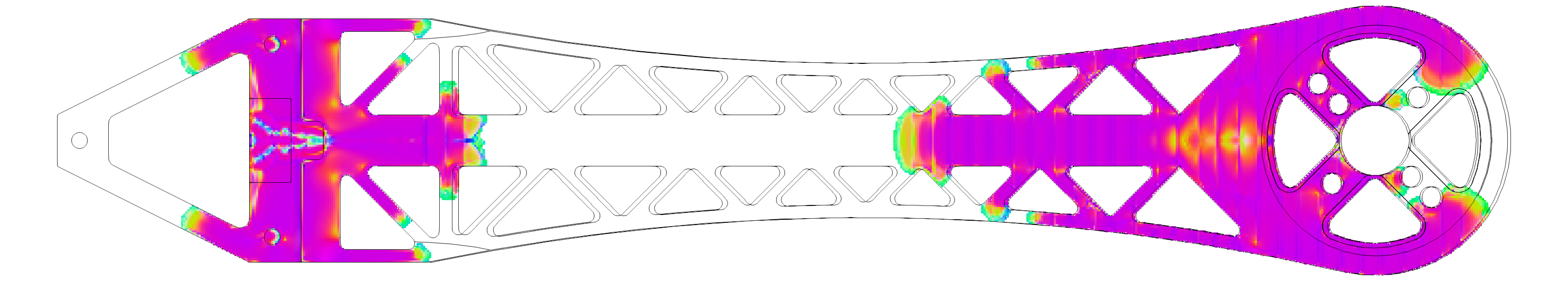}
		\caption{Volume coverage of $50\%$}
		\label{fig:injection_molding_1}
	\end{subfigure}
	\begin{subfigure}{0.49\textwidth}
		\centering
		\includegraphics[width=\textwidth]{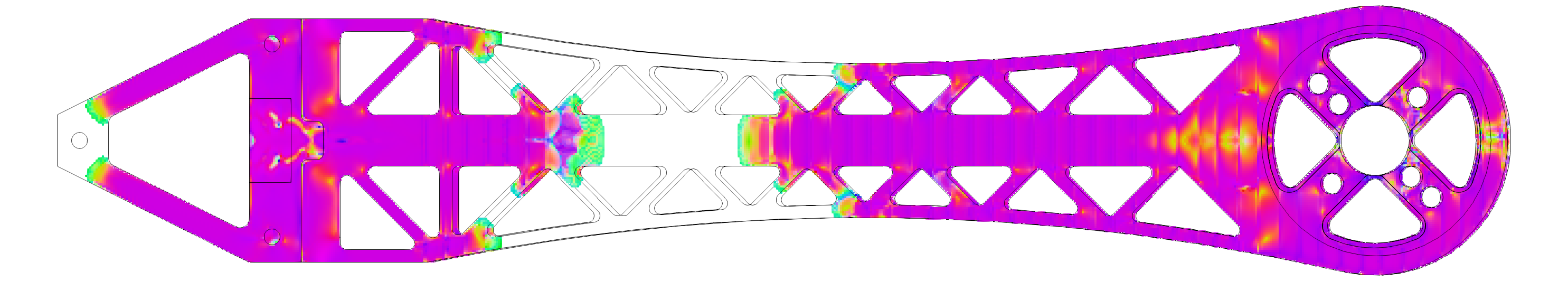}
		\caption{Volume coverage of $75\%$}
		\label{fig:injection_molding_2}
	\end{subfigure}
	\begin{subfigure}{0.49\textwidth}
		\centering
		\includegraphics[width=\textwidth]{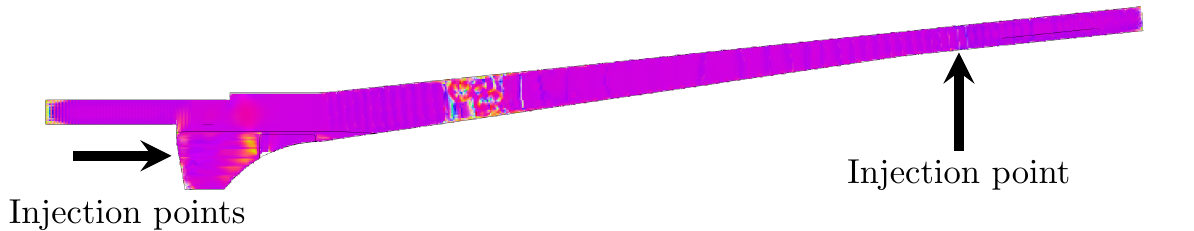}
	\end{subfigure}
	\begin{subfigure}{0.49\textwidth}
		\centering
		\includegraphics[width=\textwidth]{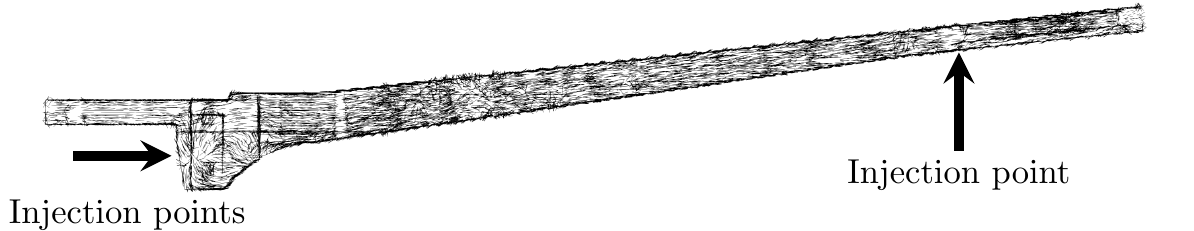}
	\end{subfigure}
	\begin{subfigure}{0.49\textwidth}
		\centering
		\includegraphics[width=\textwidth]{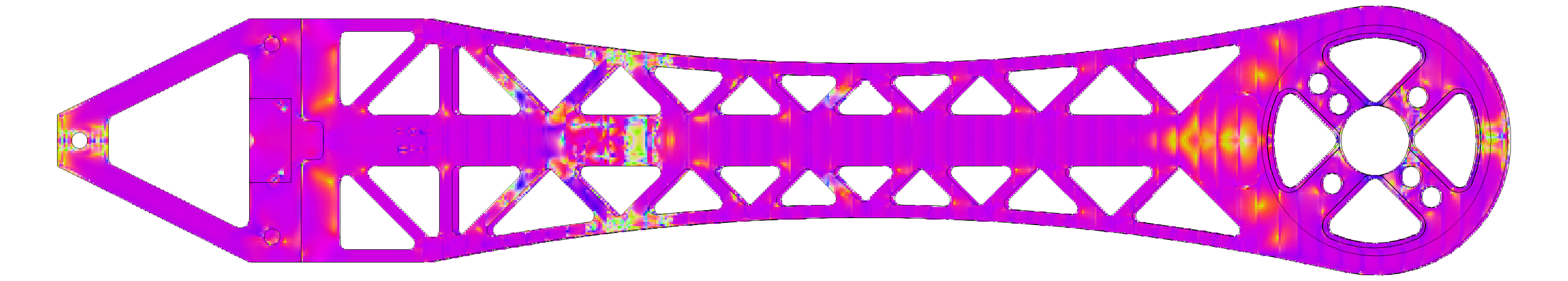}
		\caption{Volume coverage of $100\%$}
		\label{fig:injection_molding_3}
	\end{subfigure}
	\begin{subfigure}{0.49\textwidth}
		\centering
		\includegraphics[width=\textwidth]{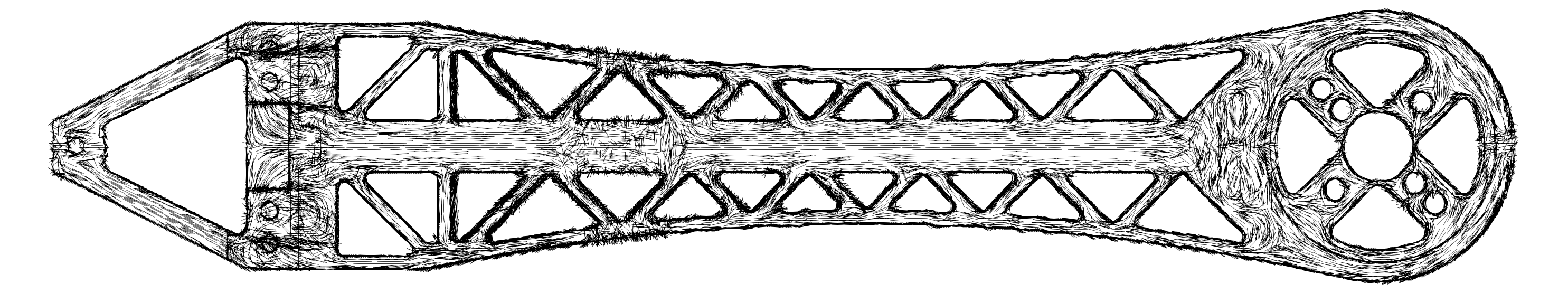}
		\caption{Principal fiber orientation}
		\label{fig:injection_molding_ori}
	\end{subfigure}
	\caption{Injection molding simulation with volume coverage of $50\%$, $75\%$ and $100\%$ and principal fiber orientation after filling}
	\label{fig:injection_molding}
\end{figure}

\begin{figure}[H]
	\centering
	\begin{subfigure}{0.7\textwidth}
		\centering
		\includegraphics[height=10mm]{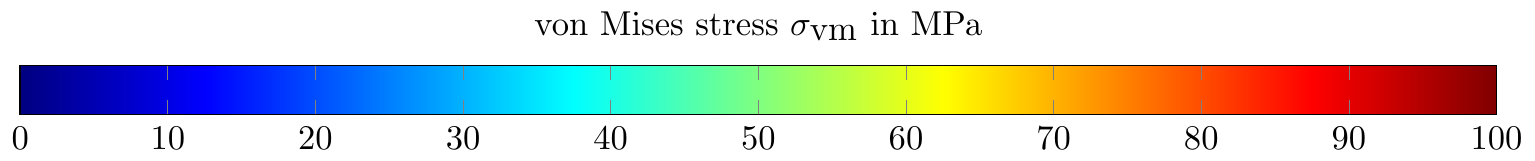}
	\end{subfigure}
	\begin{subfigure}[b]{0.49\textwidth}
		\begin{subfigure}{\textwidth}
			\centering
			\includegraphics[width=\textwidth]{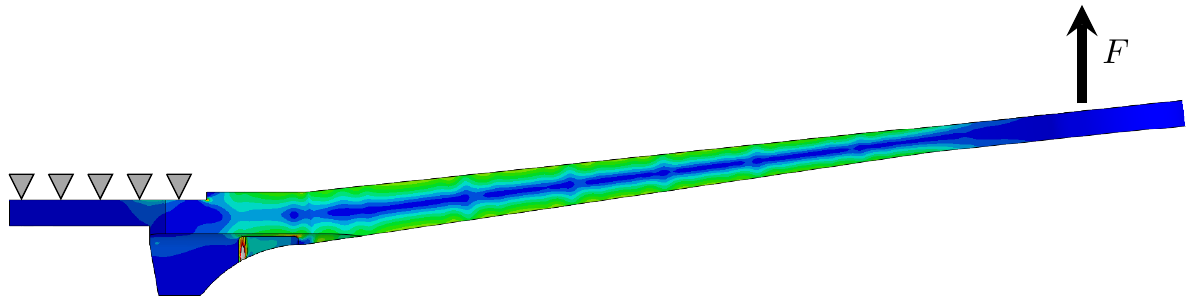}
		\end{subfigure}
		\begin{subfigure}{\textwidth}
			\centering
			\includegraphics[width=\textwidth]{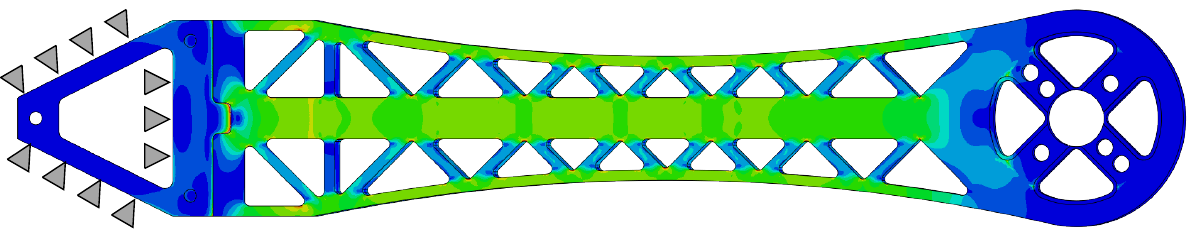}
		\end{subfigure}
		\caption{Homogeneous, isotropic fiber orientation}
	\end{subfigure}
	\begin{subfigure}[b]{0.49\textwidth}
		\begin{subfigure}{\textwidth}
			\centering
			\includegraphics[width=\textwidth]{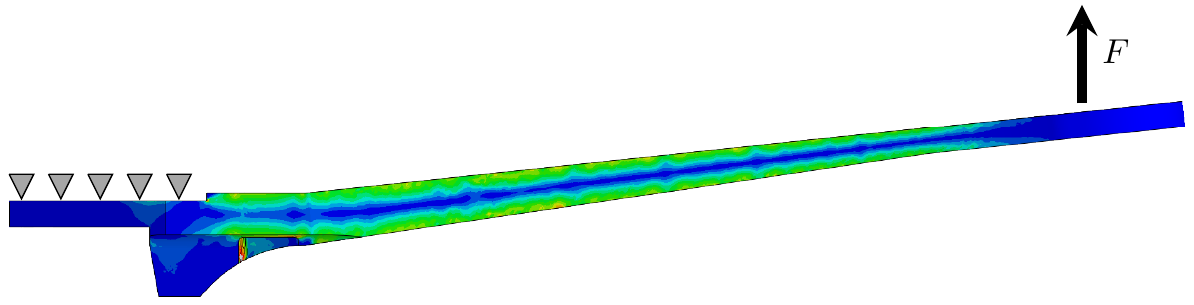}
		\end{subfigure}
		\begin{subfigure}{\textwidth}
			\centering
			\includegraphics[width=\textwidth]{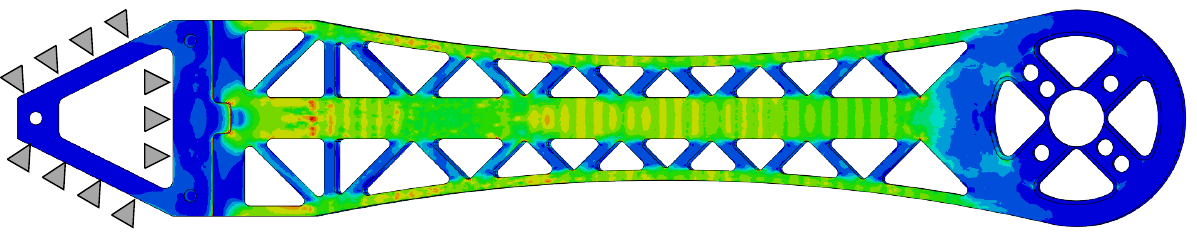}
		\end{subfigure}
		\caption{Mapped anisotropic fiber orientation}
	\end{subfigure}
	\caption{Side and top view of the simulated drone arm}
	\label{fig:drone_arm_abaqus}
\end{figure}

\begin{figure}[H]
	\centering
	\begin{subfigure}{\textwidth}
		\centering
		\includegraphics[height=10mm]{./results/Abaqus/droneFrame/legend.pdf}
	\end{subfigure}
	\begin{subfigure}{0.9\textwidth}
		\centering
		\includegraphics[width=\textwidth]{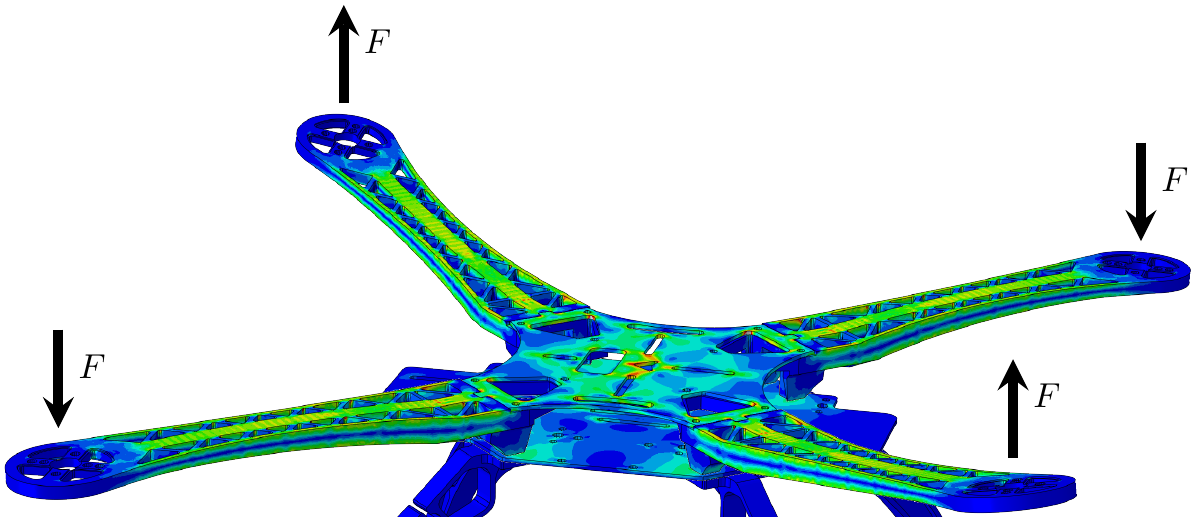}
	\end{subfigure}
	\caption{Simulated quadcopter drone frame}
	\label{fig:drone_frame_abaqus}
\end{figure}

After validating that the DMN is able to to provide sufficiently accurate results, we turn our attention to a component of industrial complexity. We choose a quadcopter frame, \cf{} Fig.~\ref{fig:motivation_quadcopter}, whose CAD geometry is publicly available~\cite{droneFrame}. We assume that the arms of the quadcopter are manufactured by injection molding. To obtain realistic fiber orientation data, we conducted a mold filling simulation for a single quadcopter arm (and utilize these data for all four identical quadcopter arms). We use the publicly available software InjectionMoldingFoam~\cite{ospald2014numerical}, which is based upon the two-phase, incompressible flow solver of OpenFOAM~\cite{OpenFOAMoriginal}. 
We choose identical settings and material parameters as K{\"o}bler et al.~\cite{Kobler2018}, \ie we assume a homogeneous fiber volume fraction and select the following Carreau-WLF equation~\cite{Kennedy2013}
\begin{equation}
	\eta(\theta, \dot{\gamma}) = \eta_0 \frac{e^{ -A_2 \left( \theta - \theta_\textrm{ref}\right) }}{\left(1 + \left(A_0 \dot{\gamma}\right)^2\right)^{\frac{1 - A_1}{2}}}.
\end{equation}
Here, $\theta$ denotes the absolute temperature and $\dot{\gamma}$ refers to the norm of the strain-rate tensor. The parameters for the injection molding simulation are summarized in Table~\ref{tab:parameters_injection_molding} and originally stem from Bhat et al.~\cite{Bhat2014ANALYSISAD}.
\begin{table}[h!]
	\begin{center}
		\begin{tabular}{l l l l}
			\hline
			Density: & $\numprint{1410}$ $\unit{kg/m^3}$ & 			Folger-Tucker coefficient: & $C_\textrm{i}=0.01$\\
			Injection temperature: & $548.15$ $\unit{K}$ & 			Particle number: & $N_\textrm{p}=0$\\
			Mold temperature: & $313.15$ $\unit{K}$ & 			Glass transition temperature: & $\theta_\textrm{ref}=503.15$ $\unit{K}$\\
			Specific heat: & $\numprint{2400}$ $\unit{J/K}$ & 			$A_0$: & $0.1$ $\unit{s}$\\
			Thermal conductivity: & $0.25$ $\unit{W/mK}$ & 			$A_1$: & $0.65$\\
			Initial orientation: & Isotropic & 			$A_2$: & $0.021$ $\unit{1/K}$\\
			Fiber aspect ratio: & $r_a=20$ & 			$\eta_0$: & $100$ $\unit{Pa s}$\\
			\hline
		\end{tabular}
	\end{center}
	\caption{Parameters used in the injection molding simulation~\cite{Bhat2014ANALYSISAD}}
	\label{tab:parameters_injection_molding}
\end{table}
The results of the injection molding simulation are shown in Fig.~\ref{fig:injection_molding}, both for a top and a side view, and at three distinct instances of time, corresponding to a volume coverage of $50\%$, $75\%$ and $100\%$, respectively. The computed fiber orientations are represented by the color scale shown in Fig.~\ref{fig:orienation_triangle_microstructures}. We observe that, at the flow fronts, planar and isotropic fiber orientations dominate. For the rest of the drone arm, fiber orientations close to the unidirectional case are prevalent. Fig.~\ref{fig:injection_molding_ori} illustrates the principal fiber orientations, \ie the eigenvector corresponding to the largest eigenvalue of $\tuckersecond$, after the mold is filled.
As a result of the location of the injection points, pronounced weld lines are formed on the left of the center of the drone arm, see Fig.~\ref{fig:injection_molding_3}.\\
First, we perform a structural simulation on a single drone arm using the FE software \abq. We mesh the drone arm by quadratic tetrahedron elements and investigate five different mesh densities ranging from \numprint{63580} up to \numprint{1005862} elements in order to analyze convergence, wall-clock times and memory requirements, \cf{} Table~\ref{tab:wall_time_online_evaluation_discretization}. The computed fiber orientation tensors serve as the input for the simulation, \ie the eigenvectors of $\tuckersecond$ are mapped onto the \abq mesh and determine the material orientation. The eigenvalues $\lambda_1$ and $\lambda_2$ are provided to the DMN subroutine via pre-defined fields. We apply a loading of $F = 80 \unit{N}$ on the motor mount via a surface force and fix the left side of the drone arm, \cf{} Fig.~\ref{fig:drone_arm_abaqus}.
The loading is applied in ten equidistantly spaced time increments. Fig.~\ref{fig:drone_arm_abaqus} shows the results for the finest discretization of about one million tetrahedron elements. The former took about $165 \unit{min}$ to complete on $96$ threads and required $133 \unit{GB}$ of DRAM. On the left hand side, the von Mises stress distribution for the last time increment and for an assumed homogeneous and isotropic fiber orientation is shown. The right hand side of Fig.~\ref{fig:drone_arm_abaqus} shows the computed stress for the mapped anisotropic, inhomogenous fiber orientation. For the mapped anisotropic fiber orientation, stress fluctuations, especially in the vicinity of weld lines are clearly visible. In contrast, stress and strain concentrations at weld lines cannot be predicted for a homogeneous fiber orientation. Accounting for the entire process chain appears imperative in order to exploit the full lightweight potential of injection-molded fiber-reinforced components, as becomes evident when comparing the predicted total deflections. Indeed, for the assumed isotropic fiber orientation, the macro simulation predicts a deflection of $6.13\unit{mm}$. A deflection of $3.99\unit{mm}$ is predicted for the anisotropic fiber distribution. Thus, the isotropic variant underestimates the actual stiffness of the component by a factor of two.\\
To demonstrate the capabilities of the introduced multiscale method, we investigate the entire drone frame in a mechanical simulation, \cf{} Fig.~\ref{fig:drone_frame_abaqus}. The four drone arms are manufactured from injection molded, short fiber reinforced polyamide with mapped anistotropic fiber orientation. A deep material network is integrated at every Gauss point. Both, the upper and lower plates, which the drone arms are attached to, are made of aluminum. For this material, we use a $J_2$-elastoplasticity model with power law hardening
\begin{equation}
	\sigma_\textrm{Y} = \sigma_0 + k\, \varepsilon_\textrm{p}^m.
\end{equation}
The material parameters are taken from Segurado et al.~\cite{Segurado2002} and summarized in Table~\ref{tab:materialParameters_aluminium}.
\begin{table}[h!]
	\begin{center}
		\begin{tabular}{l l l l l l}
			\hline
			Aluminium & $E=75\unit{GPa}$ & $\nu=0.3$ &$\sigma_Y=75 \unit{MPa}$ & $k=416 \unit{MPa}$ &$m=0.3895$\\
			\hline
		\end{tabular}
	\end{center} 
	\caption{Material parameters of the aluminum plate~\cite{Segurado2002}}
	\label{tab:materialParameters_aluminium}
\end{table}
We assume the drone legs to be made of pure polyamide. For the latter, we assign a linear elastic material behavior, \cf{} Table~\ref{tab:materialParameters_FRP}. The simulation model consists of about two million elements with almost ten million degrees of freedom, \cf{} Table~\ref{tab:wall_time_online_evaluation_abaqus_full_drone}. The drone arms are loaded as shown in Fig.~\ref{fig:drone_frame_abaqus} with a force of $F=80\unit{N}$. The loading is applied in ten equidistant load steps. The simulation took about $4.5 \unit{h}$ wall-clock time on a consumer grade workstation, running on $96$ threads in parallel. In particular, deep material networks enable high-fidelity two-scale simulations with reasonable effort. Indeed, by the results of Section~\ref{sec:res_online_evaluation}, the maximum error of DMNs, evaluated for all investigated fiber orientations and load cases, does not exceed $5.5\%$. For most engineering applications, such an error appears reasonable.
\section{Computational cost}\label{sec:computationl_cost}
Last but not least, we discuss the computational cost of deep material networks accounting for the offline training and online evaluation separately. The material sampling is performed in parallel, \ie we compute six load steps in parallel using $16$ threads for each individual simulation. All deep material networks are trained in parallel on $4$ threads each. The wall-clock times of the material sampling and the offline training are summarized in Table~\ref{tab:wall_time_offline_training}. Apparently, the sampling and offline training effort increases linearly with the number of samples. Incidentally, the number of fitting parameters, which varies depending on the type of the orientation interpolation, has no significant influence on the runtime of the offline training.\\
\begin{table}[h!]
	\begin{center}
		\begin{tabular}{l l c c c c}
			\hline
			& & \multicolumn{3}{c}{Wall-clock time} & \#\textrm{Fitting parameters}\\
			\hline
			 & & \discone & \disctwo & \discthree & \\
			\hline
			\multirow{2}{*}{Sampling} & &  $21.48 \unit{h}$ & $27.02 \unit{h}$ & $42.43 \unit{h}$\\
			                          & &  ($800$ samples) & ($\numprint{1000}$ samples) & ($\numprint{1550}$ samples)\\
			\hline
			\multirow{3}{*}{Training} & Linear & $2.43 \unit{h}$ & $3.18 \unit{h}$ & $4.60 \unit{h}$ & $\numprint{1020}$\\
			& Tri-linear & $2.58 \unit{h}$ & $3.30 \unit{h}$ & $4.60 \unit{h}$ & $\numprint{1275}$\\
			& Quadratic & $2.50 \unit{h}$ & $3.18 \unit{h}$ & $4.62 \unit{h}$ & $\numprint{1785}$\\
			\hline
		\end{tabular}
	\end{center}
	\caption{Wall-clock times for sampling, training and number of fitting parameters}
	\label{tab:wall_time_offline_training}
\end{table}\\
Turning our attention to the online evaluation, we focus on the computational costs of the DMN evaluated at a single Gauss point. Integrating a deep material network at a single Gauss point for a prescribed macro strain increment takes about $2 \unit{ms}$ on a single thread. This is about \numprint{120000} times faster than a full-field simulation of the micro problem using an FFT-based computational mechanics solver on a microstructure discretized by $256^3$ voxels, also running on a single thread. For the work at hand, we exclusively consider DMNs with eight layers. For applications which permit using DMNs with a smaller number of layers, even higher speed-up factors can be reached.
\begin{table}[h!]
	\begin{center}
		\begin{tabular}{l c c c c c c }
			\hline
			& FFT ($1$ thread) & DMN ($1$ thread) \\
			\hline
			Wall-clock time		& $242.69 \unit{s}$ & $2.03 \unit{ms}$\\
			Speed-up		& $-$ & $\numprint{119552}$\\
			\hline
			\#DOF		& $6 \times 256^3$ & $513$ \\
			\hline
		\end{tabular}
	\end{center}
	\caption{Wall-clock times and speed-up (compared to an FFT-base computational micromechanics solver) for a single time step of the inelastic micro simulation}
	\label{tab:wall_time_online_evaluation_gauss_point}
\end{table}\\
Next, we focus on the component scale simulation of the quadcopter arm. The wall-clock times of all five examined discretizations ranging from \numprint{63580} up to \numprint{1005862} quadratic tetrahedron elements and, computed on $96$ threads, are summarized in Table~\ref{tab:wall_time_online_evaluation_discretization}. 
\begin{table}[h!]
	\begin{center}
		\begin{tabular}{l r r r r r}
			\hline
			& \multicolumn{5}{c}{\abq ($96$ threads)}\\
			\hline
			Elements	& \numprint{63580} & \numprint{121416} & \numprint{247444} & \numprint{488689} & \numprint{1005862}\\
			\#DOF		& \numprint{308987} & \numprint{572688} & \numprint{1134597} & \numprint{2194091} & \numprint{4418695}\\
			\hline
			Wall-clock time		& $9\unit{min}$ & $15\unit{min}$ & $32\unit{min}$ & $63\unit{min}$ & $165\unit{min}$\\
			Memory consumption		& $9 \unit{GB}$ & $15 \unit{GB}$ & $33 \unit{GB}$ & $62 \unit{GB}$ & $133 \unit{GB}$\\
			\hline
		\end{tabular}
	\end{center}
	\caption{Wall-clock time and memory consumption of the single drone arm for different mesh sizes}
	\label{tab:wall_time_online_evaluation_discretization}
\end{table}
We observe that the DRAM footprint is roughly proportional to the number of elements. The wall-clock times, however, increase super-linearly. We attribute this effect to the complexity of the direct solver used by ABAQUS. Apparently, the applicability of the method is more restricted by the memory requirements, and the computational effort plays a minor role. The required DRAM depends on the number of internal variables to be stored. For a DMN of eight layers, linear elastic fibers and an elastoplastic matrix, $128 \times (1 + 5)=768$ floating-point numbers need to be stored for every Gauss point. To improve the convergence of Newton's method, the displacement jumps of the last converged time step are stored as well, \ie  $768+ 255 \times 3 = \numprint{1533}$ scalars need to be kept in memory. Since we rely upon the thinned binary tree as introduced in Section~\ref{sec:online_evaluation}, $\numprint{1533}$ serves as an upper bound. For the application at hand, the actual number of internal variables of the DMN surrogate model is $\numprint{1297}$.\\
To analyze the drone arm, using less than one million elements for the discretization would be sufficient. Rather, by choosing such a fine discretization, we demonstrate that deep material networks easily scale to component scale simulations with higher complexity. The hardware requirements implicated by Table~\ref{tab:wall_time_online_evaluation_discretization} can be provided by any state-of-the-art workstation.\\
Computing all ten load steps on $96$ threads for the entire quadcopter took $267\unit{min}$ and required $252 \unit{GB}$ or DRAM, \cf{} Table~\ref{tab:wall_time_online_evaluation_abaqus_full_drone}. This corresponds to over two million elements and about ten million degrees of freedom. In our opinion, this clearly shows that DMNs are a promising technique to enable two-scale simulations of industrial complexity with manageable resource expenditure. 
\begin{table}[h!]
	\begin{center}
		\begin{tabular}{l c r c}
			\hline
			Part & Materials & \multicolumn{2}{c}{Discretization}\\
			\hline
			Arms	& DMN & $ 4 \times \numprint{488689}$ & Quadratic tetrahedron elements\\
			Bottom plate & Aluminum  & $\numprint{39422}$ & Quadratic hexahedron elements\\
			Top plate & Aluminum & $\numprint{19028}$ & Quadratic hexahedron elements\\
			Legs & Polyamide & $ 4 \times \numprint{20054}$ & Quadratic tetrahedron elements\\
			\hline
			Total & - & $\numprint{2093422}$ & -\\
			\hline
			\#DOF		& \multicolumn{3}{r}{$\numprint{9378683}$}\\
			Wall-clock time		& \multicolumn{3}{r}{$267 \unit{min}$}\\
			Memory consumption		& \multicolumn{3}{r}{$252\unit{GB}$}\\
			\hline
		\end{tabular}
	\end{center}
	\caption{Wall-clock time and memory consumption for the simulation of the entire quadcopter frame}
	\label{tab:wall_time_online_evaluation_abaqus_full_drone}
\end{table}

\section{Conclusion}

In this work, we investigated the capabilities of deep material networks to provide a digital twin for short fiber reinforced plastic microstructures, which can be used in concurrent multiscale simulations. 
To realize the full lightweight potential of short fiber reinforced components, it is imperative to account for the locally varying fiber orientation in mechanical simulations on component scale. Building upon the work of Köbler et al.~\cite{Kobler2018}, we proposed a robust and computationally efficient approach to utilize direct deep material networks for variable fiber orientations. Instead of identifying multiple deep material networks and interpolating the effective stress, we interpolated the DMN's microstructure characteristics on the fiber orientation triangle. Assuming that the local fiber volume fractions of the individual laminates in the hierarchy are independent of the local fiber orientation, it suffices to fix the fiber volume fraction and to interpolate the lamination directions only. This procedure gives rise to a single DMN surrogate model covering all fiber orientations. Presumably, the scheme easily extends to incorporating local variations in the fiber volume fraction by interpolating the DMN's volume fractions as well. By sampling the training data from up to $31$ microstructure realizations with different fiber orientation, we fitted the DMN to multiple fiber orientations simultaneously. Subsequently, we showed that the DMN generalizes to the entire fiber orientation triangle with small error, also for the inelastic regime.\\
To evaluate the ensuing performance of our approach, we simulated the entire process chain of a quadcopter frame starting from an injection molding simulation. We mapped the computed fiber orientations upon a finite element mesh of the complete quadcopter frame and conducted a two-scale simulation of the full component. Our results indicate that deep material networks enable two-scale simulations of structures with industrial complexity with moderate hardware requirements. In this way, the \textrm{FE}-\textrm{DMN} method finally realizes the promise of concurrent multiscale simulations and constitutes a powerful piece of technology which promises to become a standard tool for industrial applications.\\
It should be interesting to extend the \textrm{FE}-\textrm{DMN} method to problems involving damage or fracture~\cite{PhasefieldFracture2019,HomFrac2019,SandMultiPhysics}, finite strains~\cite{fiberOrientationIntegration,compositeVoxelHyperelastic} or thermomechanical coupling~\cite{Chatzi2016,ThermoWizard2020} and to extend the range of applicability, for instance in the context of polycrystalline materials~\cite{Tessellator2020,DualCrystalPlasticity2019} or sheet molding compound composites~\cite{Gorthofer2019,SMC2019}.\\
Despite the apparent success in practice, there is still a need for theoretical results which shed light on the approximation capabilities (and the limitations) of DMNs. Indeed, whether every fixed two-phase microstructure has a microstructure twin of hierarchical laminate with identical effective properties, appears to be unresolved, \cf{} Problem $4$ in Milton~\cite{milton2020open}. Interestingly, there exist counter examples for five-phase composites where the former is false, \cf{} Milton~\cite{Milton1986}. 

\section*{Acknowledgements}
{SG}, {MS} and {TB} acknowledge financial support by the German Research Foundation (DFG) within the International Research Training Group “Integrated engineering of continuous-discontinuous long fiber reinforced polymer structures” (GRK 2078/2). The support by the German Research Foundation (DFG) is gratefully acknowledged.\\

\bibliographystyle{ieeetr}
{\footnotesize\bibliography{literature}}

\end{document}